\documentclass[fleqn,usenatbib,useAMS]{mnras}

\usepackage{newtxtext,newtxmath}

\usepackage[T1]{fontenc}

\DeclareRobustCommand{\VAN}[3]{#2}
\let\VANthebibliography\thebibliography
\def\thebibliography{\DeclareRobustCommand{\VAN}[3]{##3}\VANthebibliography}

\usepackage{graphicx}	
\usepackage{subfigure}
\usepackage{amsmath}	
\usepackage{amssymb}	

\usepackage{gensymb}    

\usepackage[thicklines]{cancel}
\newenvironment{sequation}{\begin{equation}\scriptsize}{\end{equation}}

\title[Steepness statistics]{Weak lensing peak statistics -- steepness versus height}

\author[Z.W.Li et al.]{Ziwei Li,$^{1}$
Xiangkun Liu,$^{1}$\thanks{E-mail: liuxk@ynu.edu.cn}
Zuhui Fan$^{1}$\thanks{E-mail: zuhuifan@ynu.edu.cn}
\\
$^{1}$South-Western Institute for Astronomy Research, Yunnan University, Kunming 650500, China
}

\date{Accepted XXX. Received YYY; in original form ZZZ}

\pubyear{2015}

\begin{document}
\hypersetup{draft}
\label{firstpage}
\pagerange{\pageref{firstpage}--\pageref{lastpage}}
\maketitle

\begin{abstract}
In weak-lensing cosmological studies, peak statistics is sensitive to nonlinear structures and thus complementary to cosmic shear two-point correlations. In this paper, we explore a new approach, namely, the peak steepness statistics, with the overall goal to understand the cosmological information embedded there in comparison with the commonly used peak height statistics. We perform the analyses with ray-tracing simulations considering different sets of cosmological parameters $\Omega_{\rm m}$ and $\sigma_8$. A theoretical model to calculate the abundance of high peaks based on steepness is also presented, which can well describe the main trend of the peak distribution from simulations. We employ $\Delta\chi^2$ and Fisher analyses to study the cosmological dependence of the two peak statistics using our limited sets of simulations as well as our theoretical model. Within our considerations without including potential systematic effects, the results show that the steepness statistics tends to have higher sensitivities to the cosmological parameters than the peak height statistics and this advantage is diluted with the increase of the shape noise. Using the theoretical model, we investigate the physical reasons accounting for the different cosmological information embedded in the two statistics. Our analyses indicate that the projection effect from large-scale structures plays an important role to enhance the gain from the steepness statistics. The redshift and cosmology dependence of dark matter halo density profiles also contributes to the differences between the two statistics. 
\end{abstract}




\section{Introduction}

Weak lensing (WL) effects result from gravitational deflections of light paths by large-scale structures in the Universe, and carry important physical information \citep[e.g.,][]{2001PhR...340..291B,2014RAA....14.1061F,2015RPPh...78h6901K}.   
However cosmic shear signals, the WL induced shape distorsions of background galaxies, are only at the percent level or less, much smaller than their intrinsic ellipticities. Thus WL analyses can only be done statistically demanding large shear samples \citep[e.g.,][]{2018ARA&A..56..393M}.  

The rapid development of large photometric surveys for the past two decades has made WL cosmology possible. From Stage II surveys such as the Canada–France–Hawaii Telescope Lensing Survey \citep[CFHTLenS;][]{2012MNRAS.427..146H} to the current Stage III surveys of the Kilo Degree Survey \citep[KiDS;][]{2015MNRAS.454.3500K}, 
the Dark Energy Survey \citep[DES;][]{2016MNRAS.460.1270D} and the Hyper Suprime-Cam Subaru Strategic Program \citep[HSC-SSP;][]{2018PASJ...70S...4A}, WL analyses have derived important cosmological constraints 
and become one of the major probes in cosmology \citep{2013MNRAS.430.2200K,2017MNRAS.465.1454H,2021A&A...646A.140H,2018PhRvD..98d3528T,2022PhRvD.105b3520A,2019PASJ...71...43H}. 
With the next generation of Stage IV surveys being in operation in a few years, including the Vera Rubin Observatory Legacy Survey of Space and Time \citep[LSST;][]{2019ApJ...873..111I}, 
{\it Euclid} \citep{2011arXiv1110.3193L}, the Roman Sapce Telescope \citep[Roman;][]{2015arXiv150303757S} and the China Space Station Telescope \citep[CSST;][]{2011SSPMA..41.1441Z,2019ApJ...883..203G,2023A&A...669A.128L}, 
WL studies will enter the era of precision cosmology aiming to deliver critical constraints on 
fundamental quantities, such as the nature of dark energy and the physical properties of neutrinos.       

For WL statistics, the most extensively applied analyses are power spectra or cosmic shear two-point correlations \citep[e.g.,][]{2013MNRAS.430.2200K,2020A&A...633A..69H}. On the other hand, the nonlinear formation process of large-scale structures leads to non-Gaussianity of the cosmic matter distribution, especially on small scales. 
Thus additional statistical means are needed in order to fully uncover the cosmological information in WL data. Higher-order correlation studies have been utilised in some observational analyses \citep[e.g.,][]{2013MNRAS.433.3373V,2014MNRAS.441.2725F,2022PhRvD.105j3537S}. Recently, peak statistics have drawn increasing attention, and been carried out
using data from different surveys \citep[e.g.,][]{2012ApJ...748...56S,2014MNRAS.442.2534S,2015PhRvD..91f3507L,2015MNRAS.450.2888L,2016MNRAS.463.3653K,2016PhRvL.117e1101L,2018MNRAS.474.1116S,2018MNRAS.474..712M,2021PASJ...73..817O,2022MNRAS.511.2075Z,2023MNRAS.519..594L}. Minkowski functionals have also been proposed to extract non-Gaussian information 
\citep[e.g.,][]{2012PhRvD..85j3513K,2013PhRvD..88l3002P,2015PhRvD..91j3511P,2022arXiv220603877G}. Besides these summary statistics, convolutional neural networks (CNN) have also been applied to simulated WL maps to explore the power of machine learning in extracting cosmological information 
\citep[e.g.,][]{2018PhRvD..97j3515G,2019NatAs...3...93R,2019MNRAS.490.1843R,2019PhRvD.100f3514F,2020PhRvD.102l3506M,2021MNRAS.504.1825S,2022PhRvD.105h3518F}.
Recently, \cite{2020MNRAS.499.5902C} put forwards the analyses of scattering transform and in \cite {2021A&A...645L..11A} a statistics called the starlet $l_1$-norm was introduced. They can both generate a set of coefficients as summary statistics by processing WL maps for cosmological studies. 
 
For peak analyses, most of the studies concern peak abundances in different bins based on the signal-to-noise ratio of peak heights. By performing CNN investigations, \cite{2019NatAs...3...93R} show that the profile gradients around peaks, referred to as the steepness, 
carry important cosmological information, and the peak statistics based on the steepness    
can give better cosmological constraints than that based on the peak height. Along this line, in this paper we carry out systematic studies to compare the two peak statistics using numerical simulations. We further present a theoretical model to calculate the peak abundances in terms of the steepness for high peaks based on our
previous studies \citep{2010ApJ...719.1408F,2018ApJ...857..112Y}. Importantly, using the model, we investigate the physical causes that lead to the differences of the two peak statistics. 

The paper is organized as follows. In Sec. \ref{chapter2}, we present the basics of the WL theory and our model for high peak steepness statistics. Sec. \ref{chapter3} describes the simulations used in the analyses, and the corresponding results are shown in Sec. \ref{chapter4}. In Sec. \ref{chapter5}, we compare our model predictions with that from simulations aiming to understand the differences between the two peak statistics. Summary and discussions are given in Sec. \ref{chapter6}. 

\section{Theoretical aspects}
\label{chapter2}
\subsection{Weak lensing effects}\label{sec:Weak lensing convergence field}
Large-scale structures in the Universe gravitationally deflect light rays from distant sources inducing lensing effects. The mapping between the position of a source in the source plane $\boldsymbol \beta$ and that in the observed image plane $\boldsymbol \theta$ is described by the lens equation given by
\begin{equation}
\boldsymbol{\beta}=\boldsymbol{\theta}-\boldsymbol{\alpha},
\label{eq:lenseq}
\end{equation}
where $\boldsymbol{\alpha}$ is the reduced deflection angle and $\boldsymbol{\alpha}=\nabla \psi$ with $\psi$ being the lensing potential. For an extended galaxy, the weak-lensing effects can be characterised by the Jocobian matrix 
\begin{equation}
\mathbf{A}=\left(\begin{array}{cc}{1-\kappa-\gamma_{1}} & {-\gamma_{2}} \\ {-\gamma_{2}} & {1-\kappa+\gamma_{1}}\end{array}\right),
\label{eq:Jacobian}
\end{equation}
where $\kappa$ is the convergence resulting in isotropic changes of lensed images, and $\gamma_1$ and $\gamma_2$ are the two components of the shear leading to image shape distortions. The shear is often written in the complex form 
as $\boldsymbol \gamma=\gamma_1+\mathrm{i} \gamma_2$. They are calculated from the second derivatives of the lensing potential. Specifically, we have
\begin{equation}\label{eq:convergence and shear}
\begin{aligned} \kappa=\frac{1}{2} \nabla^{2} \psi ;\quad  
\gamma_{1}=\frac{1}{2}(\psi_{,11}-\psi_{,22}); \quad \gamma_{2}=\psi_{,12} . \end{aligned}
\end{equation}

Under the Born approximation, the lensing potential for a source at the comoving radial distance $\chi$ and the angular position $\boldsymbol {\theta}$ can be written as \citep[e.g.][]{2015RPPh...78h6901K}
\begin{equation}\label{eq:lensing potential}
\psi(\boldsymbol{\theta}, \chi)=\frac{2}{c^{2}} \int_{0}^{\chi} \mathrm{d} \chi^{\prime} \frac{f_{K}(\chi-\chi^{\prime})}{f_{K}(\chi) f_{K}(\chi^{\prime})} \Phi\left(f_{K}(\chi^{\prime}) \boldsymbol{\theta}, \chi^{\prime}\right)
\end{equation}
where $c$ is the speed of light, $f_{K}$ the comoving angular diameter distance, and $\Phi$ is the 3-D gravitational potential satisfying $\nabla^{2} \Phi=4 \pi G a^{2} \bar{\rho} \delta$ with $\delta$ being the 3-D matter density perturbation and $a$ the cosmic scale factor. 
Therefore the convergence $\kappa$ is given by
\begin{equation}\label{eq:lensing convergence}
\kappa(\boldsymbol{\theta}, \chi)=\frac{3H_0^2\Omega_{\rm m}}{2c^2} \int_{0}^{\chi} \mathrm{d} \chi^{\prime} \frac{f_{K}(\chi^{\prime})f_{K}(\chi-\chi^{\prime})}{f_{K}(\chi)} \frac{\delta[f_{K}(\chi^{\prime}) \boldsymbol{\theta}, \chi^{\prime}]}{a(\chi^{\prime})},
\end{equation}
where $H_0$ and $\Omega_{\rm m}$ are the Hubble constant and the present dimensionless matter density, respectively. It is seen that the convergence $\kappa$ is the projected matter density fluctuations weighted by the combination of comoving 
angular diameter distances to the source, to the lens and between the lens and the source, as well as the cosmic scale factor.   

Observationally, the measured ellipticity of a galaxy $\boldsymbol {\epsilon}$ consists of the contributions from its intrinsic ellipticity $\boldsymbol {\epsilon_s}$ and the lensing distortion characterised by the reduced shear $\boldsymbol{g}=\boldsymbol{\gamma}/(1-\kappa)$. It is given by \citep{1997A&A...318..687S}
\begin{equation}
\boldsymbol{\epsilon}=\left\{\begin{array}{ll}
\displaystyle{\frac{\boldsymbol{\epsilon}_{\mathrm{s}}+\boldsymbol{g}}{1+\boldsymbol{g}^{*}\boldsymbol{\epsilon}_{\mathrm{s}}}} ;& \text { for }|\boldsymbol{g}| \leqslant 1 \\ 
\displaystyle{\frac{1+\boldsymbol{g} \boldsymbol{\epsilon}_{\mathrm{s}}^{*}}{\boldsymbol{\epsilon}_{\mathrm{s}}^{*}+\boldsymbol{g}^{*}}}, & \text { for }|\boldsymbol{g}|>1
\end{array}\right.,
\end{equation}
where the asterisk denotes the complex conjugate operation and the complex ellipticity is defined as 
\begin{equation}
\boldsymbol{\epsilon}=\displaystyle{\frac{1-r}{1+r}}e^{2i\phi},
\end{equation}
with $r=b/a$ being the minor-to-major axial ratio and $\phi$ being the position angle. In the case of no intrinsic alignments of galaxies, it is shown that $\left\langle\boldsymbol{\epsilon}\right\rangle$ can give rise to an unbiased estimate of the reduced shear $\boldsymbol{g}$ \citep{1997A&A...318..687S}.

From the shear measurements, we can construct the convergence map using the relation between $\boldsymbol {\gamma}$ and $\kappa$ as shown in Eq.(\ref{eq:convergence and shear}) \citep[e.g.,][]{1993ApJ...404..441K, 1995ApJ...449..460K, 2013MNRAS.433.3373V, 2015MNRAS.450.2888L}. Alternatively, one can construct aperture mass maps directly
from the shear fields \citep[e.g.,][]{1996MNRAS.283..837S, 1998MNRAS.296..873S,2012MNRAS.423.3405L,2016MNRAS.463.3653K, 2018MNRAS.474..712M, 2022MNRAS.511.2075Z}. In the limit of ${\boldsymbol g}\approx {\boldsymbol \gamma}$, aperture mass maps with a filter $Q$ are equivalent to convergence maps with a corresponding filter $U$ \citep{1998MNRAS.296..873S}. 

For a reconstructed convergence field, the existence of intrinsic ellipticities of galaxies leads to a residual shape noise field $N$, which can be approximated as a Gaussian random field in the case with a large enough number of galaxies in the smoothing kernel. Thus we have 
\begin{equation}
K_N(\boldsymbol{\theta})=K(\boldsymbol{\theta})+N(\boldsymbol{\theta}),
\end{equation}
where $K_N$ and $K$ are the reconstructed and the true convergence fields smoothed  with a proper filter, respectively. 

For the purpose of this study to analyse the differences between peak height and steepness statistics, here we directly use convergence maps from simulations and add shape noises according to the number density of galaxies. The reconstruction from shears and its impacts on peak steepness statistics are left to our future investigations. 

\subsection{A theoretical model for high peak abundances based on steepness}\label{sec:Convergence peak abundance model}
To better understand the differences between the peak statistics based on height and steepness, here we describe a theoretical model for high peaks. It is based on our previous studies that put forwards a halo-based model to calculate high peak abundances with respect to peak height \citep{2010ApJ...719.1408F, 2018ApJ...857..112Y}.
Validated with simulation mocks, it has been applied to observational analyses to derive important cosmological constraints \citep{2015MNRAS.450.2888L,2016PhRvL.117e1101L,2018MNRAS.474.1116S}. 
This model can be readily extended to calculate high peak steepness statistics as shown in the following.    

Assuming high peaks dominantly come from the WL effects of individual massive haloes along lines of sight, we can divide the convergence field into three parts written as \citep{2018ApJ...857..112Y}
\begin{equation}\label{eq:K_N model}
K_N=K_H+K_{\rm{LSS}}+N,
\end{equation}
where $K_H$ is the WL signal from massive haloes, $K_{\rm{LSS}}$ is the contribution from large-scale structures excluding the haloes considered in $K_H$, and $N$ is the shape noise field. 

For $K_H$, we consider haloes with mass above a threshold $M_{*}$. Simulation analyses show that $M_{*}=10^{14}h^{-1}M_{\odot}$ is a proper choice and those massive haloes do correspond well with high WL peaks \citep{2018ApJ...857..112Y,2018MNRAS.478.2987W}.  
For each of these haloes, its $K_H$ is calculated by employing the Navarro–Frenk–White (NFW) density profile \citep{1996ApJ...462..563N,1997ApJ...490..493N}.

For $K_{\rm{LSS}}$, we regard it as a random field. It is noted that the field of large-scale structures is non-Gaussian due to the nonlinear nature of gravity. Excluding haloes with $M\ge M_{*}$ reduces the non-Gaussianity but $K_{\rm{LSS}}$ still contains non-Gaussian information. On the other hand, 
in our model for high peaks, their signals are mainly from $K_H$ and the effect of $K_{\rm{LSS}}$ is a minor correction. In \cite{2018ApJ...857..112Y}, we thus approximate $K_{\rm{LSS}}$ as a Gaussian random field, and test the model extensively with numerical simulations. 
It is emphasized that our model is only for high peaks. For low peaks where $K_{\rm{LSS}}$ plays major roles, its non-Gaussianity needs to be taken into account carefully. 

Under the above approximation, the convergence field $K_N$ in a massive halo region is the sum of two Gaussian random fields $K_{\rm{LSS}}+N$ modulated by the halo profile $K_H$. Thus the peak statistics for the field $K_N$ can be calculated using the Gaussian random field theory \citep{1986ApJ...304...15B,1987MNRAS.226..655B}.   
Specifically, peaks are defined by requiring the first derivatives $K^i_N=\partial_i K_N=0$ and the second derivatives $K_N^{ij}=\partial_i\partial_j K_N$ to be negative definite with $i(j)=1,2$. With the modulation of $K_H$, the number density distribution of peaks at given $\nu_N=K_N/\sigma_0$ and $x_N=-(K_N^{11}+K_N^{22})/\sigma_2$
can be written as \citep{2010ApJ...719.1408F, 2018ApJ...857..112Y}
\begin{equation}
\begin{aligned}
&\hat{n}_{\text {peak }}\left(\nu_{N}, x_{N}\right)= \\
& \exp \left[-\frac{1}{2} \frac{\left(K_H^{11}-K_H^{22}\right)^{2}}{\sigma_{2}^{2}}-\frac{\left(K_H^{11}\right)^{2}+\left(K_H^{22}\right)^{2}}{\sigma_{2}^{2}}-4 \frac{\left(K_H^{12}\right)^{2}}{\sigma_{2}^{2}}\right] \\
& \times \exp \left[-\frac{\left(K_H^{1}\right)^{2}+\left(K_H^{2}\right)^{2}}{\sigma_{1}^{2}}\right]\left\{\frac{1}{2 \pi \theta_{N*}^{2}} \frac{1}{\left[2 \pi\left(1-\gamma_{N}^{2}\right)\right]^{1 / 2}}\right\} \\
& \times \exp \left\{-\frac{\left[\left(\nu_{N}-K_H / \sigma_{0}\right)-\gamma_{N} x_{N}-\gamma_{N}\left(K_H^{11}+K_H^{22}\right) / \sigma_{2}\right]^{2}}{2\left(1-\gamma_{N}^{2}\right)}\right\}  \\
& \times \exp \left[-\frac{x_{N}^{2}}{2}-x_{N}\left(\frac{K_H^{11}+K_H^{22}}{\sigma_{2}}\right)\right] \frac{1}{(2 \pi)^{1 / 2}}\\
& \times \int_{0}^{1 / 2} d e_{N} \frac{8}{\pi}\left(x_{N}^{2} e_{N}\right) x_{N}^{2}\left(1-4 e_{N}^{2}\right) \exp \left(-4 x_{N}^{2} e_{N}^{2}\right) \\
& \times \int_{0}^{\pi} d \theta_{N} \exp \left[-4 x_{N} e_{N} \cos 2 \theta_{N}\left(\frac{K_H^{11}-K_H^{22}}{\sigma_{2}}\right)\right] \\
& \times \exp \left(-8 x_{N} e_{N} \sin 2 \theta_{N} \frac{K_H^{12}}{\sigma_{2}}\right),
\end{aligned}
\label{2DPeak}
\end{equation}
where $K_H$, $K_H^{i}$ and $K_H^{ij}$ are the convergence value at a position contributed from the considered massive halo, and its first and second derivatives, respectively. The quantities $\sigma_i^2$ ($i=0,1,2$) are the moments of the Gaussian field $K_{\rm{LSS}}+N$, and $\sigma_i^2=\sigma_{\text{LSS},i}^2+\sigma_{\text{ran},i}^2$.
Here $\sigma_{\text{LSS},i}^2$ and $\sigma_{\text{ran},i}^2$ are the respective moments of $K_{\rm{LSS}}$ and the shape noise field $N$.
The parameters $\gamma_N=\sigma_1^2/(\sigma_0\sigma_2)$ and $\theta_{N*}=\sqrt{2}\sigma_1/\sigma_2$, with the first being the spectral parameter determining the shape of the peak number density distribution for a Gaussian random field and the second being 
the effective angular scale with $\theta_{N*}^2$ the overall multiplicative surface area term in the peak density distribution \citep{1986ApJ...304...15B,1987MNRAS.226..655B}. 

For the shape noise field without considering intrinsic alignments of galaxies and under the Gaussian smoothing $W(\boldsymbol{\theta})=\exp(-\theta^2/\theta_G^2)/(\pi \theta_G^2)$, we have $\sigma_{\rm{ran},0}^2=\sigma^2_{\epsilon}/(4\pi n_g\theta_G^2)$ where $\sigma_{\epsilon}$ is the root mean square (r.m.s.) of the total intrinsic ellipticities of galaxies, 
and $\sigma_{\rm{ran},0}:\sigma_{\rm{ran},1}:\sigma_{\rm{ran},2}=1:\sqrt{2}/\theta_G:2\sqrt{2}/\theta_G^2$ \citep{2000MNRAS.313..524V}.  

For $K_{\rm{LSS}}$, its moments $\sigma_{\rm{LSS}, i}^2$ are calculated following the procedures described in \cite{2018ApJ...857..112Y}. Specifically, the power spectrum of $K_{\rm{LSS}}$ is computed by subtracting the one-halo term contribution from haloes with $M\ge M_*$ from the fully nonlinear power spectrum. Thus the leftover power spectrum of $K_{\rm{LSS}}$ contains the contributions from one-halo term with $M<M_*$ and the full two-halo term in the language of the halo model \citep[e.g.,][]{2002PhR...372....1C}. With the power spectrum, the moments $\sigma_{\text{LSS}, i}^2$ can be readily calculated taking into account the smoothing kernel. It is clear that $\sigma_{\text{LSS},i}^2$ are cosmology-dependent quantities. 

In the two-dimensional peak number distribution of Eq.(\ref{2DPeak}), $\nu_N$ is the signal-to-noise ratio of peak height, and $x_N$ is the signal-to-noise ratio of the steepness $-(K_{N}^{11}+K_{N}^{22})$ reflecting the profile of a peak. 
The steepness defined here corresponds to the Laplacian operator used in \cite{2019NatAs...3...93R}. 

From Eq.(\ref{2DPeak}), we can then derive the peak distribution in terms of either $\nu_N$ or $x_N$ by integrating the other. They are given as follows. 
\begin{equation}\label{eq:height number density}
\begin{aligned}
\hat{n}_{\text {peak }}\left(\nu_N\right)=& \exp \left[-\frac{\left(K_H^{1}\right)^{2}+\left(K_H^{2}\right)^{2}}{\sigma_{1}^{2}}\right]\left\{\frac{1}{2 \pi \theta_{N*}^{2}} \frac{1}{(2 \pi)^{1 / 2}}\right\} \\
& \times \exp \left(-\frac{1}{2}u(\nu_N)^{2}\right) d \nu_{N}\int_0^{\infty} \frac{d x_{N}}{\left[2 \pi\left(1-\gamma_{N}^{2}\right)\right]^{1 / 2}} \\
& \times \exp \left[-\frac{\left(m(x_N)-\gamma_{N}u(\nu_N)\right)^{2}}{2\left(1-\gamma_{N}^{2}\right)}\right] \times F\left(x_{N}\right),
\end{aligned}
\end{equation}
and
\begin{equation}\label{eq:steepness number density}
\begin{aligned}
\hat{n}_{\text {peak }}\left(x_N\right)\bigg{|}_{\nu_N\ge\nu_{\text{cut}}}=& \exp \left[-\frac{\left(K_H^{1}\right)^{2}+\left(K_H^{2}\right)^{2}}{\sigma_{1}^{2}}\right]\left\{\frac{1}{(2 \pi \theta_{N*})^{2}} \frac{(2\pi)^{1/2}}{2}\right\} \\
& \times \exp \left(-\frac{1}{2}m(x_N)^{2}\right) \times \text{erfc}(t(\nu_{\text{cut}},x_N))\\
& \times F\left(x_{N}\right)d x_{N},
\end{aligned}
\end{equation}
where erfc$(x)$ is the complementary error function. For convenience, we define
\begin{equation}\label{eq:fx}
\begin{aligned}
& u(\nu_N) = \nu_N-\displaystyle{\frac{K_H}{\sigma_0}}, \\
& m(x_N) = x_N+\displaystyle{\frac{K_H^{11}+K_H^{22}}{\sigma_2}}, \\
& t(\nu_N,x_N) = \displaystyle{\frac{u(\nu_N)-\gamma_Nm(x_N)}{[2(1-\gamma_N^2)]^{1/2}}},\\
& \hbox{and}\\
&
\begin{aligned}
F\left(x_N\right)=& \exp \left[-\left(\frac{K_H^{11}-K_H^{22}}{\sigma_{2}}\right)^{2}-4 \frac{\left(K_H^{12}\right)^{2}}{\sigma_{2}^{2}}\right] \\
& \times \int_{0}^{1 / 2} d e_{N} \  8\left(x_N^{2} e_{N}\right) x_N^{2}\left(1-4 e_{N}^{2}\right) \exp \left(-4 x_N^{2} e_{N}^{2}\right) \\
& \times \int_{0}^{\pi} \frac{d \theta_{N}}{\pi} \exp \left[-4 x_N e_{N} \cos 2 \theta_{N}\left(\frac{K_H^{11}-K_H^{22}}{\sigma_{2}}\right)\right] \\
& \times \exp \left(-8 x_N e_{N} \sin 2 \theta_{N} \frac{K_H^{12}}{\sigma_{2}}\right).
\end{aligned}
\end{aligned}
\end{equation}
Because the model here is only valid for high peaks, we apply a lower cut on the peak height, denoted as $\nu_{\rm {cut}}$, in Eq.(\ref{eq:steepness number density}). A numerical method to calculate $F(x_N)$ approximately within a spherical halo is provided in Appendix \ref{sec:fx series}.

With the above formulae for peak statistics within a massive halo and a further assumption that a field can be divided into halo regions occupied by massive haloes with $M\ge M_*$ and the rest of non-halo regions, we can write down the peak number distribution over a field by
\begin{equation}
n_{\text {peak }}(y) d y=\left[n_{\text {peak }}^{H}(y) +n_{\text {peak }}^{N}(y)\right] d y,
\end{equation}
where $n_{\text {peak }}^{H}$ and $n_{\text {peak }}^{N}$ refer to the distributions in halo and non-halo regions, respectively, and the argument $y$ can be either $\nu_N$ or $x_N$. 

For $n_{\text {peak }}^{H}$, we have
\begin{equation}\label{eq:Peakhalo}
\begin{aligned}
n_{\text {peak }}^{H}(y)=& \int d z \frac{d V(z)}{d z d \Omega} \int_{M_{*}}^{\infty} d M n(M, z) \\
& \times \int_{0}^{\theta_{\text {vir }}} d \theta(2 \pi \theta) \hat{n}_{\text {peak }}^{H}(y, M, z, \theta),
\end{aligned}
\end{equation}
where $dV$ and $d\Omega$ are the cosmic volume and solid angle elements, respectively, $n(M, z)$ is the halo mass function, and ${\theta_{\text {vir }}}$ is the angular virial radius of a halo with mass $M$ at redshift $z$. The quantity $\hat{n}_{\text {peak }}^{H}(y, M, z, \theta)$ is the peak number density at an angular position $\theta$ from the centre of a halo with mass $M$, which can be calculated by equations (\ref{eq:height number density}) and (\ref{eq:steepness number density}) for height and steepness cases respectively. For the density profile of haloes, we adopt the NFW form with the mass-concentration relation from \cite{2008MNRAS.390L..64D}. The halo mass function is modelled based on \cite{2013MNRAS.433.1230W}. 

For the peak distribution $n_{\text {peak }}^{N}$ outside halo regions, it is given by
\begin{equation}
\begin{array}{l}
n_{\text {peak }}^{N}(y)=\frac{1}{d \Omega} \\
\times\left\{\hat{n}_{\text {\rm{ran} }}(y)\left[d \Omega-\int d z \frac{d V(z)}{d z} \int_{M_{*}}^{\infty} d M n(M, z)\left(\pi \theta_{\text {vir }}^{2}\right)\right]\right\},
\end{array}
\end{equation}
where $\hat{n}_{\text {\rm{ran} }}(y)$ is the mean number density of the Gaussian random field $K_{\rm{LSS}}+N$, which can be calculated by equations (\ref{eq:height number density}) and (\ref{eq:steepness number density}) but with the halo terms $K_H$, $K_H^i$ and $K_H^{ij}$ setting to be zero.

\section{Simulated convergence maps}\label{chapter3}
To investigate and compare systematically the peak statistics based on height and steepness, we carry out N-body simulations and perform ray-tracing calculations to obtain WL convergence maps with the source redshift set to be $z_s=1$. 

\subsection{N-body simulations and ray-tracing}\label{sec:N-body simulations and ray-tracing}
Five cosmological models within the flat $\Lambda$CDM framework are considered in our studies. Their specific cosmological parameters are shown in Table\,\ref{table:cosmological parameters}. Here $\Omega_\text{m}$, $\Omega_{\Lambda}$, and $\Omega_\text{b}$ are the present dimensionless cosmic density parameters of matter, dark energy in the form of cosmological constant
and the baryonic matter, respectively, $h$ and $n_s$ are the Hubble constant in units of $100$km/s/Mpc and the power index of initial density perturbations, and $\sigma_8$ is the amplitude of the extrapolated-to-present linear density fluctuations over a top-hat scale of $8h^{-1}\hbox{Mpc}$. 
We regard Cos0 as the fiducial model, and change $\sigma_8$ and $\Omega_\text{m}$ in (Cos1, Cos2) and (Cos3, Cos4), respectively. 

\begin{figure*}
	\centering
	\includegraphics[width=1.0\linewidth]{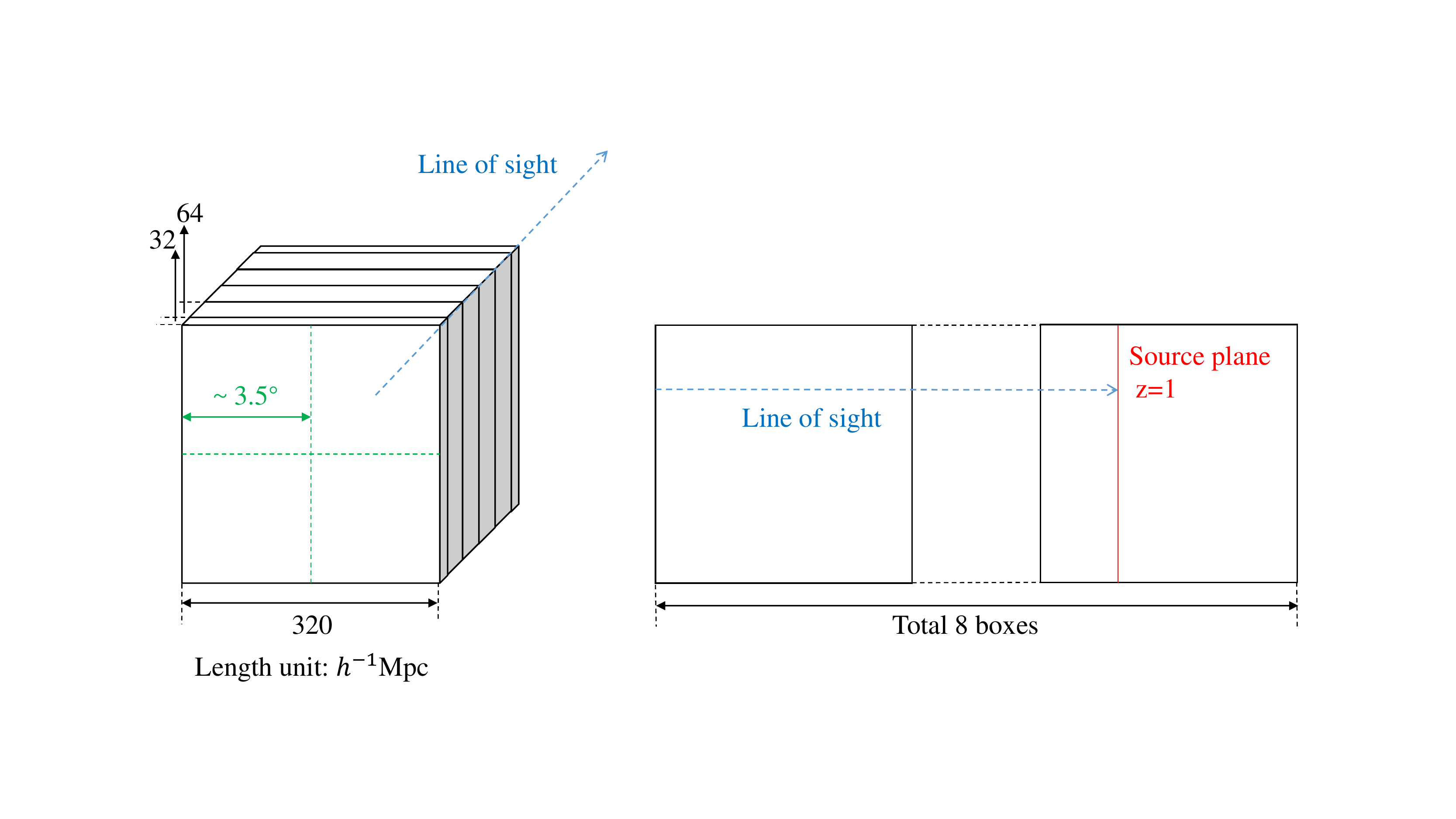}
	\caption{Schematic illustrations of stacking N-body simulation boxes for ray-tracing. The left and right panels show different views along different directions.  For each set of stacked simulation boxes, four WL maps are obtained each with an area of $\sim 3.5\degree\times3.5\degree$.}
	\label{fig:N-body boxes}
\end{figure*}

For each model, we run N-body simulations with the box size of $320h^{-1}$Mpc and the number of particles of $640^3$ using GADGET-2 (\cite{2005MNRAS.364.1105S}). The force softening length is $\sim 20h^{-1}\hbox{kpc}$. The simulations start at $z=50$ and the initial conditions are generated using 2LPTic \citep{2012ascl.soft01005C}. 
For ray-tracing calculations to $z_s=1$ with the comoving angular diameter distance of $\sim 2.4h^{-1}\hbox{Gpc}$, we run eight simulations with independent initial condition realizations, and pad them together to build a light cone to $z_s=1$, as illustrated in Figure \ref{fig:N-body boxes}. 

From the matter distribution within the light cone, we perform multiple-lens-plane ray tracing calculations with the methodology in accord with \cite{2009A&A...499...31H} and the coding described in detail in
\cite{2014ApJ...784...31L}. For the lower-redshift seven boxes, the width of a lens plane is $64h^{-1}\hbox{Mpc}$, and thus each simulation box is divided into five planes using the snap shots at the corresponding redshifts. For the eighth box in the light cone reaching $z_s=1$, the number of lens planes and their width are $\Omega_{\rm m}$ dependent. The ray-tracing computations result in four convergence maps each with an area of $3.5^{\circ}\times 3.5^{\circ}$ as shown in Figure \ref{fig:N-body boxes}.
Each convergence map is sampled $1024\times 1024$ pixels with the pixel scale of $\sim0.2\hbox{ arcmin}$. Correspondingly, we build four convergence maps using eight N-body simulation boxes for each of the other four cosmological models. To reveal the cosmology-dependence cleanly,
for each model, we use exactly the same random seeds as those in fiducial simulations in generating initial conditions for the N-body simulations.

To increase the simulation area, for each model, we build 24 sets of light cones leading to the total area of convergence maps of $24\times4\times 3.5^{\circ}\times 3.5^{\circ}=1176\deg^2$, which is order-of-magnitude similar to the current WL survey coverages \citep[e.g.,][]{2021A&A...646A.140H}. 

We note that in our analyses, we are interested in WL peaks which physically consist of high peaks dominantly arising from clusters of galaxies, and relatively low peaks attributed largely by the collective WL effects from large-scale structures. Furthermore, in WL peak analyses, depending on the galaxy number density, certain smoothing operations are necessary to suppress the shape noise with the typical smoothing scale of a few arcminutes \citep[e.g.,][]{2004MNRAS.350..893H,2015MNRAS.450.2888L}.
Given the WL surveys with the median redshift typically around $0.7$ and above, the mass and force resolutions adopted here are sufficient. We note that the resolution parameters are the same as those in our previous studies where the simulations are validated by showing good agreements of the halo mass function and the WL power spectrum between the simulation results and the theoretical predictions \citep{2014ApJ...784...31L,2018ApJ...857..112Y, 2019ApJ...884..164Y}, and are used to build mock data for different observational applications of WL peak analyses \citep{2015MNRAS.450.2888L, 2018MNRAS.474.1116S, 2023MNRAS.519..594L}.

\begin{table}
\centering
\caption{Cosmological parameters of simulations.} 
\label{table:cosmological parameters}
$\begin{array}{|l|l|l|l|l|l|}
\hline
\text{Name} & {\text{Cos0}} & {\text{Cos1}} & {\text{Cos2}} & {\text{Cos3}} & {\text{Cos4}} \\ \hline \sigma_{8} & {0.82} & {0.77} & {0.87} & {0.82} & {0.82} \\ {\Omega_{\mathrm{m}}} & {0.28} & {0.28} & {0.28} & {0.25} & {0.31} \\ {\Omega_{\Lambda}} & {0.72} & {0.72} & {0.72} & {0.75} & {0.69} \\ {\Omega_{\mathrm{b}}} & {0.046} & {0.046} & {0.046} & {0.046} & {0.046}\\ {h} & {0.7} & {0.7} & {0.7} & {0.7} & {0.7}\\ {n_{s}} & {0.96} & {0.96} & {0.96} & {0.96} & {0.96}\\ \hline \text{Sets} & {24} & {24} & {24} & {24} & {24} \\
\hline
\end{array}$
\end{table}

\subsection{Shape noise and smoothing}\label{sec:Shape noise and smoothing}
In our studies here, we focus on comparisons of the peak statistics based on the height and the steepness and their cosmological inferences. We therefore perform analyses on the simulated convergence maps directly and add the shape noise as a Gaussian random field. The issues related to
constructing either convergence maps or aperture mass maps from shear samples and how such an operation can affect the steepness studies are left to our future investigations. 

Following \cite{2004MNRAS.350..893H}, we add a Gaussian shape noise with a zero mean on pixels. Its variance is given by
\begin{equation}\label{eq:sigma_pix}
\sigma_{\rm{ran},\mathrm{pix}}^{2}=\frac{\sigma_{\epsilon}^{2}}{2} \frac{1}{n_{{g}} \theta_{\mathrm{pix}}^{2}},
\end{equation}
where $\sigma_{\epsilon}$ is the root mean square (r.m.s.) amplitude of the total intrinsic ellipticities of galaxies, taken to be $\sigma_{\epsilon}=0.4$, $n_g$ is the source galaxy number density, and $\theta_{\mathrm{pix}}\approx 0.205$ arcmin is the pixel size of our convergence maps. For $n_g$, we consider different values with $n_g=10, 20, 30$ and $50\hbox{ arcmin}^{-2}$, respectively, corresponding to roughly the number density achievable by current WL surveys and the expected capability of future surveys. The value of $\sigma_{\epsilon}=0.4$ is taken to be similar to that from surveys such as CFHTLenS and KiDS \citep{2012MNRAS.427..146H, 2017MNRAS.465.1454H}. It is noted that $\sigma_{\epsilon}$ contains contributions from the true intrinsic ellipticity of galaxies and the shear measurement uncertainty \citep{2013MNRAS.429.2858M}. The latter can be reduced with the improvement of data quality and the measurement methods. We also note that in our analyses, it is the residual shape noise that matters the most, which is a combination of $\sigma_{\epsilon}$, $n_g$ and the smoothing scale $\theta_G$ with $\sigma_{\rm{ran},0}^2=\sigma^2_{\epsilon}/(4\pi n_g\theta_G^2)$. The parameters we take result in different $\sigma_{\rm{ran},0}$ representative for different WL surveys, and allow us to investigate systematically the effects of the shape noise on our studies.

For each convergence map of the fiducial model with a given $n_g$, we produce $10$ random noise realizations using different random seeds. The same sets noise are added to the corresponding convergence maps of the other four cosmological models. 
For different $n_g$, the random seeds for a map are all the same and the noise variance is adjusted according to Eq.(\ref{eq:sigma_pix}). Therefore for each map, the noise pattern is the same but only the amplitude changes with $n_g$.

In peak analyses, a smoothing is normally applied to suppress the shape noise. Here we adopt a Gaussian smoothing function given as 
\begin{equation}
W(|\boldsymbol{\theta}|)=\frac{1}{\pi \theta_{G}^{2}} \exp \left(-\frac{|\boldsymbol{\theta}|^{2}}{\theta_{G}^{2}}\right).
\end{equation} 
For the smoothing scale, two cases with $\theta_G=2.0$ and $3.0$ arcmin, respectively, are analysed. 

After the Gaussian smoothing, we calculate the moments of the shape noise from noise maps. They are in excellent agreements with the theoretical calculations shown in the previous section. For $\theta_G=2 \hbox{ arcmin}$, the relative differences of $\sigma_{\rm {ran},0}$, $\sigma_{\rm {ran},1}$ and $\sigma_{\rm {ran},2}$ between the results from different individual maps and the theoretical values are all less than $3\%$. For the $1000$ bootstrap samples described later each containing $100$ maps, the corresponding relative differences are less than $1\%$, $1\%$, and $2\%$, respectively. We notice a slightly systematic bias of $-0.7\%$ for $\sigma_{\rm {ran},1}$ and $-1.7\%$ for $\sigma_{\rm {ran},2}$ from maps, which should be associated with the map pixelation effect in calculating the first and second derivatives. In comparison with our bin widths for peak statistics, the biases are negligible. For $\theta_G=3 \hbox{ arcmin}$, comparing to the theoretical values, the relative differences of the three noise parameters from the bootstrap samples are less than $\sim 1\%$. We therefore adopt the theoretical values for these moments in the following analyses.

We emphasise that our main goal here is to investigate the differences and the relations between the two peak statistics. Thus our simulation settings are somewhat simplified, e.g., by fixing $z_s=1$ without
considering redshift distributions of source galaxies, and varying only $\Omega_{\rm m}$ and $\sigma_8$ in relatively narrow ranges with all the other cosmological parameters fixed. Potential systematics, such as baryonic effects, intrinsic alignment effects, etc., are not taken into account. We leave these issues to our future studies.  

\section{Statistical analyses}\label{chapter4}
\subsection{Peak height and steepness from simulated convergence maps}\label{sec:Peak height and steepness in simulated maps}
Peaks in a smoothed convergence map are identified by finding pixels with their convergence values higher than that of their 8 nearest neighbouring pixels. 
To avoid boundary effects on peak statistics arising from the smoothing operation while having sufficient effective areas left, 
we exclude the outermost $\sim 3.5\theta_G$ region on each side of a map in peak counting, which are $\sim 7$ and $10.5\hbox{ arcmin}$, corresponding to 36 and 53 pixels 
in the cases of $\theta_G=2.0$ and $3.0$ arcmin, respectively. 

For the peak height $K_N$, its signal-to-noise ratio is defined as $\nu=K_N/\sigma_{\rm{ran},0}$ in accord with real observations where only $\sigma_{\rm{ran},0}$ can be estimated. It is noted that in Eqs.(\ref{2DPeak}) and (\ref{eq:height number density}), $\nu_N$ is defined using the total $\sigma_0$.
Thus in comparing simulation results with our model predictions, we need to take into account the difference of $\nu$ and $\nu_N$ by applying a scaling factor $\sigma_{\rm{ran},0}/\sigma_{0}$ in model calculations. 

To calculate the peak steepness corresponding to $-(K_N^{11}+K_N^{22})$ in simulated maps, we apply the following discrete operator 
\begin{equation}\label{steepness}
\begin{aligned}
[\text{S}]&= 4\times[K_N \text{ at the peak position}]\\
&-[\text{Sum of } K_N \text{ of the nearest 4 pixels}].
\end{aligned}
\end{equation}
In Appendix \ref{sec:Fisher L vs L2}, we show that mathematically Eq.(\ref{steepness}) is equal to $-(K_N^{11}+K_N^{22})$ in the unit of pixel$^{-2}$ under the second-order Taylor expansion.
Moreover, this is exactly the discrete minus Laplacian operator used in the studies of \cite{2019NatAs...3...93R}. Its matrix form is 
\begin{equation}
-L=4\left[\begin{array}{ccc}
0 & -0.25 & 0 \\
-0.25 & 1 & -0.25 \\
0 & -0.25 & 0
\end{array}\right].
\end{equation}
Similar to the signal-to-noise ratio $\nu$ for peak height, we also define $x=S/\sigma_{\rm{ran},2}$, which differs from $x_N$ in the model definition with $x_N=x(\sigma_{\rm{ran},2}/\sigma_{2})$.
We note that in Eq.(\ref{steepness}), the derivatives are taken in the unit of pixel. Thus $\sigma_{\rm{ran},2}$ needs also to be converted to the same unit when calculating $x$ consistently. 

\begin{figure}
    \includegraphics[scale=0.54]{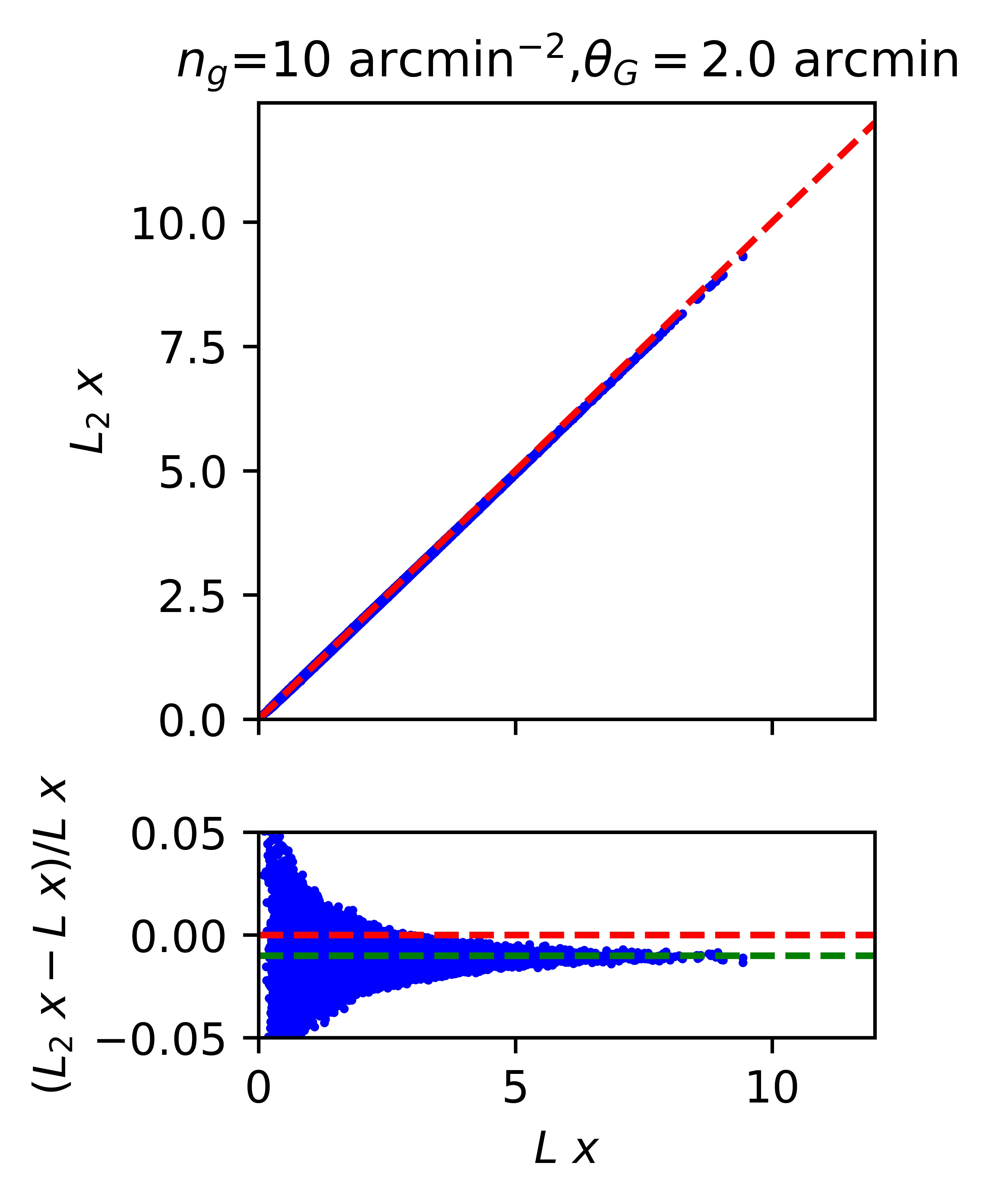}
    \includegraphics[scale=0.54]{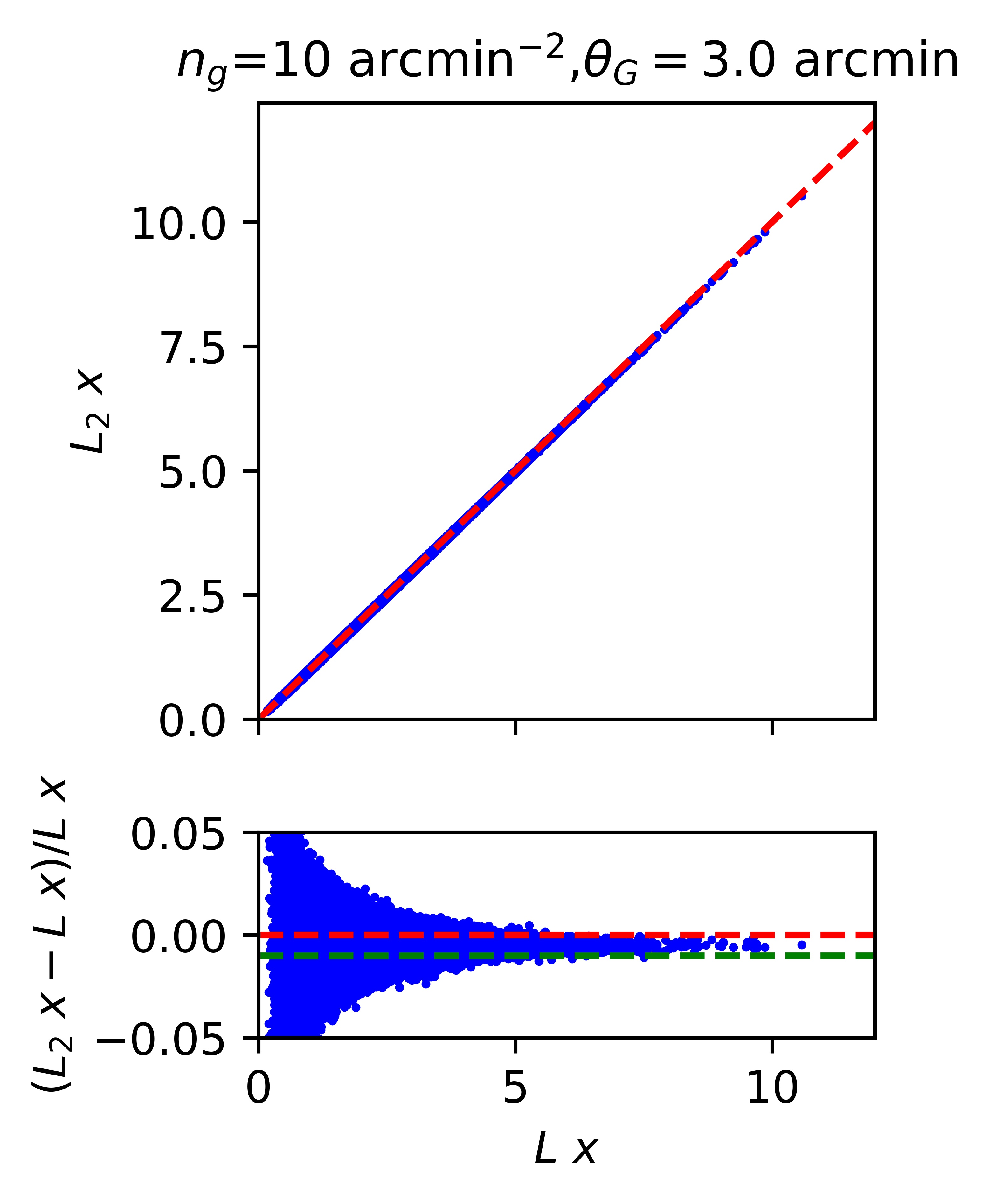}
    \caption{The comparisons between the $x$ values of peaks calculated from $-L$ and $-L_2$ for the noisy case of Cos0 with $n_g=10\hbox{ arcmin}^{-2}$. The left and right panels are for $\theta_G=2$ and $3\hbox{ arcmin}$, respectively. The upper panels are the scatter plots of the two $x$ values (blue) with the red dashed 1:1 line. The lower panels show the relative differences, and the red and green dashed lines indicate values of $0$ and $1\%$, respectively. }  
    \label{fig:vs diagram(Cos0_noisy)}
\end{figure}

It is seen that in calculating $S$ and therefore $x$, discrete differences are necessary. Thus it can suffer more from numerical artifacts, such as the pixel scale resolution effect, than the peak height. 
In our analyses, the pixel scale is $\sim 0.2\hbox{ arcmin}$, which is 10 and 15 times smaller than $\theta_G=2$ and $3\hbox{ arcmin}$, respectively. We therefore expect insignificant pixel effects on $S$. 
Nevertheless, we perform tests by using pixels further away from a peak to calculate its $S$. Specifically, we adopt
the following operator   

\begin{equation}
-L_2=\left[\begin{array}{ccccc}
0 & 0 & -0.25 & 0 & 0\\
0 & 0 & 0 & 0 & 0\\ 
-0.25 & 0 & 1 & 0 & -0.25 \\
0 & 0 & 0 & 0 & 0\\ 
0 & 0 & -0.25 & 0 & 0
\end{array}\right],
\end{equation}
i.e., we use the separation of 2 pixels to calculate $S$. The comparisons of $x$ from operators $-L$ and $-L_2$ are shown in Figure \ref{fig:vs diagram(Cos0_noisy)} for the noisy cases of $n_g=10\hbox{ arcmin}^{-2}$ and $\theta_G=2$ (left) 
and $3\hbox{ arcmin}$ (right), respectively. We can see that the overall relative differences are at the level of $\sim 1\%$. In Appendix {\ref{sec:Fisher L vs L2}}, we also show the comparison of the Fisher results obtained by using the two operators,
further demonstrating the negligible resolution effects in calculating $x$ in our studies here. 

\begin{figure}
    \centering
    \includegraphics[scale=0.5]{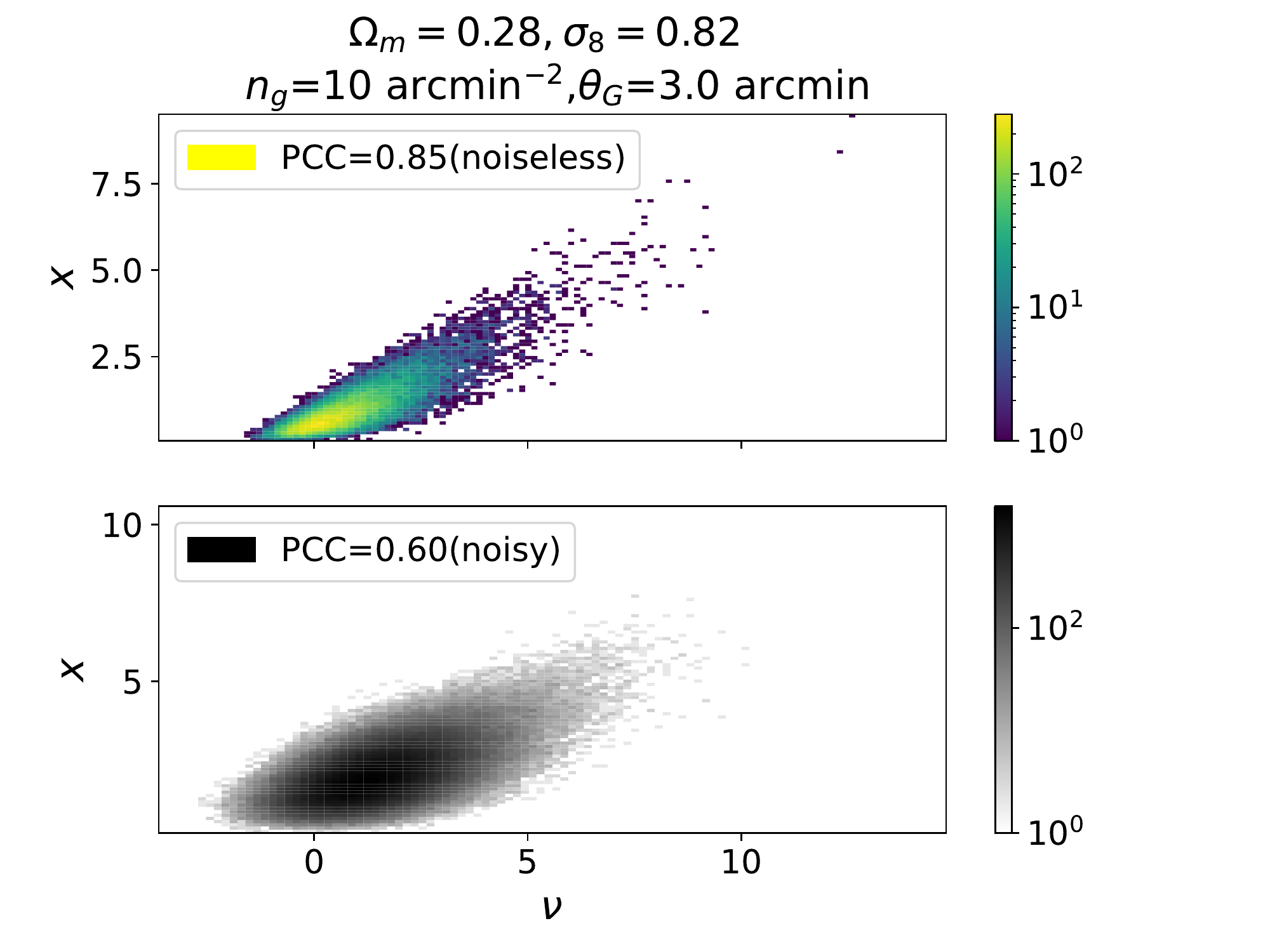}
    \caption{The peak distributions in the $\nu-x$ plane for Cos0 with $\Omega_{\rm m}=0.28$ and $\sigma_8=0.82$, and $\theta_G=3\hbox{ arcmin}$. The upper and lower panels are for the cases of noiseless and with noise under $n_g=10\hbox{ arcmin}^{-2}$, respectively. The colour bars on the right show the peak numbers.}
    \label{fig:height-steepness diagram(cos0)}
\end{figure}

\subsection{Range selection and binning of peaks}
\label{sec:Selection and binning of peaks}
In the studies here, we concern peak statistics based on two different quantities, peak height $\nu$ and the steepness $x$. 
To ensure a fair comparison of the two, we first select peaks based on their heights $\nu$ and then perform height and steepness statistics for the same set of selected peaks but with different binning in $\nu$ and in $x$, respectively. 

We consider two different peak selections. One is essentially all peaks found in simulations. The specific selection and binning schemes for the case of $n_g=10\hbox{ arcmin}^{-2}$ are listed in Table \ref{table:binning of peak height and steepness}. We refer such a sample as `{\it All}'. 
For the other $n_g$ cases, the upper limits of peaks and the bin widths, for both $\nu$ and $x$, are scaled by a factor of $\sqrt{n_g/10}$. Thus the numbers of bins used for peak statistics for different $n_g$ cases are all the same. 
This applies for both $\theta_G=2$ and $3\hbox{ arcmin}$. 

It is noted that for $n_g=10\hbox{ arcmin}^{-2}$,
the numbers of peaks at $\nu> 7$ and $x>\sim 6$ for the height and steepness cases, respectively, are very small, 
thus we combine all of these very high $\nu$ or $x$ peaks into one bin in the corresponding peak counting as 
shown in Table \ref{table:binning of peak height and steepness}. For the other $n_g$, the last bin is also relatively wide with 
a scaling described above. For high $n_g$, the number of high peaks increases and we could have more bins at high peak end. 
In our analyses here, we tend to keep the same number of bins in all cases for direct statistical comparisons. This may
underestimate the cosmological power somewhat for high $n_g$ cases. For our purpose here to compare the two peak 
statistics, the binning adopted here should not affect significantly our conclusions. 

In comparison with our high peak model presented in Sec. \ref{sec:Convergence peak abundance model}, we also select `{\it High}' peak samples from simulations. For that, we apply a lower cut $\nu_{\rm{cut}}$ on peak height. Its value in each case is chosen based on  
the comparison of the peak height distribution of the simulated peaks with our model predictions where above $\nu_{\rm{cut}}$, the two show good agreements. The specific choice of $\nu_{\rm{cut}}$ for different cases
and the corresponding peak binning are shown in Table \ref{table:binning for High sample}. 

\begin{table*}
\caption{Range and the binning schemes of the {\it All} samples with $n_g=10 \ \text{arcmin}^{-2}$.}\label{table:binning of peak height and steepness}
\centering
$\begin{array}{|c|c|}
 \hline \text{selection criteria} & \nu\in [-3.0,15.0]\ \ x\in[0.0,12.0] \\ \hline \text{height}(\nu) \text{binning} & [-3.0,-1.0,-0.5,0.0,0.5,1.0,1.5,2.0,2.5,3.0,3.5,4.0,4.5,5.0,5.5,6.0,6.5,7.0,15.0] \\ \hline \text{steepness}(x)\text{binning} &  [0.0,0.4,0.8,1.0,1.2,1.4,1.6,1.8,2.0,2.4,2.8,3.2,3.6,4.0,4.4,4.8,5.2,5.6,12.0] \\
\hline
\end{array}$
\end{table*}

\begin{table*}
\caption{Range selections and the binning schemes of the {\it High} samples in different cases.}\label{table:binning for High sample}
\centering
$\begin{array}{|c|c|}
 \hline \text{case} & n_g=10 \ \text{arcmin}^{-2} \ \ \theta_G=2.0 \ \text{arcmin}\\
 \hline \text{selection criteria} & \nu\in [4.0,15.0]\ \ x\in[0.0,12.0] \\ \hline \text{height}(\nu) \text{binning} & [4.0,4.25,4.5,4.75,5.0,5.25,5.5,5.75,6.0,6.5,7.0,15.0] \\ \hline \text{steepness}(x)\text{binning} &  [0.0,2.0,2.4,2.8,3.2,3.6,4.0,4.4,4.8,5.2,5.6,12.0] \\ \hline
 \hline \text{case} & n_g=10 \ \text{arcmin}^{-2} \ \ \theta_G=3.0 \ \text{arcmin}\\
 \hline \text{selection criteria} & \nu\in [4.5,15.0]\ \ x\in[0.0,12.0] \\ \hline \text{height}(\nu) \text{binning} & [4.5,4.75,5.0,5.25,5.5,5.75,6.0,6.25,6.5,6.75,7.0,15.0] \\ \hline \text{steepness}(x)\text{binning} &  [0.0,2.0,2.4,2.8,3.2,3.6,4.0,4.4,4.8,5.2,5.6,12.0] \\ \hline
 \hline \text{case} & n_g=20 \ \text{arcmin}^{-2} \ \ \theta_G=2.0 \ \text{arcmin}\\
 \hline \text{selection criteria} & \nu\in [3.0,15.0]\times\sqrt{2}\ \ x\in[0.0,12.0]\times\sqrt{2} \\ \hline \text{height}(\nu) \text{binning} & [3.0,3.25,3.5,3.75,4.0,4.5,5.0,5.5,6.0,6.5,7.0,15.0]\times\sqrt{2} \\ \hline \text{steepness}(x)\text{binning} &  [0.0,2.0,2.4,2.8,3.2,3.6,4.0,4.4,4.8,5.2,5.6,12.0]\times\sqrt{2} \\ \hline
 \hline \text{case} & n_g=20 \ \text{arcmin}^{-2} \ \ \theta_G=3.0 \ \text{arcmin}\\
 \hline \text{selection criteria} & \nu\in [3.25,15.0]\times\sqrt{2}\ \ x\in[0.0,12.0]\times\sqrt{2} \\ \hline \text{height}(\nu) \text{binning} & [3.25,3.5,3.75,4.0,4.25,4.5,5.0,5.5,6.0,6.5,7.0,15.0]\times\sqrt{2} \\ \hline \text{steepness}(x)\text{binning} &  [0.0,2.0,2.4,2.8,3.2,3.6,4.0,4.4,4.8,5.2,5.6,12.0]\times\sqrt{2} \\ \hline
 \hline \text{case} & n_g=30 \ \text{arcmin}^{-2} \ \ \theta_G=2.0 \ \text{arcmin}\\
 \hline \text{selection criteria} & \nu\in [2.5,15.0]\times\sqrt{3}\ \ x\in[0.0,12.0]\times\sqrt{3} \\ \hline \text{height}(\nu) \text{binning} & [2.5,2.75,3.0,3.5,4.0,4.5,5.0,5.5,6.0,6.5,7.0,15.0]\times\sqrt{3} \\ \hline \text{steepness}(x)\text{binning} &  [0.0,2.0,2.4,2.8,3.2,3.6,4.0,4.4,4.8,5.2,5.6,12.0]\times\sqrt{3} \\ \hline
 \hline \text{case} & n_g=30 \ \text{arcmin}^{-2} \ \ \theta_G=3.0 \ \text{arcmin}\\
 \hline \text{selection criteria} & \nu\in [2.75,15.0]\times\sqrt{3}\ \ x\in[0.0,12.0]\times\sqrt{3} \\ \hline \text{height}(\nu) \text{binning} & [2.75,3.0,3.5,4.0,4.25,4.5,5.0,5.5,6.0,6.5,7.0,15.0]\times\sqrt{3} \\ \hline \text{steepness}(x)\text{binning} &  [0.0,2.0,2.4,2.8,3.2,3.6,4.0,4.4,4.8,5.2,5.6,12.0]\times\sqrt{3} \\ \hline
 \hline \text{case} & n_g=50 \ \text{arcmin}^{-2} \ \ \theta_G=2.0 \ \text{arcmin}\\
 \hline \text{selection criteria} & \nu\in [2.0,15.0]\times\sqrt{5}\ \ x\in[0.0,12.0]\times\sqrt{5} \\ \hline \text{height}(\nu) \text{binning} & [2.0,2.5,3.0,3.5,4.0,4.5,5.0,5.5,6.0,6.5,7.0,15.0]\times\sqrt{5} \\ \hline \text{steepness}(x)\text{binning} &  [0.0,2.0,2.4,2.8,3.2,3.6,4.0,4.4,4.8,5.2,5.6,12.0]\times\sqrt{5} \\ \hline
 \hline \text{case} & n_g=50 \ \text{arcmin}^{-2} \ \ \theta_G=3.0 \ \text{arcmin}\\
 \hline \text{selection criteria} & \nu\in [2.5,15.0]\times\sqrt{5}\ \ x\in[0.0,12.0]\times\sqrt{5} \\ \hline \text{height}(\nu) \text{binning} & [2.5,2.75,3.0,3.5,4.0,4.5,5.0,5.5,6.0,6.5,7.0,15.0]\times\sqrt{5} \\ \hline \text{steepness}(x)\text{binning} &  [0.0,2.0,2.4,2.8,3.2,3.6,4.0,4.4,4.8,5.2,5.6,12.0]\times\sqrt{5} \\ \hline

\end{array}$
\end{table*}

\subsection{Peak height and steepness distribution}
\label{sec:Height and steepness distribution}
Before presenting peak statistics based on height and steepness, here we first show the relation of $\nu$ and $x$ for peaks. In Figure \ref{fig:height-steepness diagram(cos0)},
we present the 2-D distributions of peaks in the $\nu$-$x$ plane for the fiducial model of Cos0. The upper panel is for the noiseless case and the lower panel is for the case with noise and $n_g=10\hbox{ arcmin}^{-2}$. 
The smoothing scale for both panels is $\theta_G=3\hbox{ arcmin}$.  
Note that in the noiseless case of the upper panel, $\nu$ are $x$ are also defined using the same $\sigma_{\rm{ran},0}$ and $\sigma_{\rm{ran},2}$ as those in the case of the lower panel.

We can see that for peaks, their $\nu$ and $x$ are positively correlated. Peaks with higher height $\nu$ tend to have higher steepness $x$. However, the correlation is not perfect even in the noiseless case. We perform the Pearson Correlation analyses to 
calculate the correlation coefficient (PCC) as follows
\begin{equation}\label{eq:pcc}
r_{\nu x}=\frac{\sum_{i=1}^{n}\left(\nu_{i}-\bar{\nu}\right)\left(x_{i}-\bar{x}\right)}{\sqrt{\sum_{i=1}^{n}\left(\nu_{i}-\bar{\nu}\right)^{2}} \sqrt{\sum_{i=1}^{n}\left(x_{i}-\bar{x}\right)^{2}}}\quad ,
\end{equation}
where $i$ goes through all the peaks, and $\bar{\nu}$ and $\bar{x}$ are the respective sample mean of height and steepness. We obtain $r_{\nu x}\approx 0.85$ and $0.60$ for the upper and lower panels, respectively. The not-perfect correlation between $\nu$ and $x$ indicates that the peak statistics based on the two quantities can contain somewhat different cosmological information. 

We now proceed to analyse the peak distributions based on $\nu$ and $x$ separately for different cosmological models. To estimate the statistical fluctuations, we employ the bootstrap method to build samples from simulated maps. As described in Sec. \ref{sec:N-body simulations and ray-tracing}, from our ray-tracing simulations, we produce 96 convergence maps each with an area of $3.5^{\circ}\times 3.5^{\circ}$. For each map, we generate 10 Gaussian noise realizations for a given $n_g$ and $\theta_G$. Thus in total, we have $96$ and $96\times 10$ maps for the noiseless and noisy cases for each $n_g$ and $\theta_G$, respectively. From these maps, we build 1000 bootstrap samples each with a total area of about $1000\deg^2$ containing 100 randomly selected maps from the parent maps with replacement. The exact total area of each bootstrap sample after boundary exclusions is $1059 \hbox{ deg}^2$ for $\theta_G=2\hbox { arcmin}$ and $985  \hbox{ deg}^2$ for $\theta_G=3\hbox { arcmin}$. From the 1000 samples, we estimate the average peak numbers in each bin and the corresponding statistical fluctuations. It is admitted that the bootstrap here cannot reveal the true cosmic variance of the peak number distributions over different $\sim 1000\deg^2$ sky areas because our independent convergence maps have a total area of $1176\deg^2$. On the other hand, for peak statistics over about $1000\deg^2$, we do not expect significant cosmic variance. The statistical fluctuations obtained from our bootstrap in each bin is very close to Poisson fluctuations. 

\begin{figure*}
    \centering
    \includegraphics[scale=0.38]{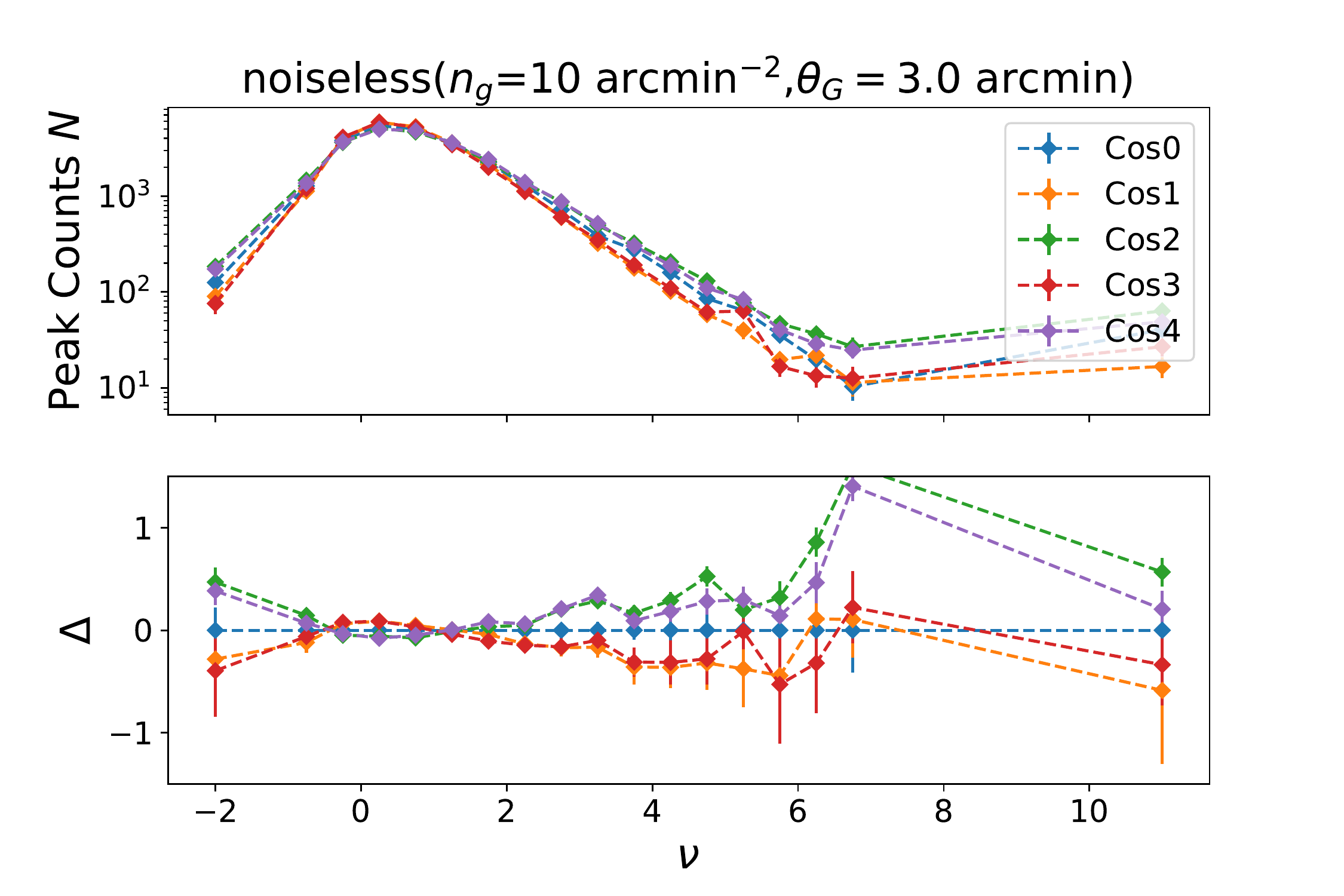}
    \includegraphics[scale=0.38]{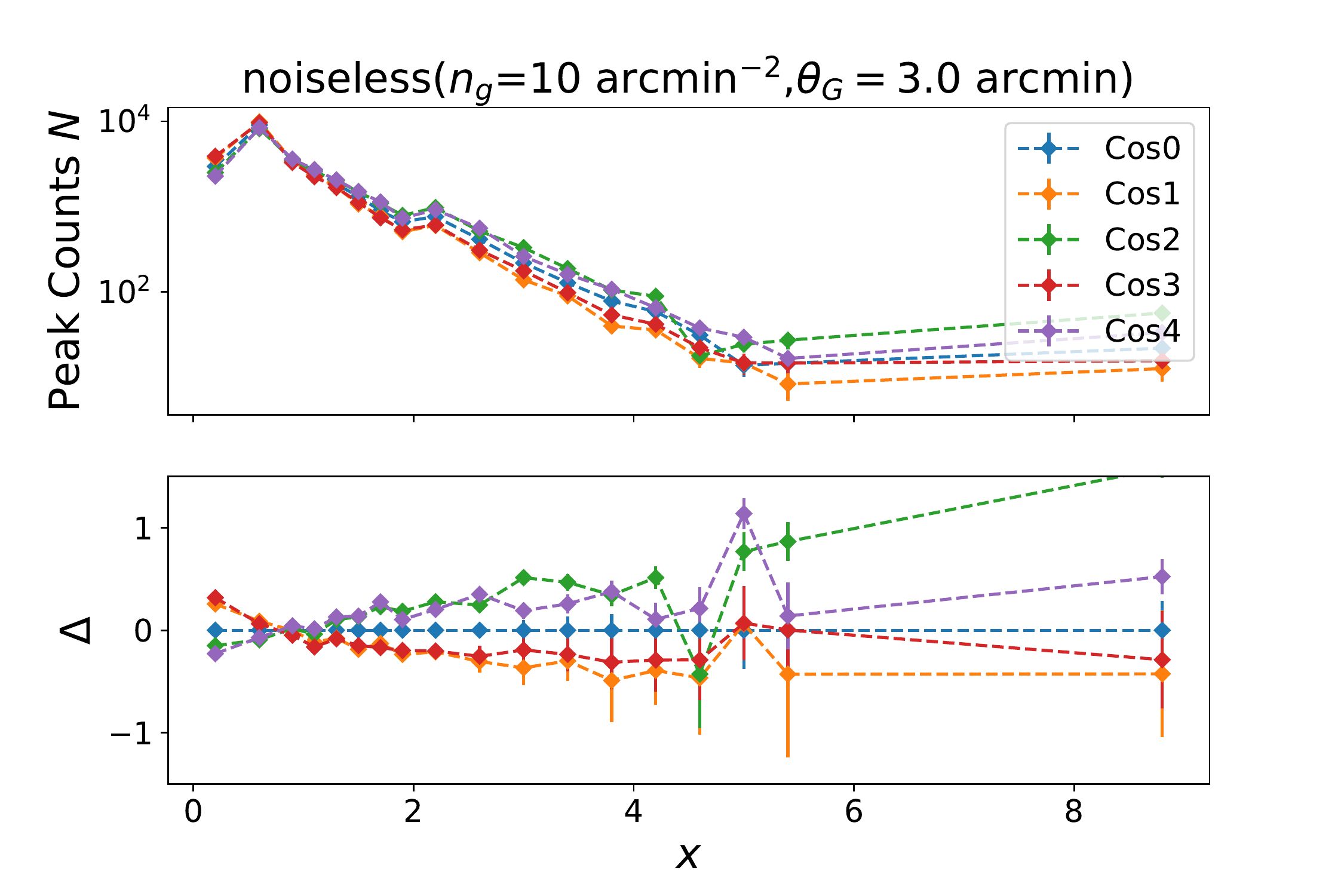}
    \includegraphics[scale=0.38]{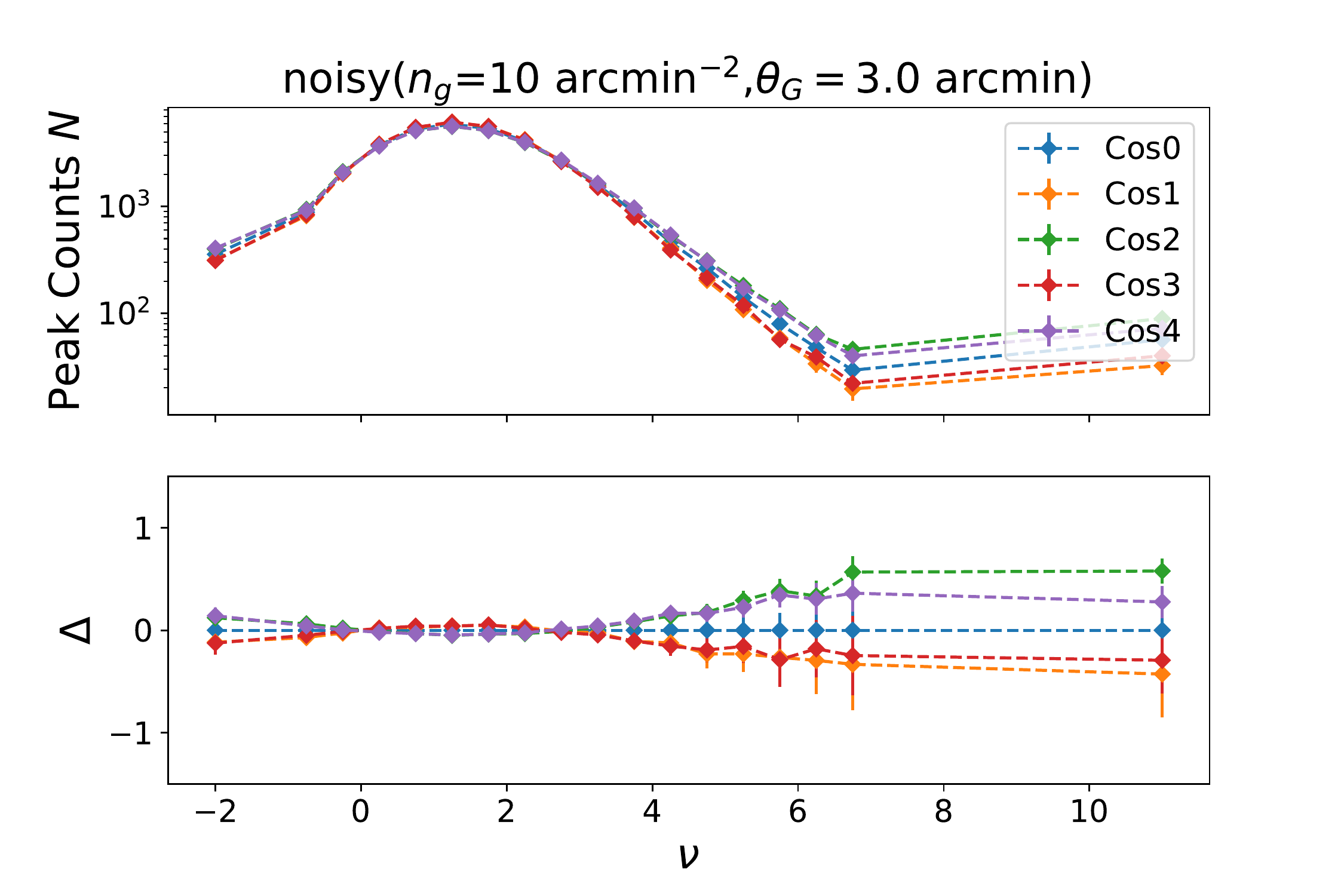}
    \includegraphics[scale=0.38]{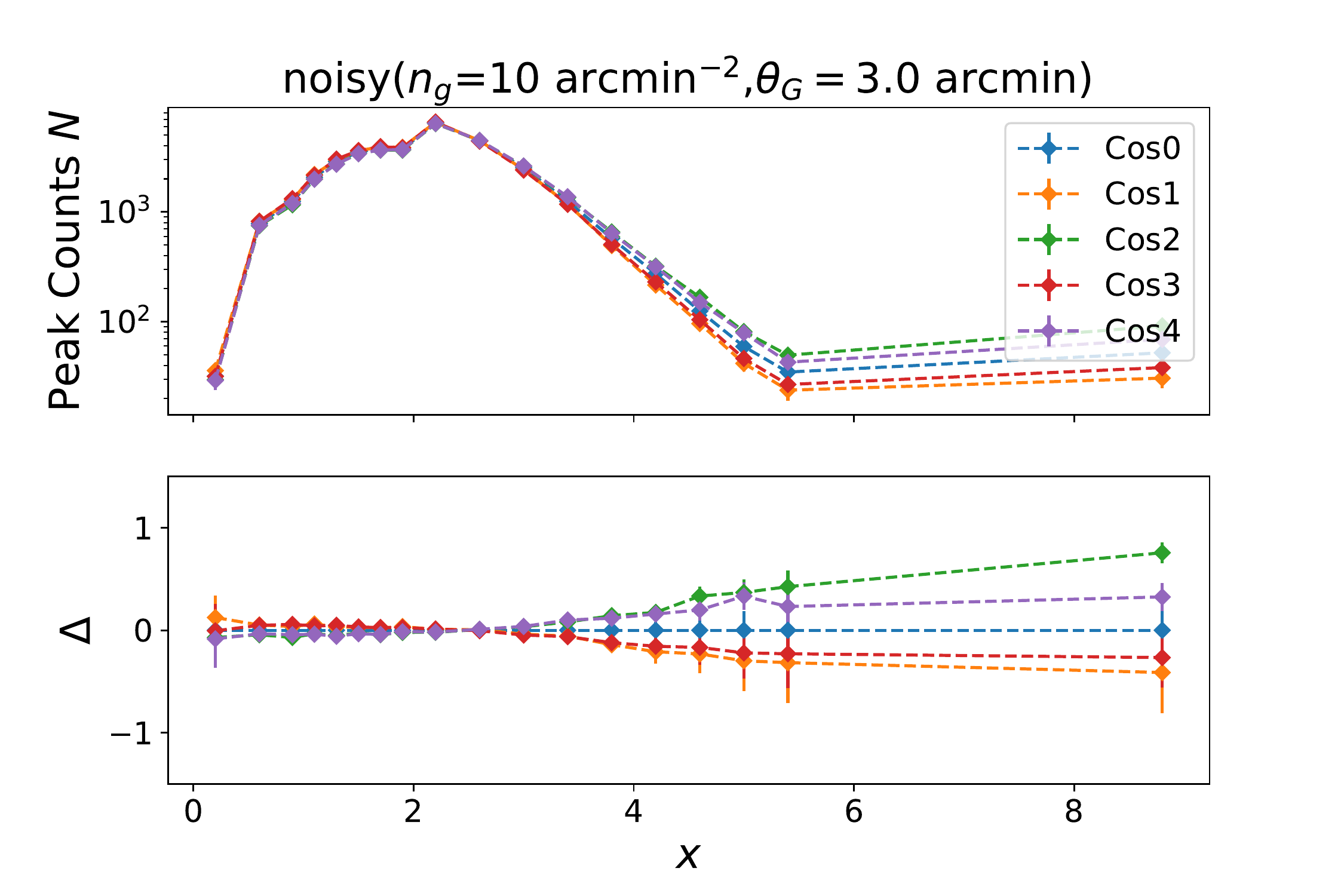}
    \caption{The peak number distributions from different cosmological models with $n_g=10\hbox{ arcmin}^{-2}$ and $\theta_G=3\hbox{ arcmin}$ with respect to the height $\nu$ (left) and the steepness $x$ (right). They are calculated from the bootstrap samples each with an area of 1000 deg$^2$.  
The top and bottom parts are for the noiseless and noisy cases, respectively. In each part, the upper ones show the mean number distributions with the errors estimated from the bootstrap samples, and the lower ones are the relative differences with respect to the fiducial model Cos0.}
    \label{fig:peak distribution}
\end{figure*}

The results of the peak distributions for different cosmological models are shown in Figure \ref{fig:peak distribution} for $n_g=10\hbox{ arcmin}^{-2}$ and $\theta_G=3\hbox{ arcmin}$. The upper and lower sets are for noiseless and noisy cases, respectively, while the left and right are for $\nu$ and $x$ distributions. For each set, $\Delta$ shows the relative differences of peak counts between the other cosmological models and the fiducial model. Again, the upper and lower panels use the same $\sigma_{\rm{ran},0}$ and $\sigma_{\rm{ran},2}$ to define $\nu$ and $x$, respectively.

It is seen that in the noiseless case, both statistics show significant cosmology dependences. For steepness, the differences between different cosmological models are apparent over the whole $x$ range. For peak height, the differences are insignificant in 
the range of $\nu\sim 0-1$, but significant for lower and higher peaks. Adding noise, the cosmology dependence is largely suppressed for low $\nu$ and low $x$ peaks, but persists for $\nu\ge 3$ and $x\ge 3$. We also note that in the noisy case, negative peaks
still show a certain level of cosmology dependence as seen in the lower left panel for the peak height statistics, while for the steepness statistics, noise dilutes strongly the cosmological information at the low $x$ side. 

In the next subsections, we will perform detailed comparisons of the two peak statistics. 

\subsection{Statistical comparisons of peak height and steepness statistics}
\label{subsection:Statistical comparisons of different cosmologies}
To investigate the differences of the two peak statistics, we first employ the $\chi^2$ approach \citep{2010PhRvD..81d3519K}. Specifically, for each statistics, we calculate $\Delta\chi^2$ between the peak counts of the fiducial cosmological model and another model. This shows to some extent the cosmology dependence  of 
the peak statistics. We then compare the reduced $\Delta\chi^2$ values of the two statistics to see the differences of their cosmological sensitivities.   

For $\Delta\chi^2$, it is defined as follows
\begin{equation}
\Delta \chi_{k}^{2}=\sum_{i j} d \bar{N}_{i}^{\left(k, f\right)}\widehat{\left(C^{(f)}\right)_{i j}^{-1}} d \bar{N}_{j}^{\left(k, f\right)},
\end{equation}
where $d \bar{N}_{i}^{\left(k, f\right)}=\bar{N}_{i}^{\left(k\right)}-\bar{N}_{i}^{\left(f\right)}$ is the difference of the mean peak number in $i$th bin between the simulation results of the cosmological model $k$ and the fiducial model $f$. The mean peak numbers are calculated by averaging over 
the 1000 bootstrap samples of a given cosmology and $n_g$ as described in the previous subsection. The covariance $C$ is computed using the bootstrap samples of the fiducial cosmological model by
\begin{equation}\label{eq:covariance matrix}
C_{i j}^{(f)}=\frac{1}{R-1} \sum_{r=1}^{R}\left(N_{i}^{(f ; r)}-\bar{N}_{i}^{(f)}\right)\left(N_{j}^{(f ; r)}-\bar{N}_{j}^{(f)}\right),
\end{equation}
where $R=1000$ is the total number of bootstrap realizations, and $N_{i}^{(f ; r)}$ and $\bar{N}_{i}^{(f)}$ are the numbers of peaks in bin $i$ from the realization $r$ and that of the mean over the $1000$ samples, respectively. Its unbiased inverse is given by \citep{2007A&A...464..399H}
\begin{equation}\label{inverse}
\widehat{C^{-1}}=\frac{R-N_{\text {bin }}-2}{R-1}C^{-1}, \quad N_{\text {bin}}<R-2,
\end{equation}
where $C^{-1}$ is the inverse of $C$ and $N_{\text {bin }}$ is the number of bins for peak counts. 

\begin{figure*}
\includegraphics[scale=0.38]{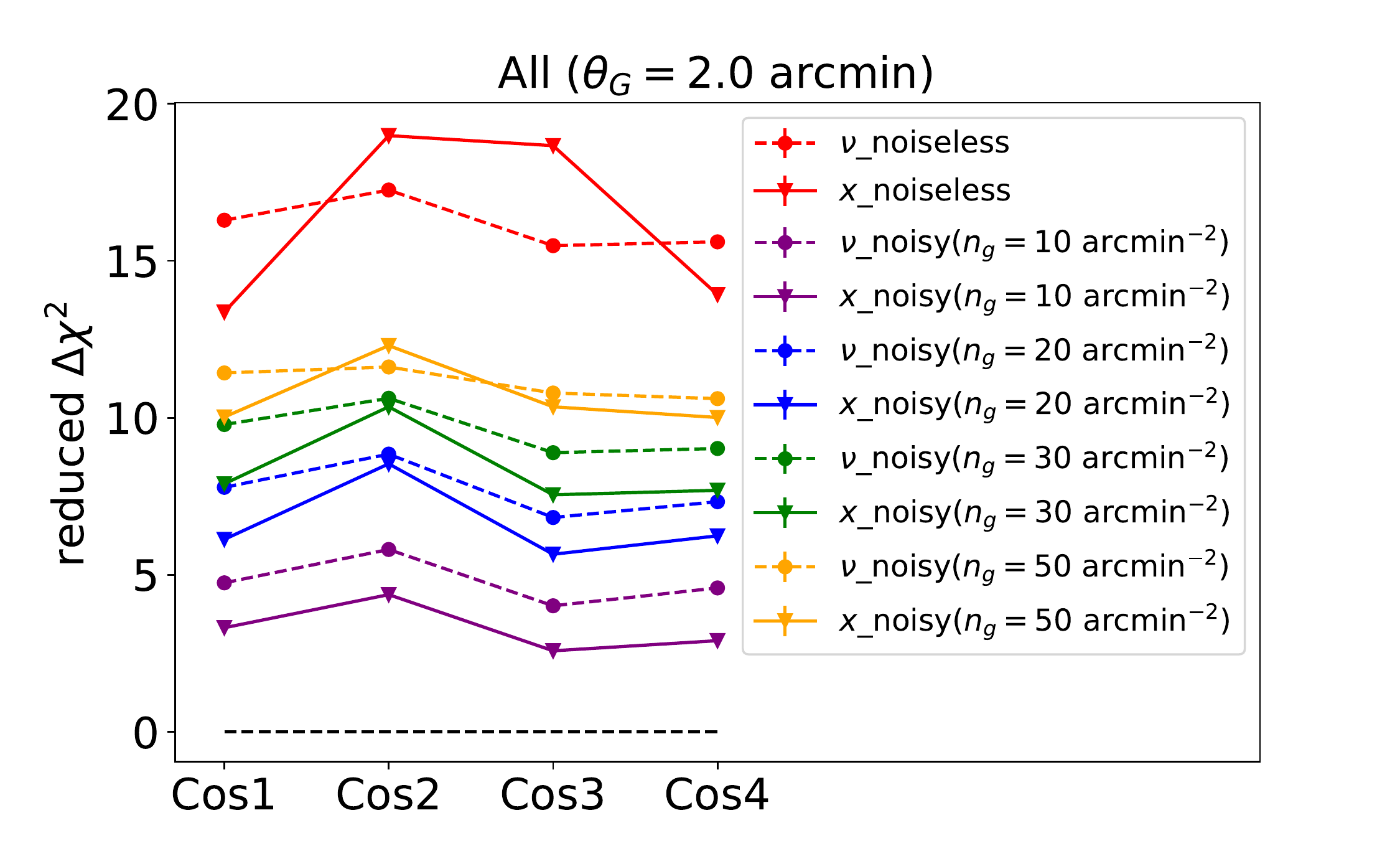}
\includegraphics[scale=0.38]{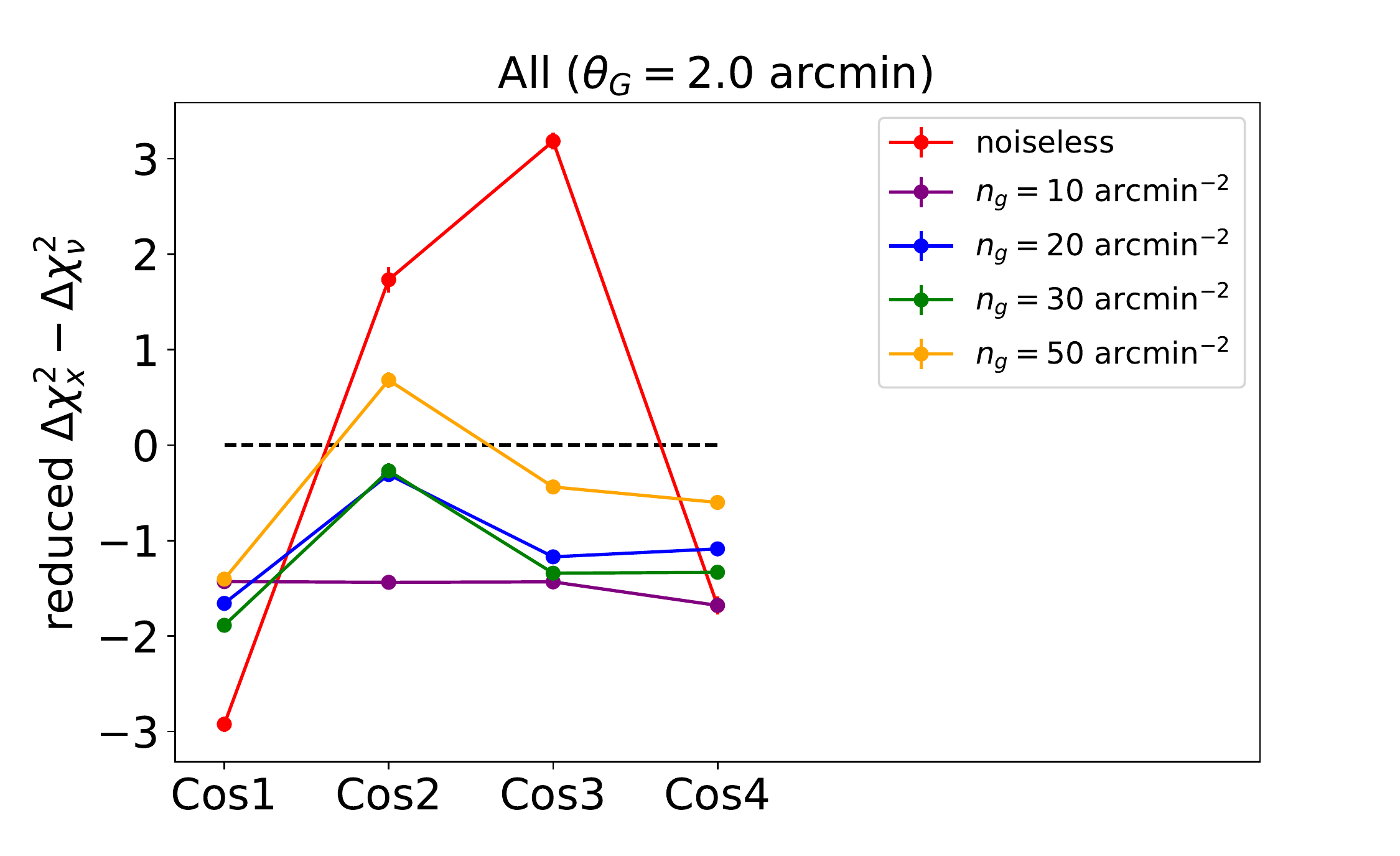}
\includegraphics[scale=0.38]{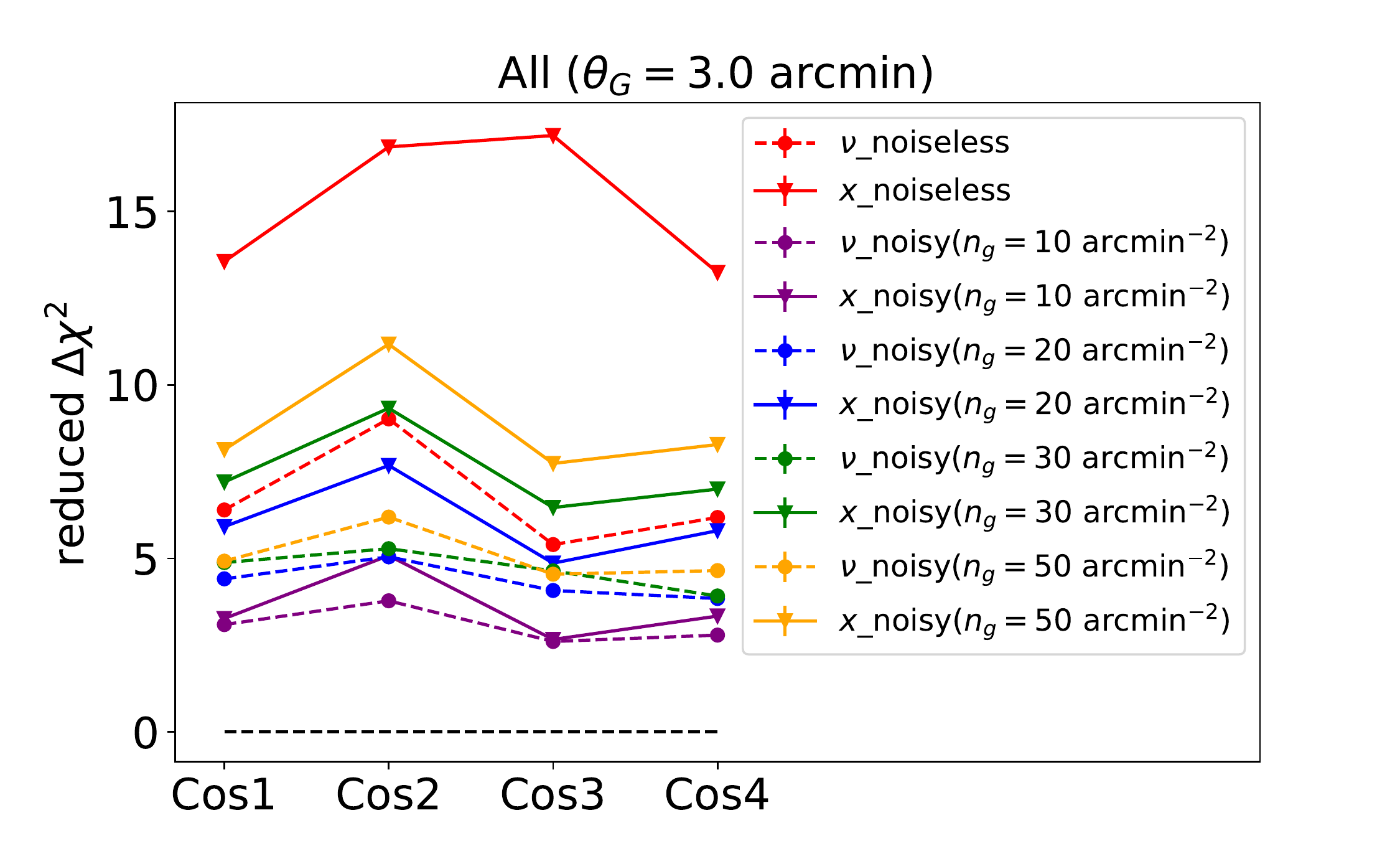}
\includegraphics[scale=0.38]{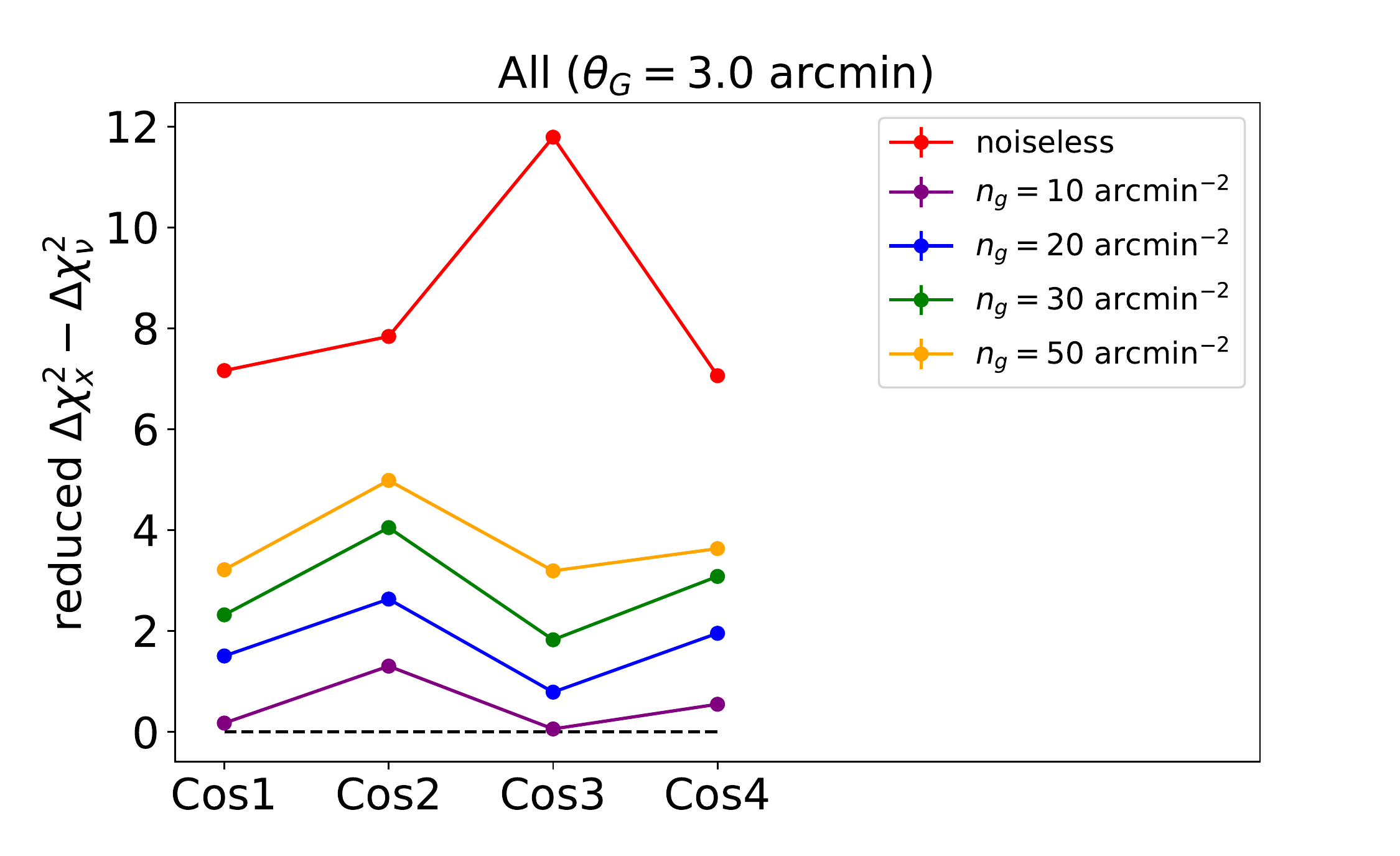}
\caption{The reduced $\Delta\chi^2$ from the {\it All} samples each considering all the peaks found in 1000 bootstrap samples of different cosmological models with respect to the fiducial one Cos0 are shown in the left panels for both the peak height (solid lines) and steepness (dashed lines) statistics with different $n_g$ and $\theta_G=2$ (upper) and $3\hbox{ arcmin}$ (lower). 
The corresponding $\Delta\chi^2$ differences between the steepness and the height statistics are shown in the right panels. The different colours refer to different shape noise levels.}
\label{fig:delta_chi2_dif}
\end{figure*}

In Figure \ref{fig:delta_chi2_dif}, we present the results of {\it All} samples, where the left panels are reduced $\Delta\chi^2$ for each statistics with different $n_g$ and the right ones are the differences of the reduced $\Delta\chi^2$ between the corresponding peak steepness and height statistics. The upper and lower panels are for 
$\theta_G=2$ and $3\hbox{ arcmin}$, respectively.

It is seen that for each of the peak statistics, the reduced $\Delta\chi^2$ increases with the increase of $n_g$ and thus the decrease of the shape noise. For $\theta_G=3\hbox{ arcmin}$, we have reduced $\Delta\chi^2_x-\Delta\chi^2_\nu>0$ for all the considered cases (lower right), and the difference increases with the decrease of the
shape noise, showing that the peak steepness statistics is more sensitive to cosmology than its height counterpart. 
For $\theta_G=2\hbox{ arcmin}$, we see negative values of $\Delta\chi^2_x-\Delta\chi^2_\nu$ for $n_g\le 30 \hbox{ arcmin}^{-2}$. 
It gradually increases to be positive for the model of Cos2 for       
$n_g=50\hbox{ arcmin}^{-2}$, and for the noiseless case, it is positive for Cos2 and Cos3. 
This should be attributed to the larger noise level in the case of $\theta_G=2\hbox{ arcmin}$ than that of $\theta_G=3\hbox{ arcmin}$, and thus the advantages of the steepness statistics are suppressed more. 
It is noted however that the $\Delta\chi^2$ comparisons here only show part of the cosmology dependences of the two peak statistics because Cos1 to Cos4 are different from the fiducial model Cos0 by only one parameter, $\sigma_8$ or $\Omega_{\rm m}$.   
Later in this section, we will investigate the differences of the two statistics from the Fisher analyses.  

\begin{figure}
    \centering
    \includegraphics[scale=0.4]{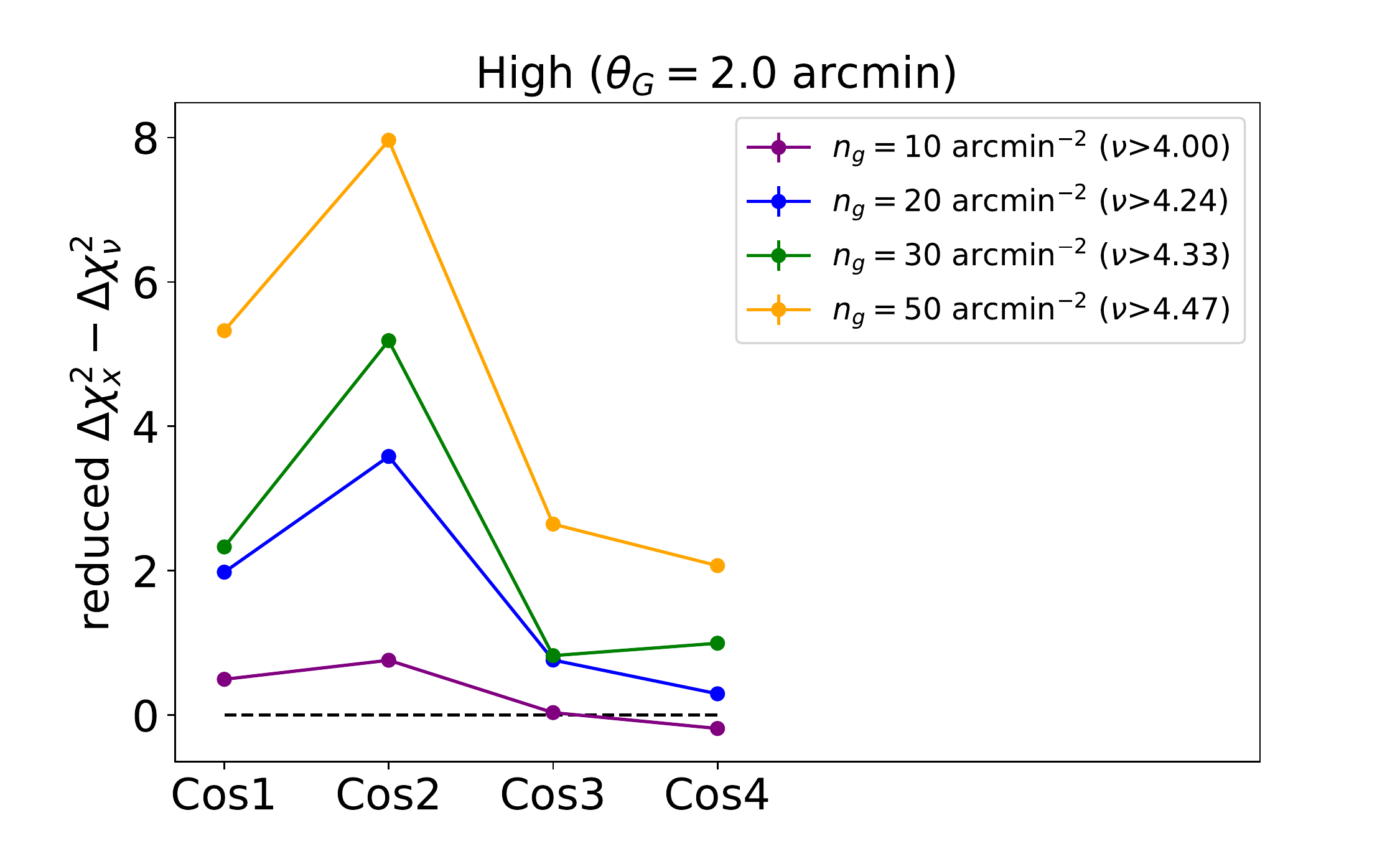}
    \includegraphics[scale=0.4]{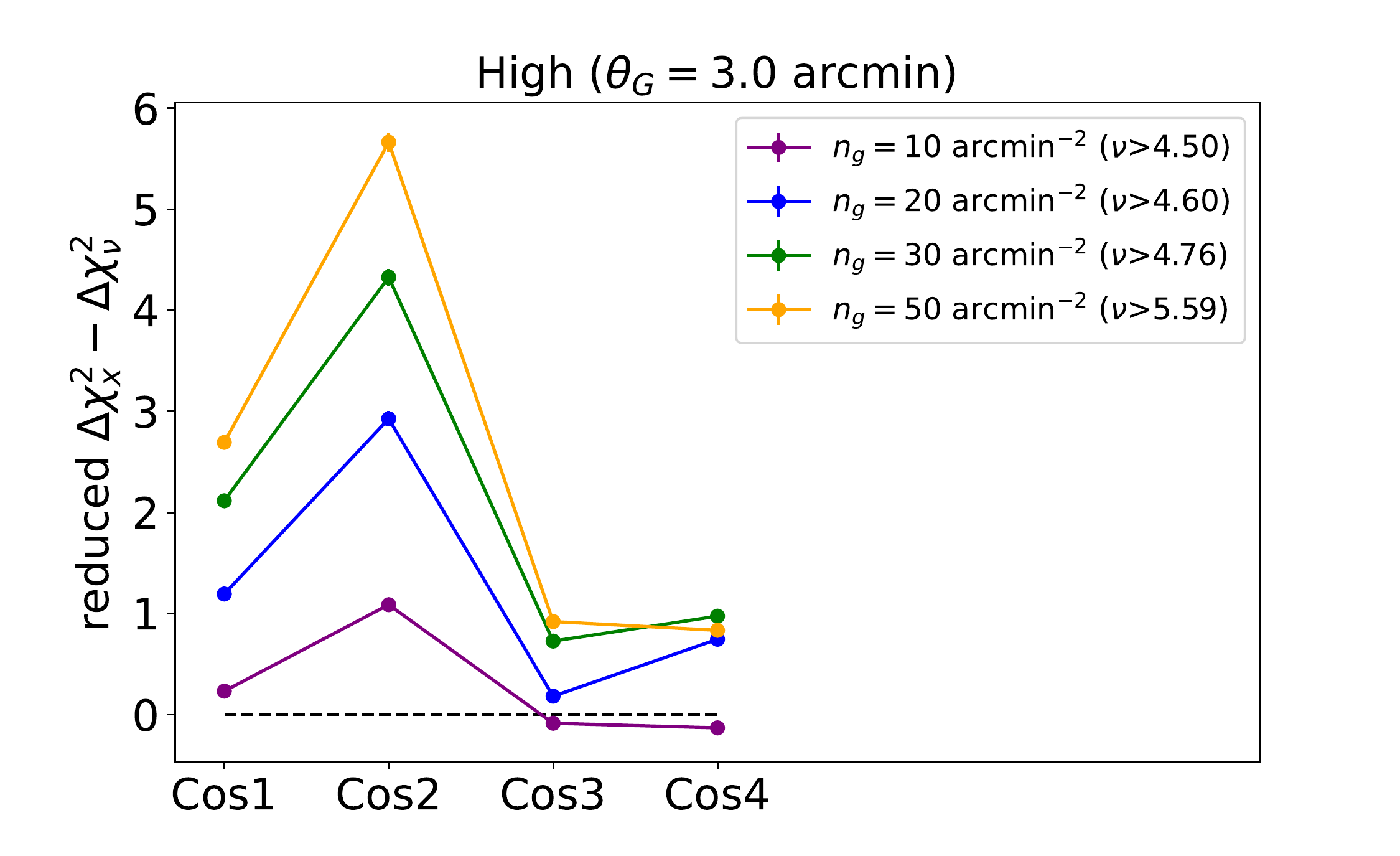}
    \caption{The reduced $\Delta\chi^2$ differences for the {\it High} samples. For these samples, we apply lower cuts on peak height with their values in different $n_g$ cases specified in the legends, corresponding to the cuts listed in Table \ref{table:binning for High sample}.}
    \label{fig:delta_chi2_diff_High}
\end{figure}

For the {\it High} samples, the $\Delta\chi^2$ comparisons are shown in Figure \ref{fig:delta_chi2_diff_High}. We see that for high peaks mostly arising from massive haloes, $\Delta\chi^2_x-\Delta\chi^2_\nu>0$ for nearly all the considered $n_g$ and for both smoothings. 
In Sec. \ref{chapter5}, we analyse the underlying physics leading to the differences of the two statistics for high peaks with the help of our theoretical model. 

We now compare the two peak statistics further using the Fisher information approach. The Fisher matrix is given by \citep[e.g.,][]{1997ApJ...480...22T,2014MNRAS.445.1687H}
\begin{equation}\label{eq:Fisher}
F_{\alpha\beta}=\sum_{i j}\left.\frac{\partial N_{i}}{\partial p_{\alpha}}\right|_{f}\widehat{\left(C^{(f)}\right)_{i j}^{-1}}\left.\frac{\partial N_{j}}{\partial p_{\beta}}\right|_{f},
\end{equation}
where the derivatives with respect to the cosmological parameters $p_{\alpha}$ are estimated from the simulated peak numbers in each bin for each statistics by the double-sided derivative estimator \citep{2013MNRAS.432.1338M}
\begin{equation}
\begin{aligned}
&\left.\frac{\partial N_{i}}{\partial p_{\alpha}}\right|_{f} \\
&=\frac{1}{R} \sum_{r=1}^{R} \frac{N_{i}^{(r)}\left(p_{\alpha}+\Delta p_{\alpha}\right)-N_{i}^{(r)}\left(p_{\alpha}-\Delta p_{\alpha}\right)}{2 \Delta p_{\alpha}}.
\end{aligned}
\end{equation}
Here again $R=1000$ is the total number of bootstrap realizations, and $\ \Delta p_\alpha=0.05$ and $0.03$ for $\sigma_8$ and $\Omega_\text{m}$, respectively, from our simulations. 
The unbiased inverse covariance matrix $\widehat{\left(C^{(f)}\right)_{i j}^{-1}}$ is calculated by Eq.(\ref{inverse}) using the bootstrap samples of the fiducial cosmological model without taking into account its cosmology dependence. 

\begin{figure*}
    \flushleft
    \includegraphics[ scale=0.285]{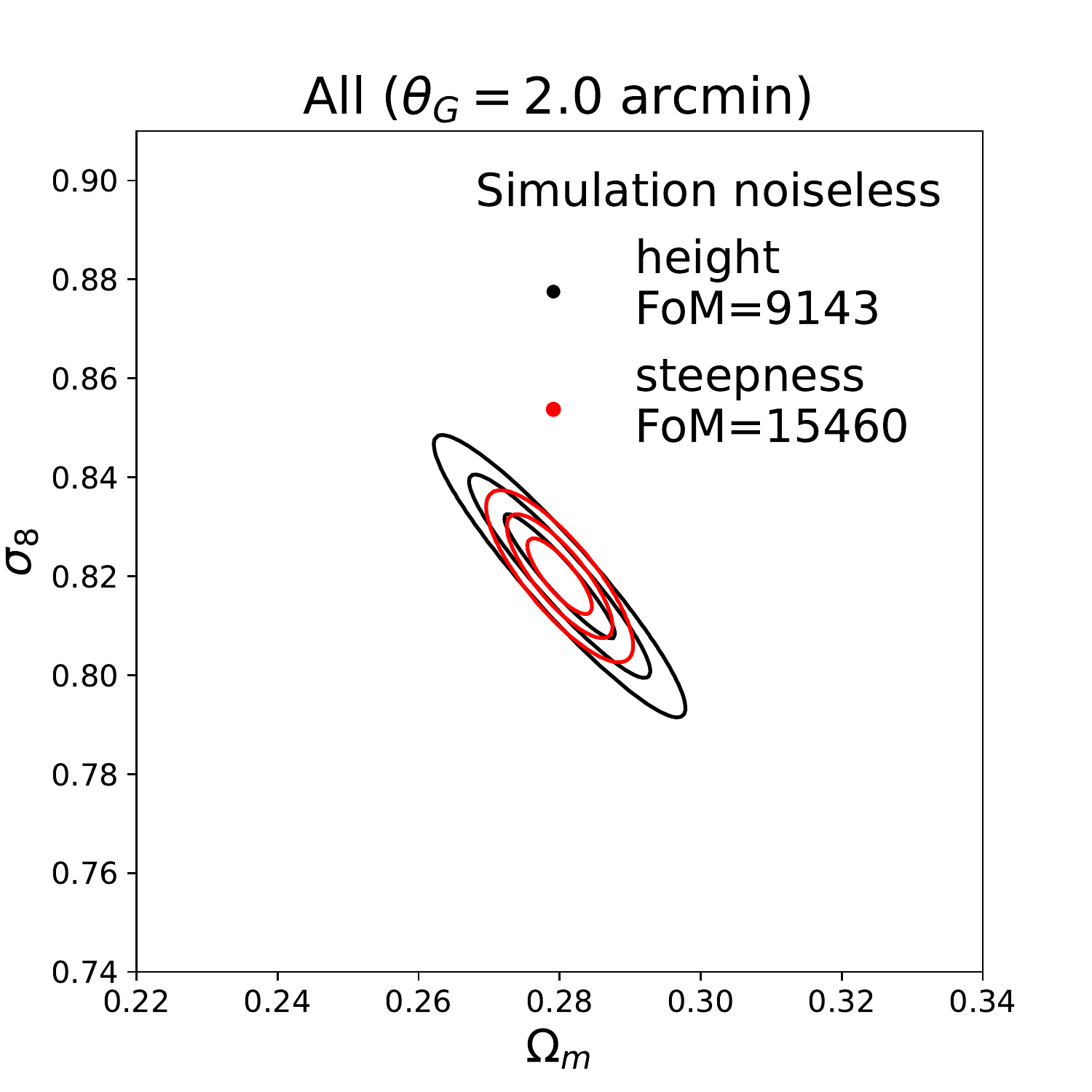}\\
    \flushleft
    \includegraphics[ scale=0.285]{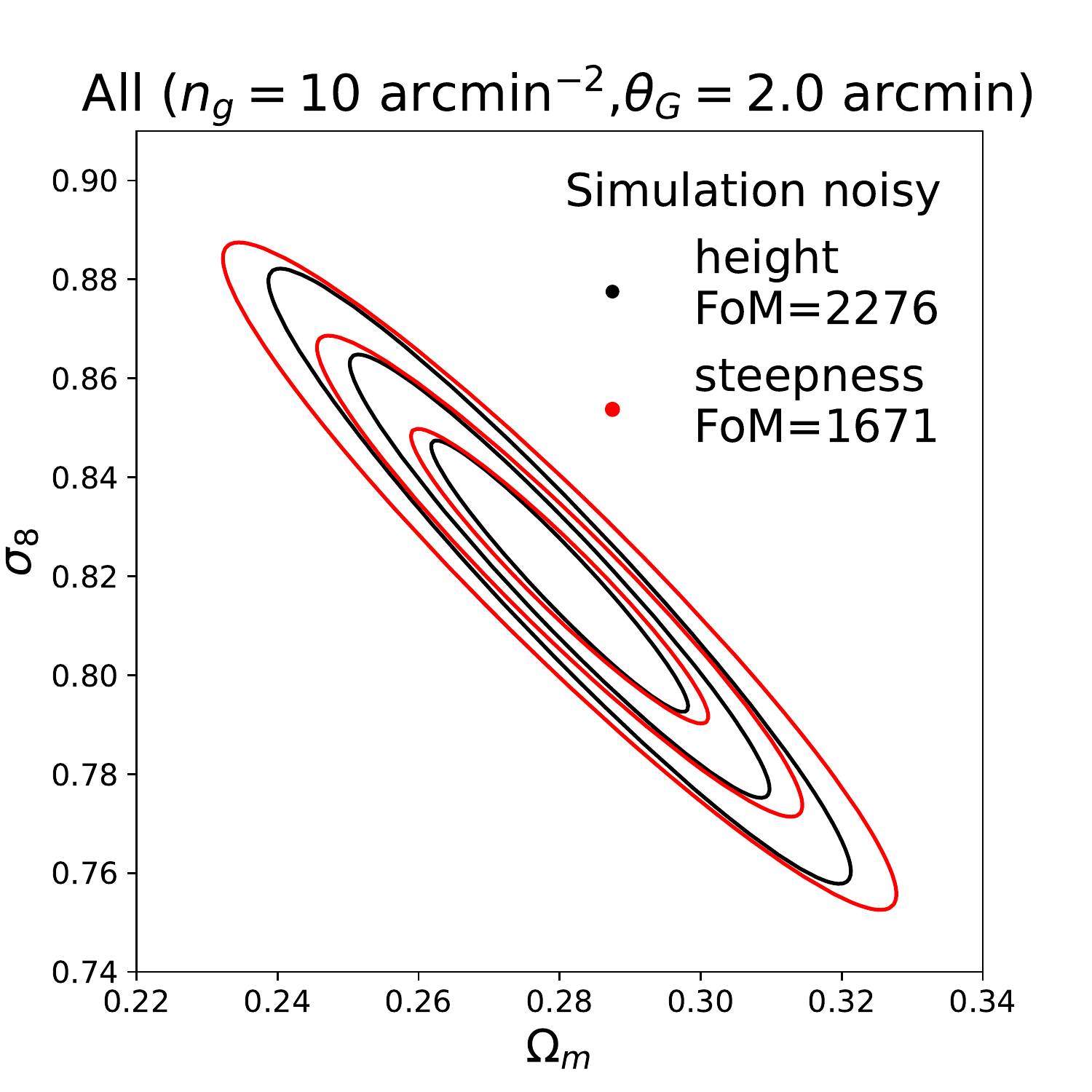}
    \includegraphics[ scale=0.285]{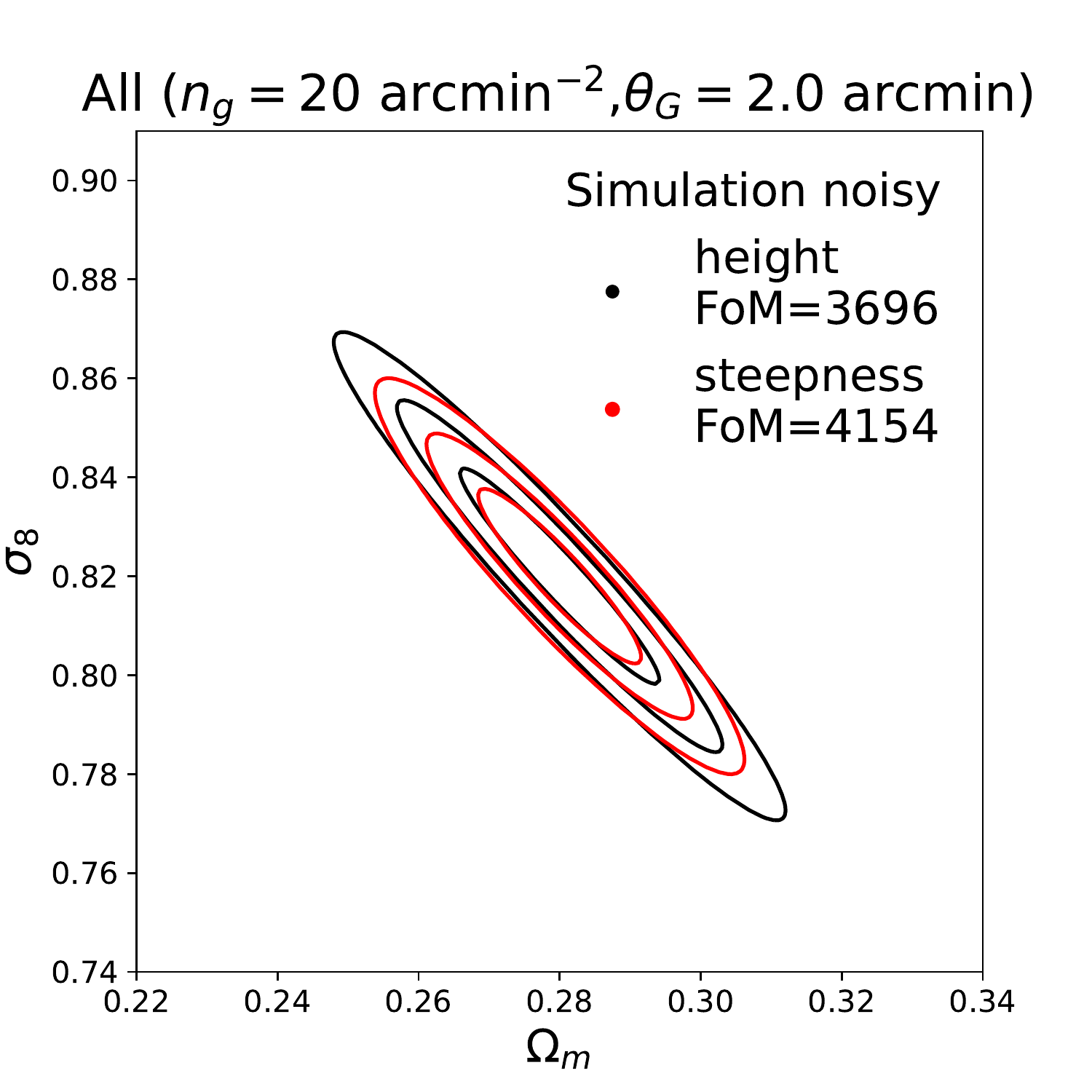}
    \includegraphics[ scale=0.285]{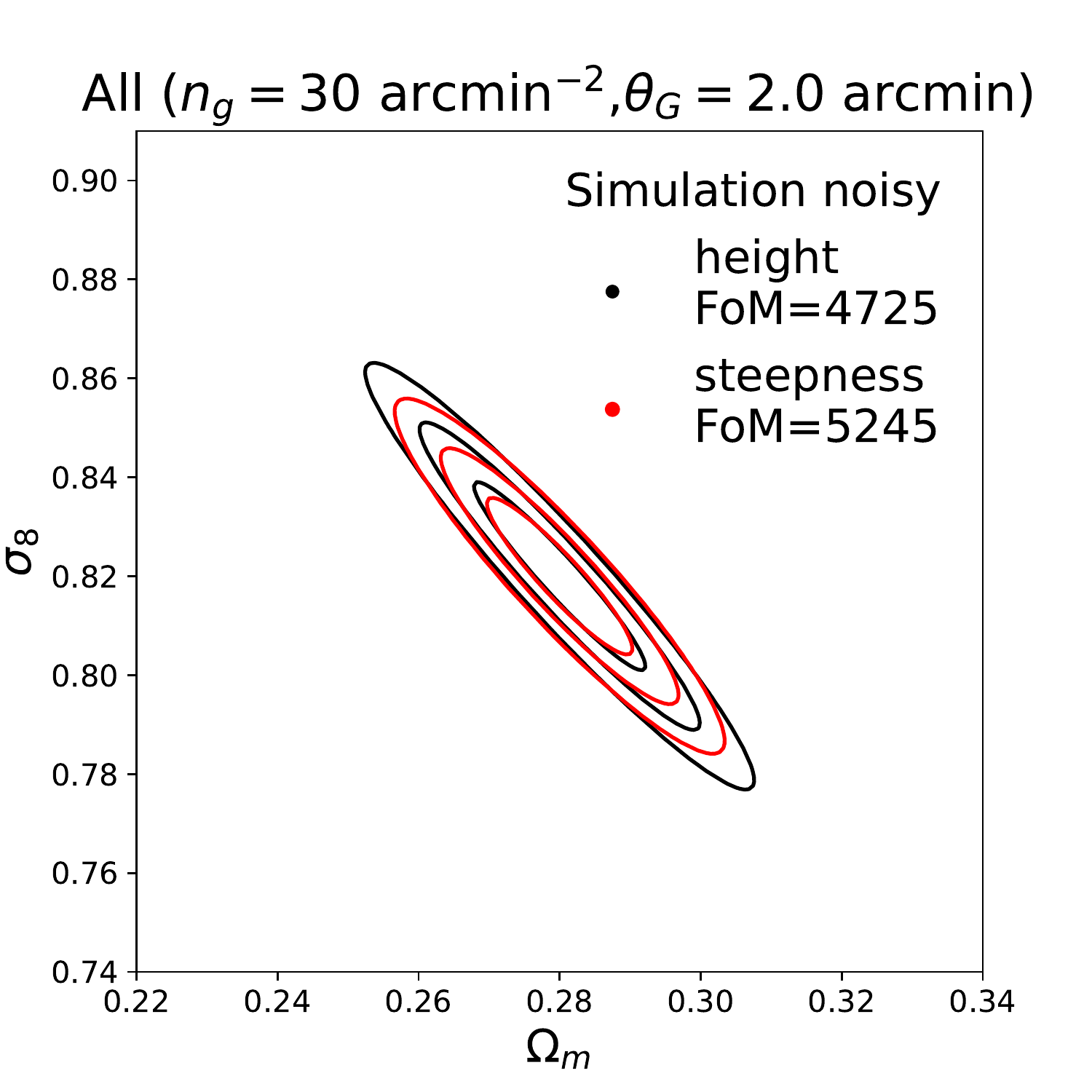}
    \includegraphics[ scale=0.285]{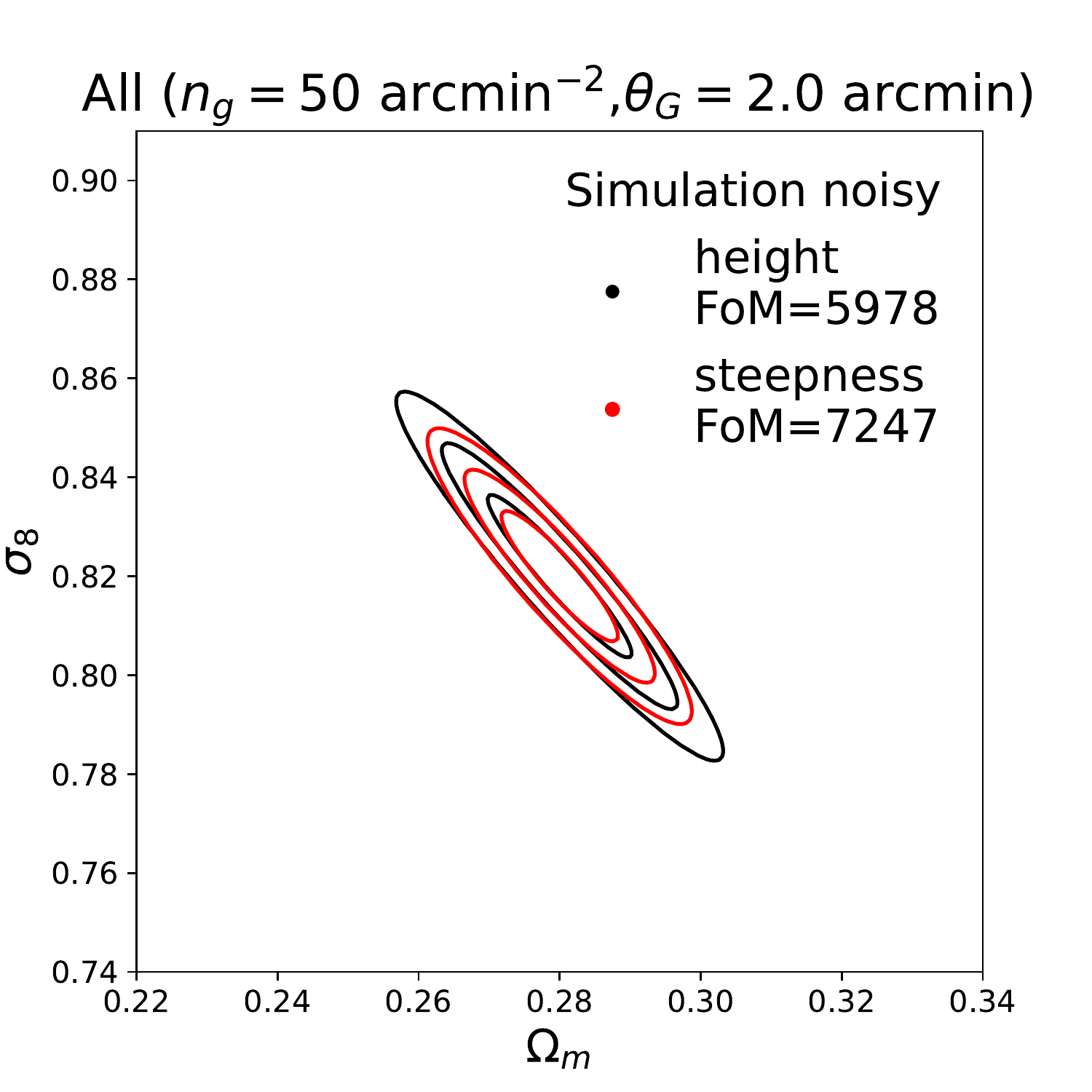}\\
    \includegraphics[ scale=0.285]{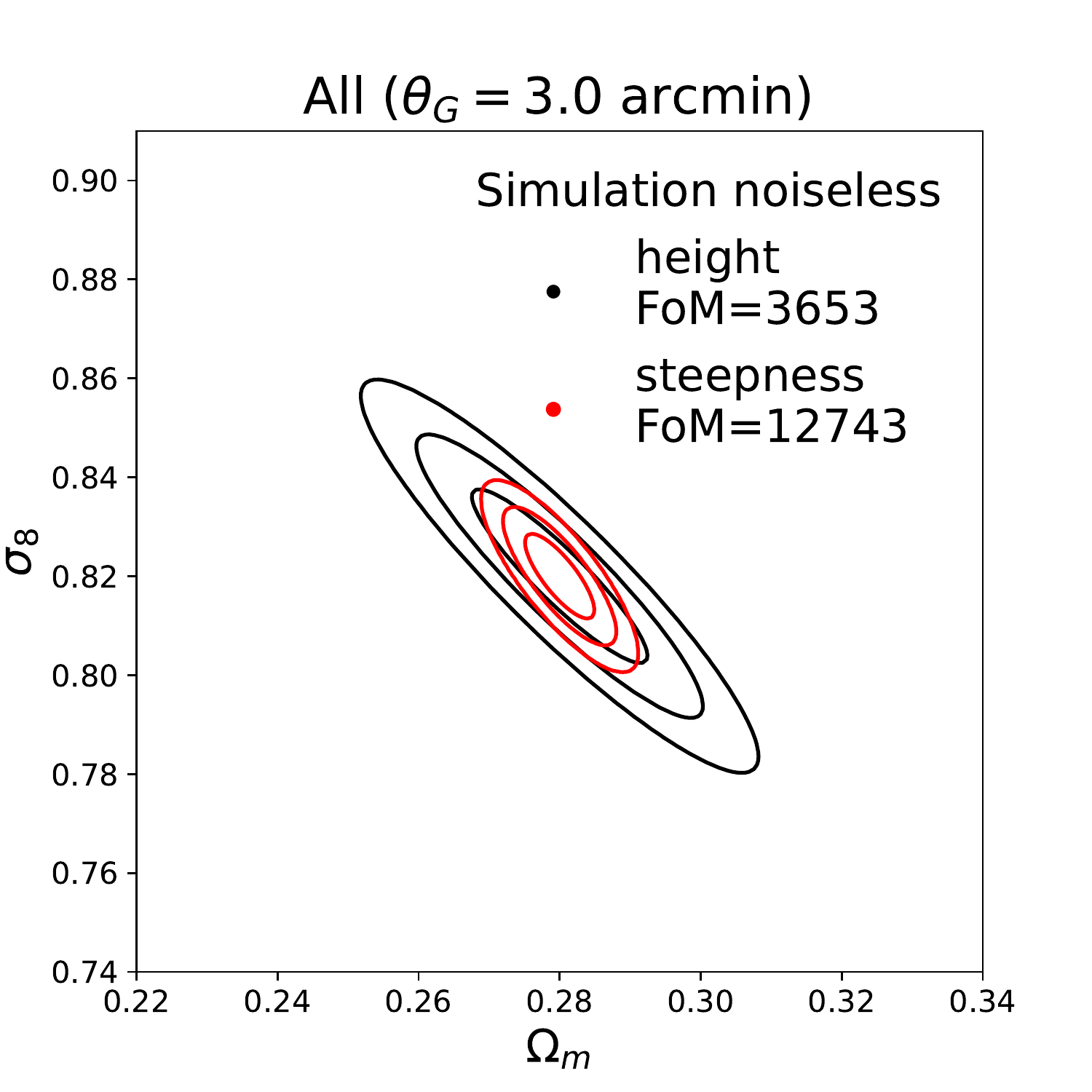}\\
    \flushleft
    \includegraphics[ scale=0.285]{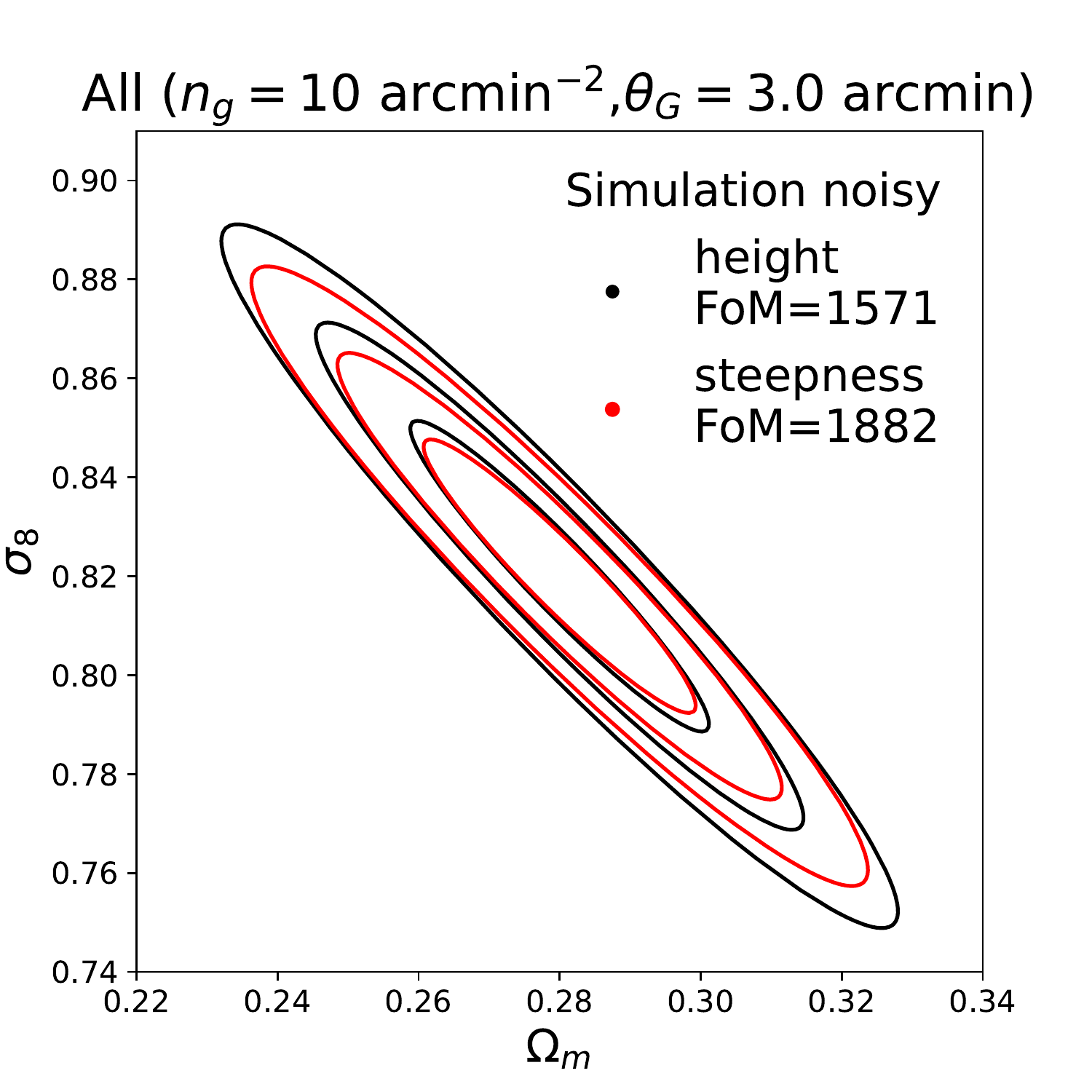}
    \includegraphics[ scale=0.285]{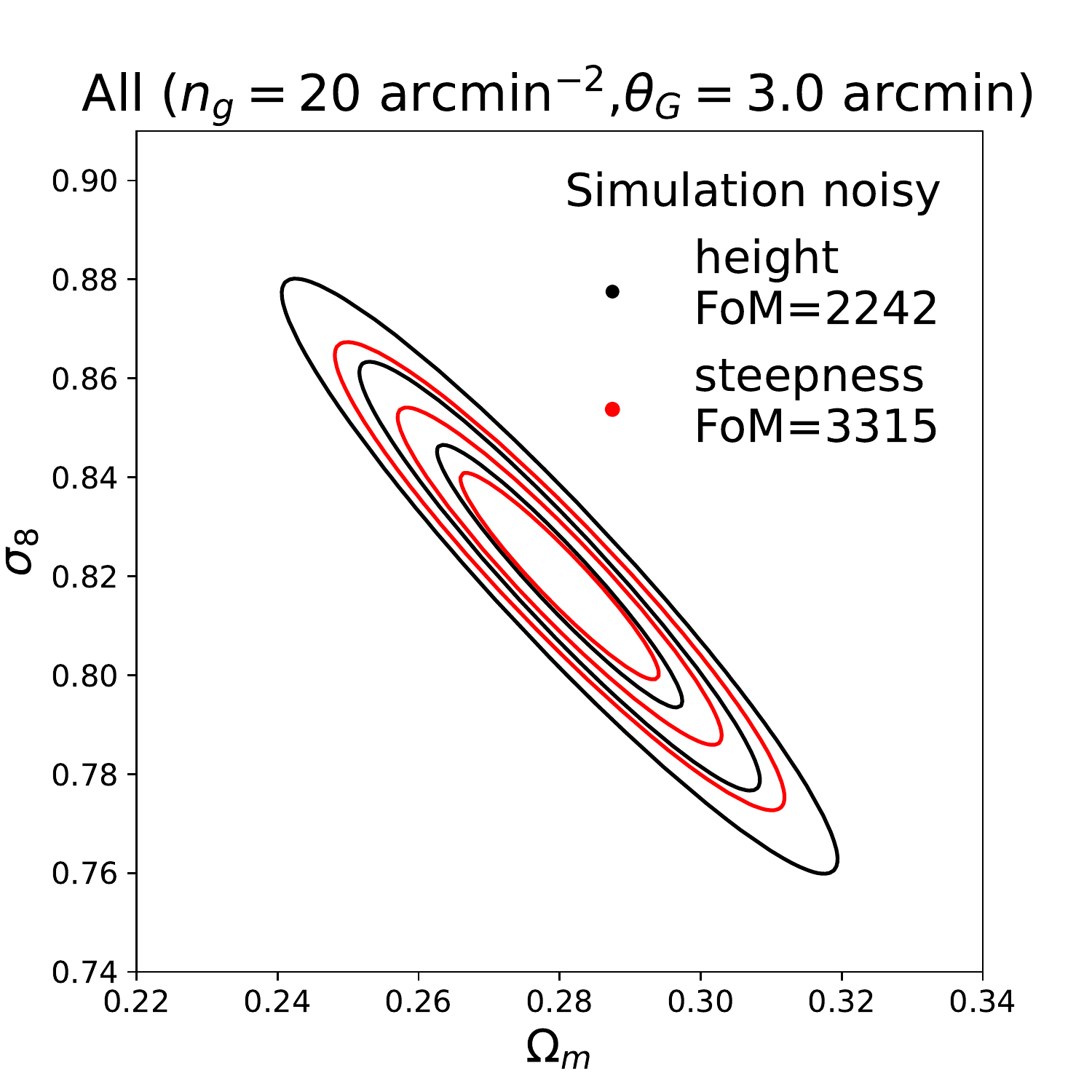}
    \includegraphics[ scale=0.285]{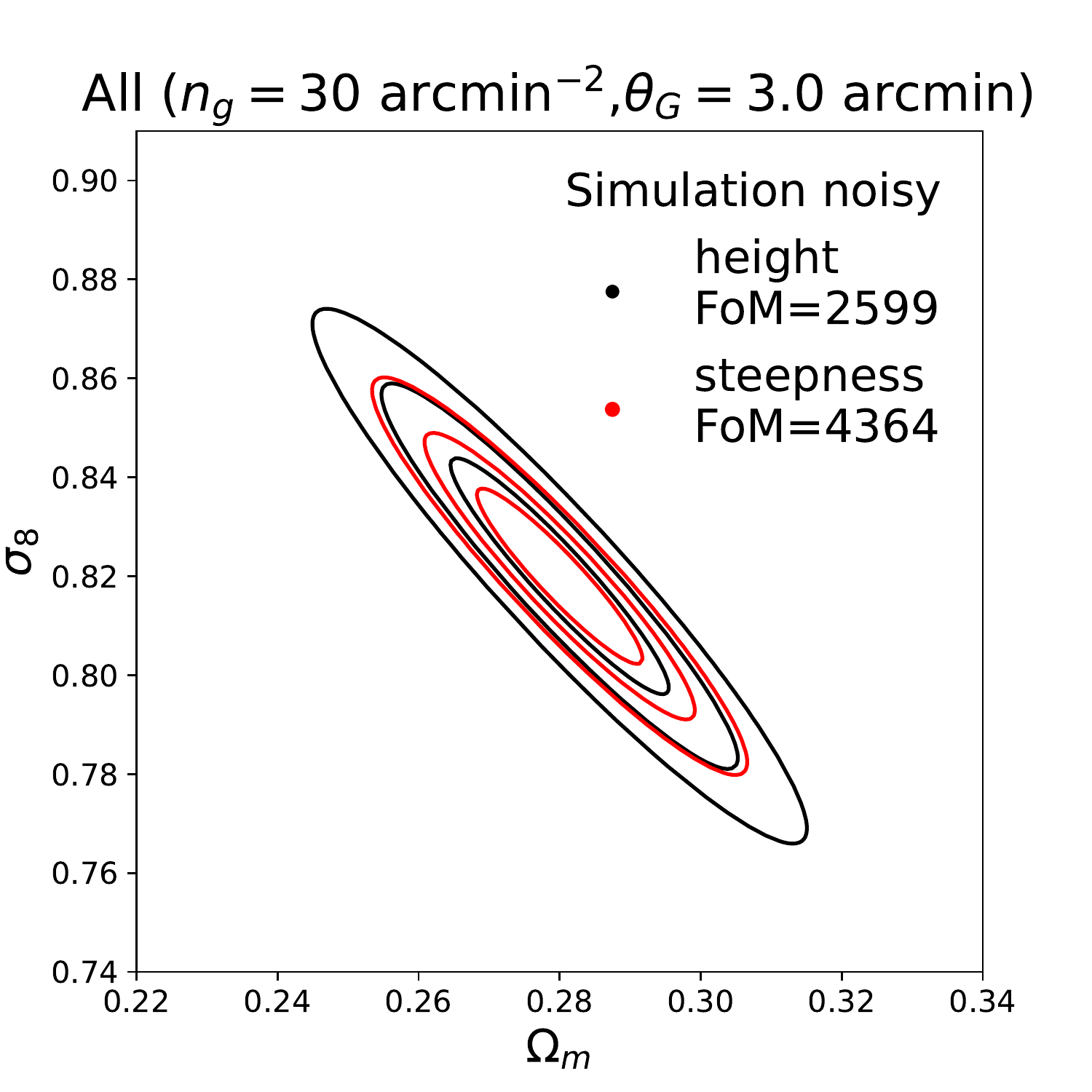}
    \includegraphics[ scale=0.285]{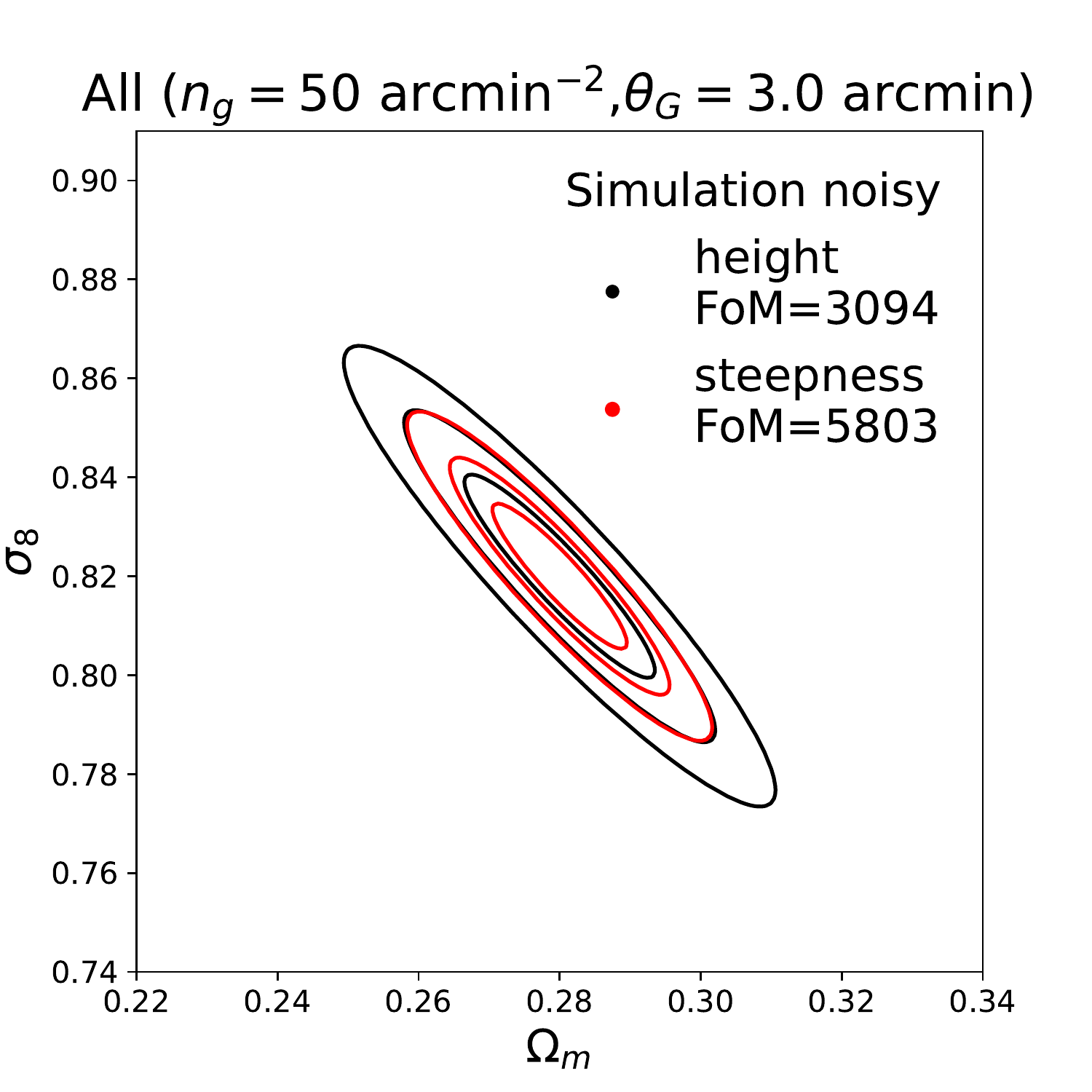}
    
    \caption{The Fisher posterior probability distributions with respect to Cos0 in the $\Omega_{\rm m}-\sigma_8$ plane from the {\it All} samples. The top and bottom parts are for $\theta_G=2$ and $\theta_G=3$ arcmin, respectively. In each part, the upper panel is the noiseless result and the lower ones are 
for the noisy cases with different $n_g$. The black and red contours are from the peak height and steepness statistics, respectively. From inner to outer contours are the 68\%, 95 \% and 99 \% confidence levels.}
    \label{fig:Fisher_all}
\end{figure*}

The resulted Fisher posterior probability distributions in $\Omega_\text{m}-\sigma_8$ space for the {\it All} samples are presented in Figure \ref{fig:Fisher_all}, where the figure of merit (FoM) is defined as the inverse of 68\% confidence area. 
For $\theta_G=3\hbox{ arcmin}$ (lower part), the peak counts based on steepness lead to tighter cosmological constraints (red contours) in all the cases than that based on peak height (black contours), in line with the trend shown in Figure \ref{fig:delta_chi2_dif}. With the decrease of the shape noise, the differences of the constraints
derived from the two statistics become more apparent. 

For $\theta_G=2\hbox{ arcmin}$ (upper part), at $n_g=10\hbox{ arcmin}^{-2}$, the peak height statistics give rise to a slightly better cosmologcial constraints than its steepness counterpart. With the increase of $n_g$, the steepness statistics tend to deliver somewhat tighter constraints, 
but the differences are less significant than that of the case with $\theta_G=3\hbox{ arcmin}$. This trend is also more or less consistent with the $\Delta\chi^2$ comparisons of Figure \ref{fig:delta_chi2_dif}. On the other hand, in Fisher analyses, the cross terms involving 
$\left.\frac{\partial N_{i}}{\partial p_{\alpha}}\right|_{f}\widehat{\left(C^{(f)}\right)_{i j}^{-1}}\left.\frac{\partial N_{j}}{\partial p_{\beta}}\right|_{f}$ with $\alpha\ne \beta$ contribute, but they are absent in $\Delta\chi^2$ calculations. Thus for $\theta_G=2\hbox{ arcmin}$, we see
slightly tighter constraints from steepness statistics based on Fisher analyses for $n_g=20, 30$ and $50\hbox{ arcmin}^{-2}$ while the corresponding $\Delta\chi^2_x-\Delta\chi^2_\nu<0$. Noting that in these cases, $\Delta\chi^2_x-\Delta\chi^2_\nu$ values are already close to zero and  
thus the effects of the cross terms are able to inverse the trend from Fisher analyses in comparison with that from $\Delta\chi^2$ studies.  

Our analyses here show that the level of shape noise can affect the differences of the cosmological inferences from the two peak statistics. Lowering the noise level leads to more significant advantages of the statistics based on steepness over that based on height.   

The Fisher results for {\it High} samples will be presented in the next section. There with the aid of our theoretical model, we will investigate the causes leading to the differences of the two peak statistics. 

\section{Understanding the differences of the two peak statistics}\label{chapter5}
Analyses in the previous section show that for a same set of peaks, statistics based on height and steepness show a certain level of 
differences in cosmological dependence. Here we employ our theoretical model for high peaks to explore the physical causes for the differences
by comparing the model predictions with the simulation results.

\subsection{The peak distributions}
\begin{figure*}
    \centering
    \includegraphics[scale=0.38]{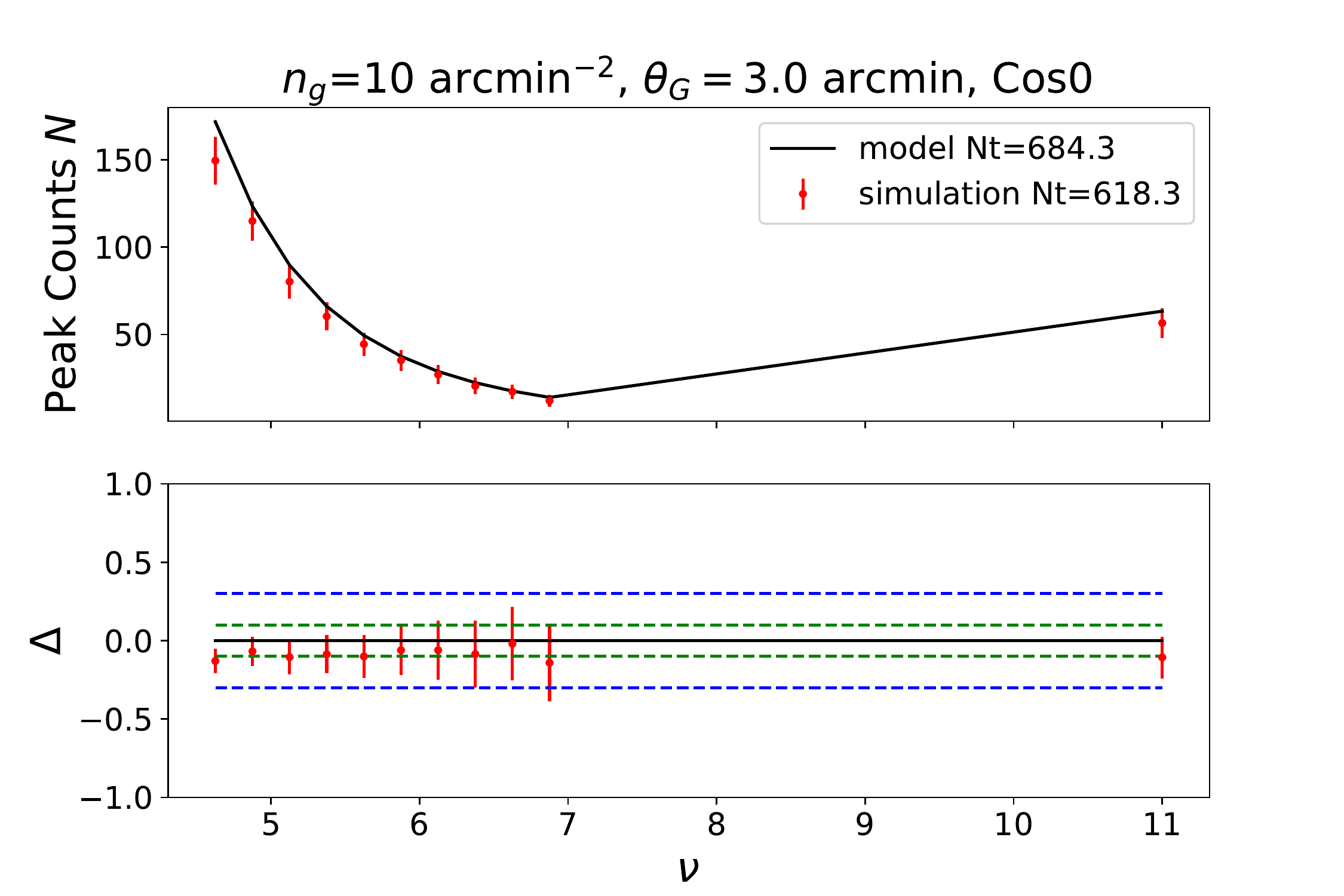}
    \includegraphics[scale=0.38]{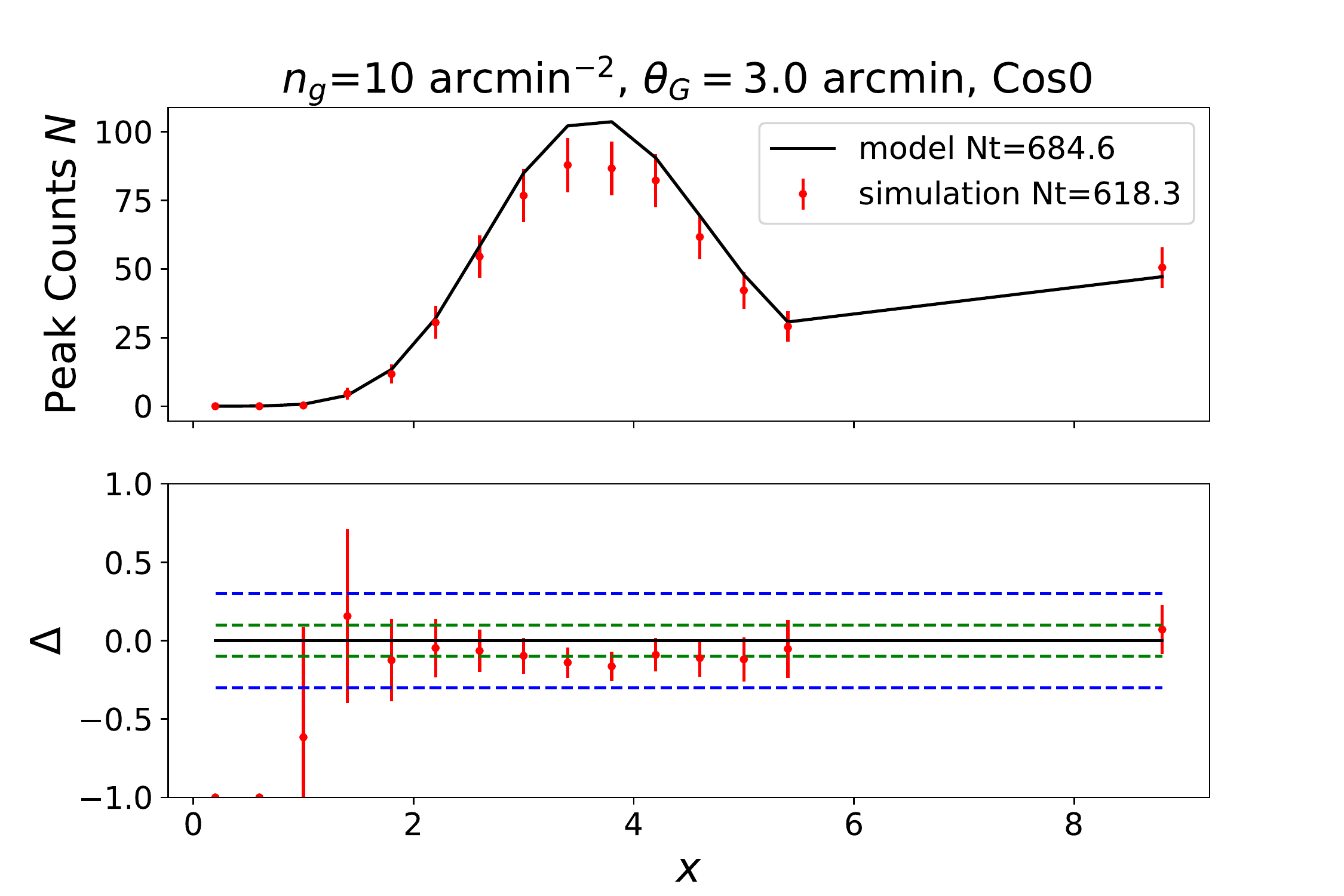}
    \includegraphics[scale=0.38]{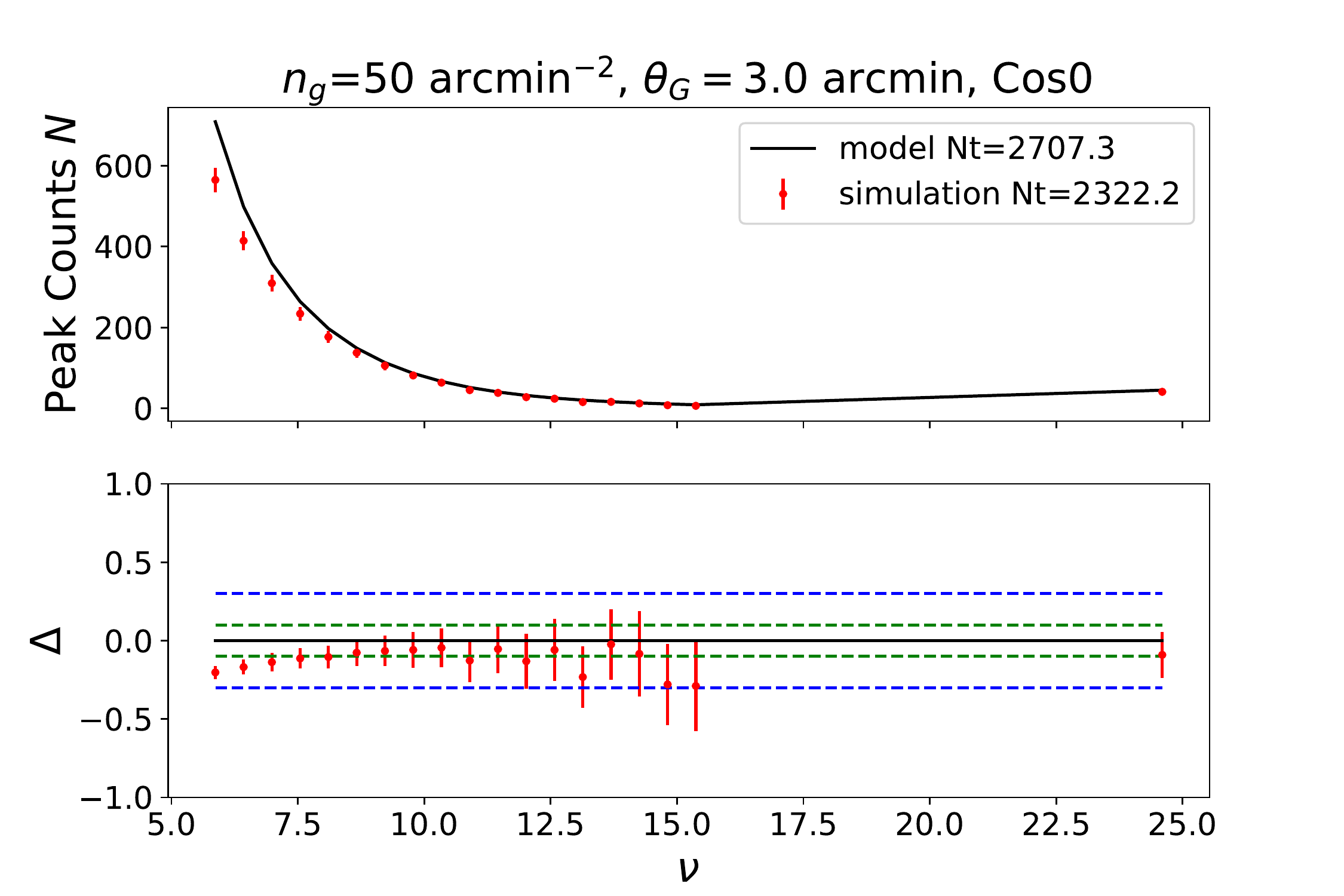}
    \includegraphics[scale=0.38]{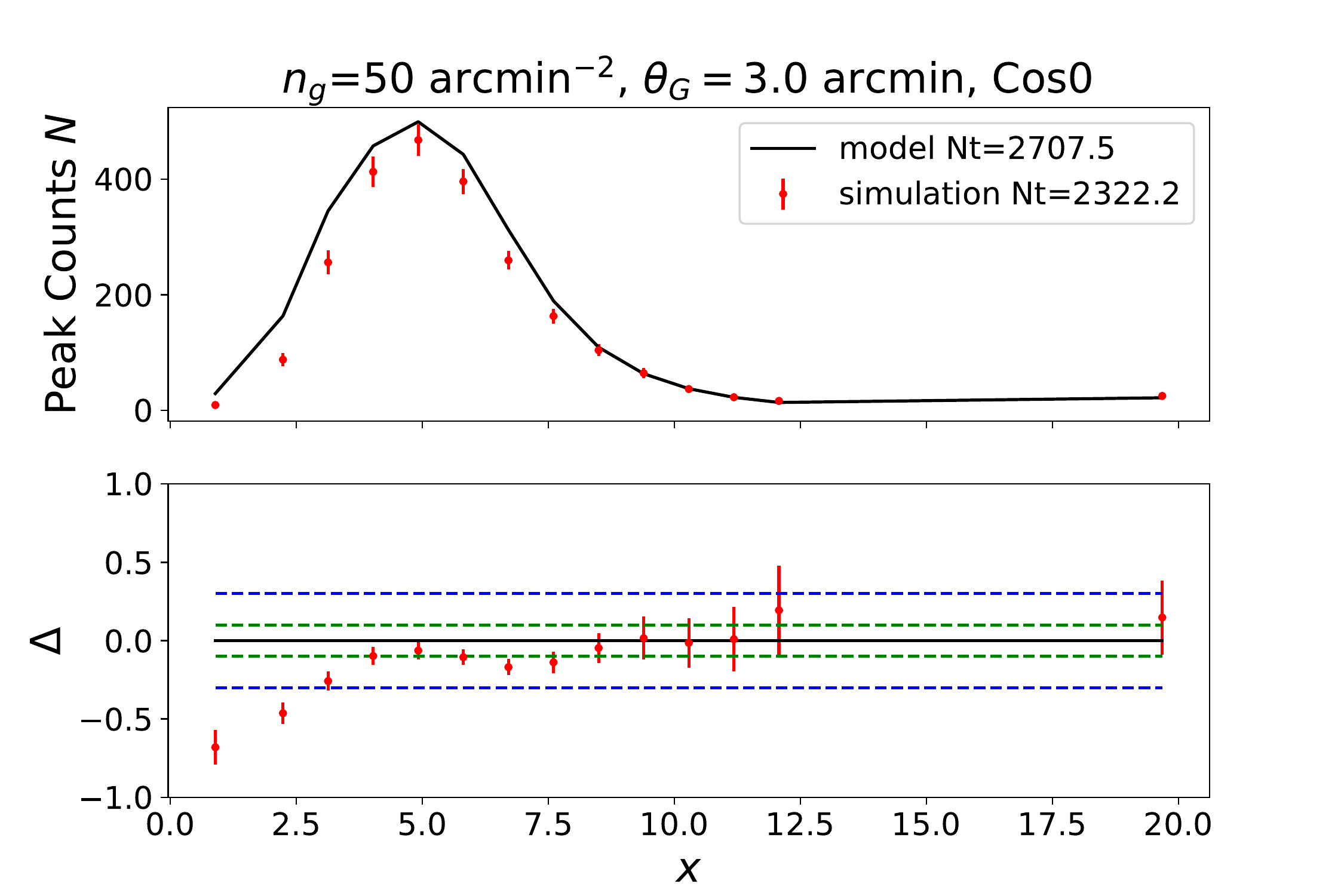}
    \caption{Comparisons of the height (left) and steepness (right) distributions for high peaks between simulation results (red points with error bars) and the predictions from our theoretical model (black lines) for Cos0 and $\theta_G=3\hbox{ arcmin}$. The results with $n_g=10$ and $50\hbox{ arcmin}^{-2}$ are shown in the top and bottom parts, respectively. The quantity $\Delta$ is the relative difference between the results of simulations and the model predictions.}
    \label{fig:peak model distribution}
\end{figure*}

We first show the comparisons of the peak distributions. 
Figure \ref{fig:peak model distribution} presents the results of the peak height (left) and steepness distributions (right) for the fiducial model Cos0 
with $\theta_G=3\hbox{ arcmin}$ and $n_g=10$ (top part) and $50\hbox{ arcmin}^{-2}$ (bottom part), respectively. 
The red data points are from simulations with the error bars estimated from the 1000 bootstrap samples described previously.
The lines are the predictions from our theoretical model discussed in Sec.\ref{sec:Convergence peak abundance model}.
In each part, the upper and lower panels are for the peak distributions and the relative differences between the results from model predictions 
and from simulations, respectively. The horizontal dashed lines in the lower panels indicate $10\%$ and $30\%$ values. 

For peak height distributions, we see that the model predictions are in good agreements with the simulation results. For the considered high peak range,  
the relative difference $\Delta<\sim 10\%$, in consistent with our previous studies \citep{2018ApJ...857..112Y}. For the steepness distributions, 
this is the first time to show such comparisons. Overall, reasonable agreements are also seen. For $n_g=10\hbox{ arcmin}^{-2}$, $\Delta<\sim 10\%$. 
For $n_g=50\hbox{ arcmin}^{-2}$ with reduced shape noise, $\Delta$ gets larger at $x<5$. 

\begin{figure}
    \centering
    \includegraphics[scale=0.325]{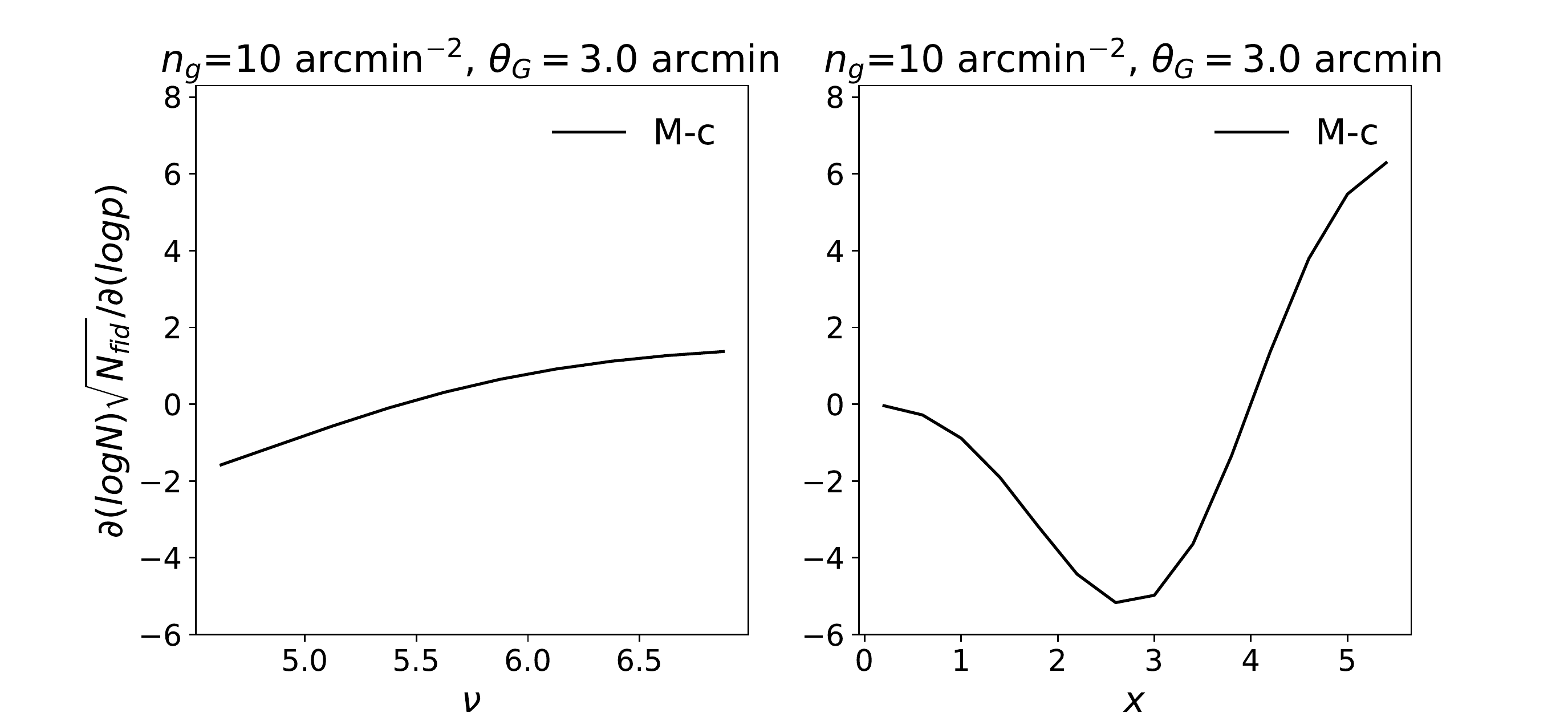}
    \includegraphics[scale=0.325]{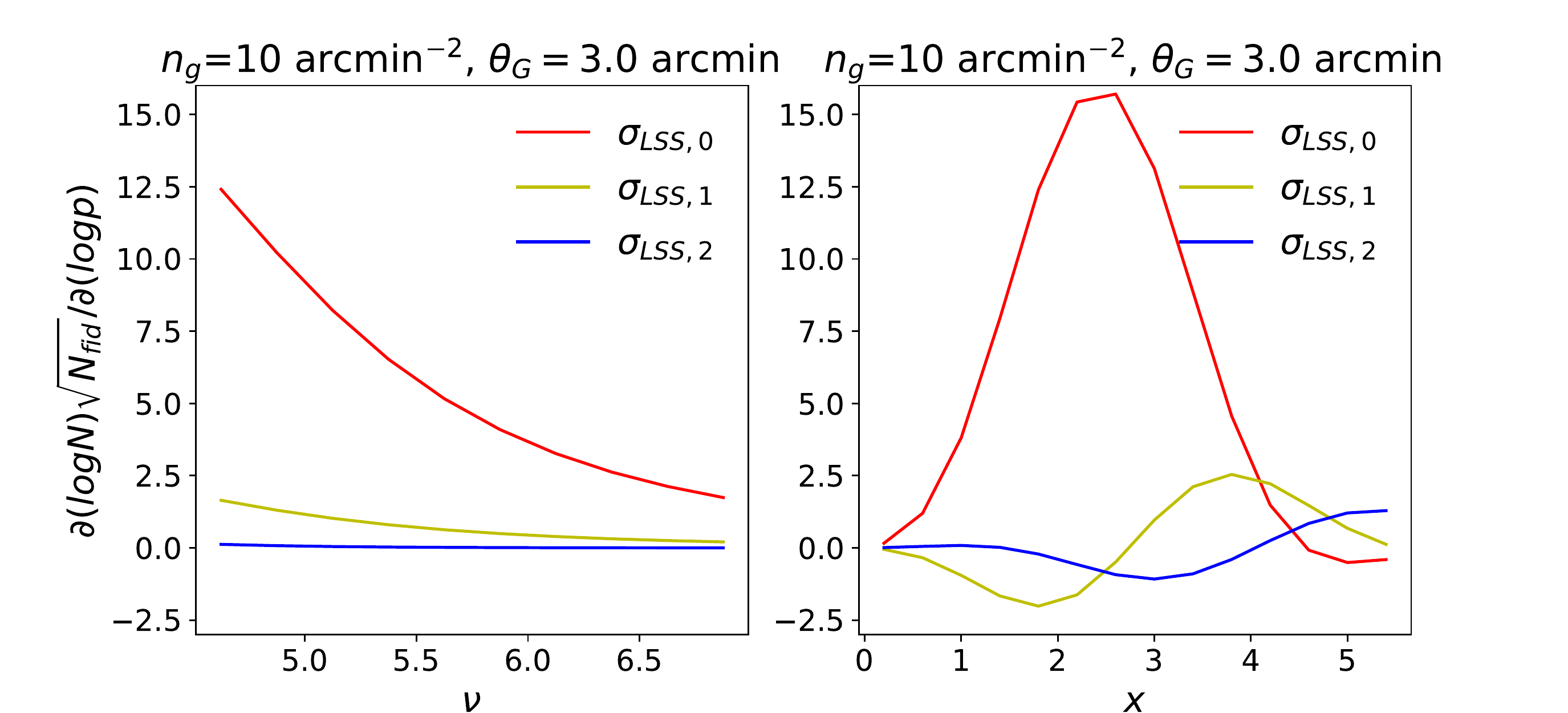}
    \includegraphics[scale=0.325]{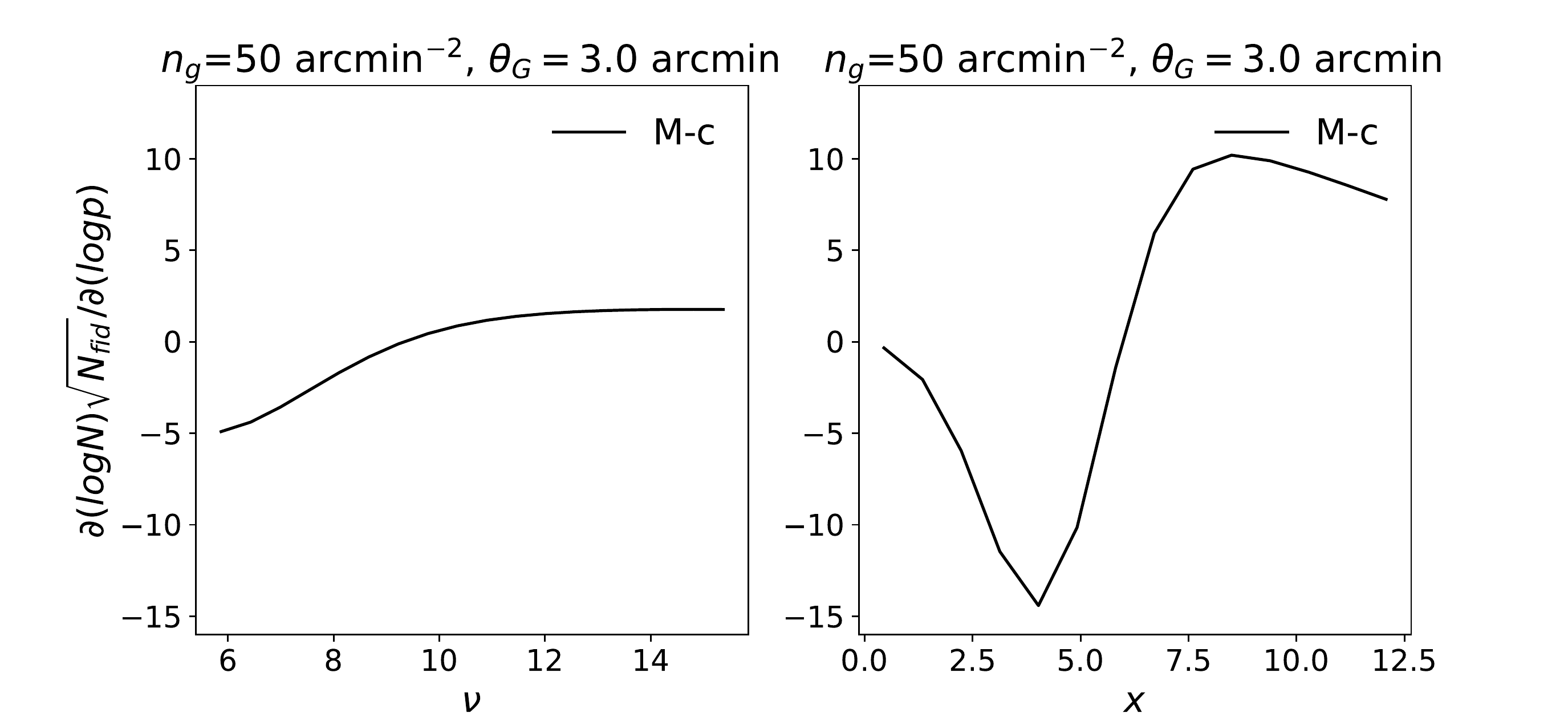}
    \includegraphics[scale=0.325]{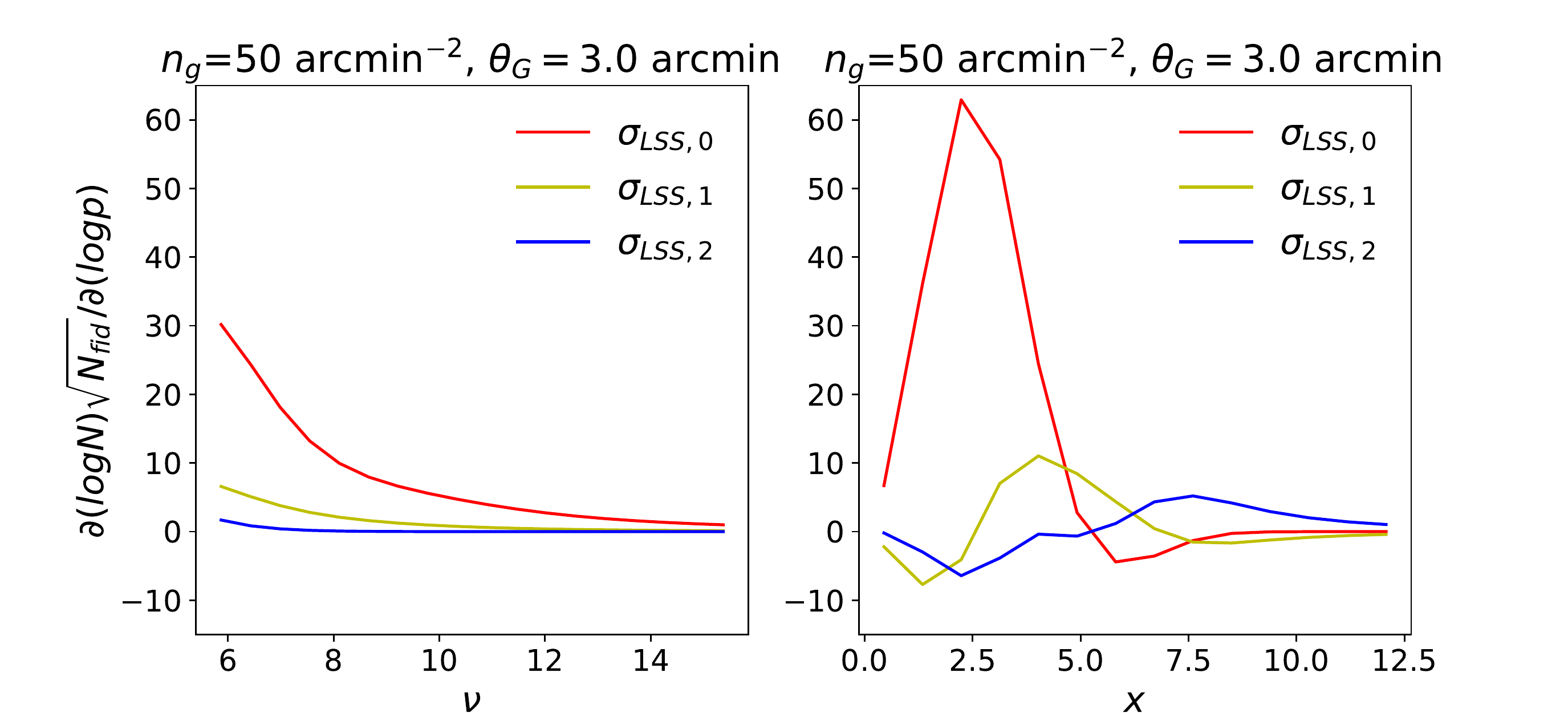}
    \caption{Theoretical derivatives of the peak height (left) and steepness (right) statistics with respect to the amplitude of the M-c relation of dark matter haloes (upper) and to $\sigma_{\rm{LSS},i}$ (lower). The top and bottom parts correspond to the two cases shown in 
Figure \ref{fig:peak model distribution}. The bins considered here are also in accord with the ones in Figure \ref{fig:peak model distribution}.} 
    \label{fig:peak model derivatives}
\end{figure}

Naively, for a same set of high peaks, we expect about the same level of agreements with theoretical model predictions for the two peak statistics. 
Here we see some differences with the peak height distributions being in line with the model predictions better than the steepness distribution. 
This hints for somewhat different sensitivities of the two peak statistics to the ingredients employed in the model, including the mass function and 
the M-c relation of dark matter haloes, and the components of large-scale structures $\sigma^2_{\rm{LSS},i}$. To see this more clearly, we calculate the parameter dependences from our theoretical model, and the results are shown
in Figure \ref{fig:peak model derivatives} with the top and bottom parts corresponding to the two cases shown in Figure \ref{fig:peak model distribution}. In each part, the left and right panels are for the peak height and steepness statistics, 
while the upper and lower panels are for dependences on the amplitude $A$ of the M-c relation using the power-law form taken from \cite{2008MNRAS.390L..64D} and on $\sigma_{\rm{LSS},i}$, respectively. 
The vertical axes are the quantities of $(\partial N_j/\partial p)(p/\sqrt{N_j})$ with $N_j$ being the peak number in a bin and $p$ for the dependent parameter considered. Note that we divide by $\sqrt{N_j}$ instead of $N_j$ 
in order to be in accord with the Fisher calculations [see Eq.(\ref{eq:Fisher})], which leads to an area dependence of $(\partial N_j/\partial p)(p/\sqrt{N_j})$. The results presented in Figure \ref{fig:peak model derivatives} correspond to the area and thus the peak numbers 
shown in Figure \ref{fig:peak model distribution}.  

We see from Figure \ref{fig:peak model derivatives} that the two peak statistics do have different dependences on these physical parameters. Overall, the peak steepness statistics are more sensitive to them, especially for $n_g=50\hbox{ arcmin}^{-2}$ with a low shape noise.   
In our model calculations in Figure \ref{fig:peak model distribution}, we use a fixed M-c relation from \cite{2008MNRAS.390L..64D} assuming spherical dark matter haloes.
This cannot be perfectly the same as that of the simulated haloes. Furthermore, for the projection effects of large-scale
structures, we calculate $\sigma^2_{\rm{LSS},i}$ approximately by subtracting the one-halo contribution from haloes with $M\ge M_*$ from the full nonlinear power spectrum.
Because of the higher sensitivities on these quantities, the peak steepness statistics are affected 
more by the inaccuracies of these approximations resulting in relatively larger differences between the model predictions and the simulation results than the height statistics.  

The above results and the discussions point to further improvements of the theoretical model 
for the steepness statistics in deriving high precision cosmological constraints. On the other hand, the sensitivity differences to different physical 
quantities for the two peak statistics also provide an explanation for the different cosmological dependences seen in our studies.   

In the following subsections, we present further analyses with the help of our theoretical model to understand the differences of 
the two peak statistics. We note that although the steepness statistics with $n_g=50\hbox{ arcmin}^{-2}$ show relatively large deviations from 
the model predictions at $x<5$, overall, we see a reasonable agreement in the trend between the simulation results and the model calculations. To quantify the effects of the bins with $x<5$ shown in the lower right panel of Figure \ref{fig:peak model distribution}, in Appendix \ref{modeltest}, we show the Fisher results for $n_g=50\hbox{ arcmin}^{-2}$ from the steepness statistics excluding some low $x$ peaks in comparison with that without exclusions. It is seen that the Fisher results, both from simulations and from the model calculations, do not change considerably. This indicates that the bins with our model showing large deviations play relatively minor roles in terms of the cosmological information content. For the purpose to understand the differences of the two peak statistics, our model should be sufficient. Nevertheless, as discussed above, the steepness statistics is more sensitive to the theoretical ingredients assumed in the model calculations than the height statistics. We will further improve our model in our future studies.

\subsection{Correlations between peak height and steepness}
We have seen from the examples shown in Figure \ref{fig:height-steepness diagram(cos0)} that there is a strong positive correlation between $\nu$ and $x$, but it is not perfect. In Table \ref{table:PCC}, we present the PCC for all the considered cases. With a slight cosmology dependence, $r_{\rm x}>\sim 0.85$ for all the five models in the noiseless case, and $r_{\nu x}\sim 0.55$ and $0.6$ at $\theta_G=2$ and $3\hbox{ arcmin}$, respectively, in the most noisy case with $n_g=10\hbox{ arcmin}^{-2}$.

\begin{figure}
    \centering
    \includegraphics[scale=0.44]{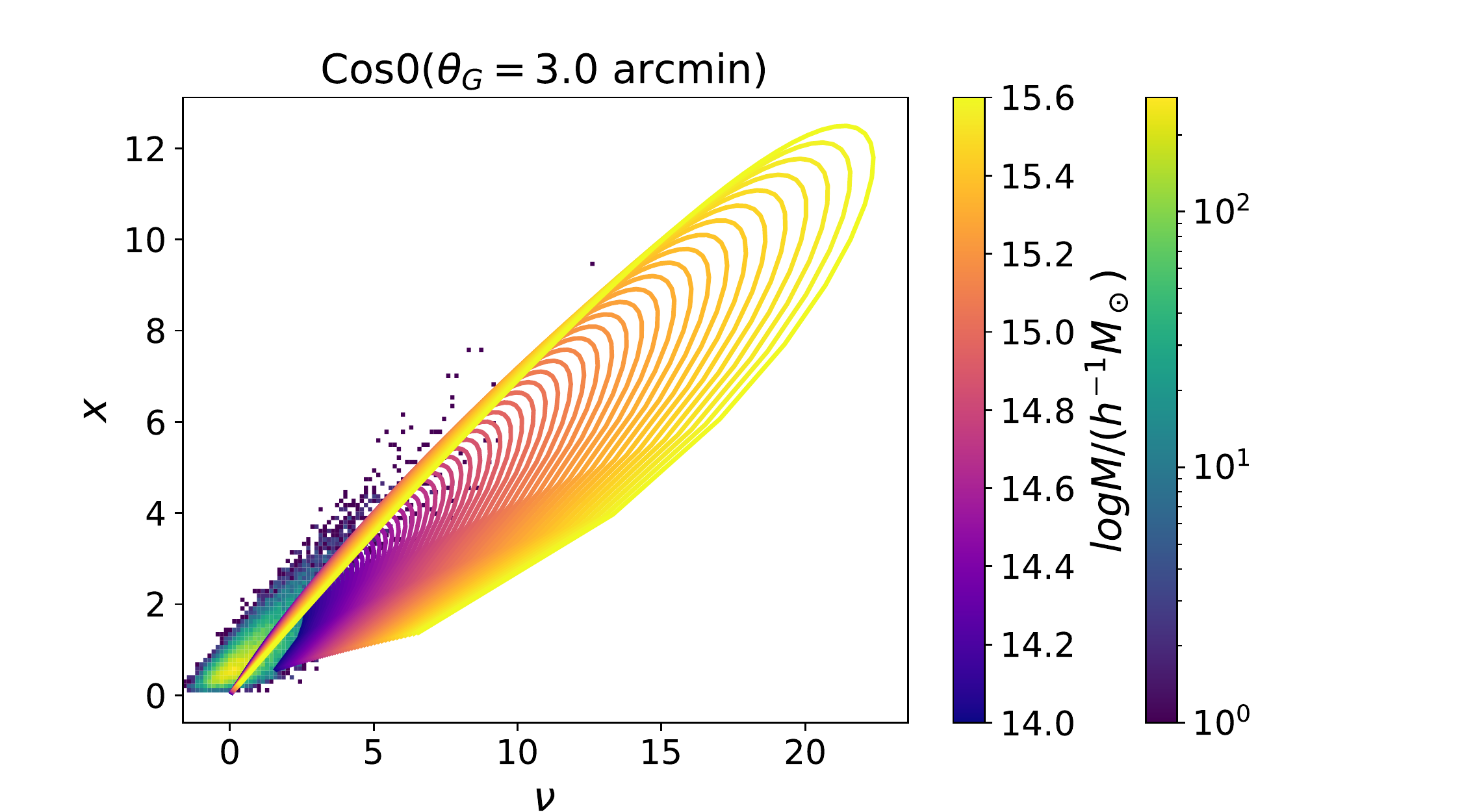}
    \caption{The relation between the peak height and the steepness for the noiseless case of Cos0 with $\theta_G=3\hbox{ arcmin}$. The data points and the outer colour bar are the same as those in the upper panel of Figure \ref{fig:height-steepness diagram(cos0)}.
Each trajectory line is calculated from a NFW halo of a given mass at different redshifts from $z=0$ to $z=1$. The halo mass of the different trajectories is indicated by the inner colour bar.}
    \label{fig:halo diagram(cos0)}
\end{figure}

To understand the not-perfect correlation even in the noiseless case, we turn to our theoretical model. In the model, massive dark matter haloes are the main sources for high peaks. We therefore calculate the $\nu-x$ trajectory from a halo at different redshifts from $z=0$ to $z=1$. The results are shown in Figure \ref{fig:halo diagram(cos0)} with different lines corresponding to the trajectories of haloes with different mass as indicated by the colour bar. Overlaid are the noiseless data from simulations of Cos0. It is seen that the trajectory from a halo is not a one-one line in the $\nu-x$ plane. Rather, at a fixed $\nu$, $x$ has two values. This can be understood as follows. Because of the lensing efficiency kernel, with the source redshift $z_s=1$, two lens redshifts can give rise to a same $\nu$ value for a halo. At these two lens redshifts, the steepness values from the halo are however different. This arises because the calculation of $x$ from the second derivatives leads to an extra redshift-dependent factor that does not show up in the peak height $\nu$ calculation. As a result, the $x$ values at the two lens redshifts are different. In other words, the redshift degeneracies of $\nu$ and $x$ are different. Combining all massive haloes taking into account the weight from the mass function result in scatters of high peaks in the $\nu-x$ plane. Additionally, the contributions from the large-scale structures other than the massive haloes are in the form of random fields in our model, which naturally give rise to scatters of the $\nu-x$ correlation. For low peaks, although our model is not valid quantitatively, the concept that in the noiseless case these low peaks arise from the projection effects of large-scale structures that are random fields in nature should hold. This randomness can explain the scatters of these peaks in the $\nu-x$ plane. 

From Eq.(\ref{eq:Peakhalo}) of our model, we can see that the different redshift dependences of $\nu$ and $x$ of a massive halo also lead to somewhat different sensitivities of the two peak statistics to the halo mass function at different redshifts, which in turn can affect the cosmological inferences from the two peak counts. The large-scale structure projection effects add additional differences as discussed previously and also to be seen in the next subsection.

\begin{table}
\centering
\scalebox{0.85}{
$\begin{array}{|c|cc|cc|}
\hline
& \theta_G=2.0 \text{ arcmin}& & \theta_G=3.0 \text{ arcmin} &  \\ \hline (\text{arcmin}^{-2}\text{)}&\text{noiseless} & \text{noisy} & \text{noiseless} & \text{noisy} \\ \hline \text{Cos0} & & & &  \\ \hline  n_g=10 & 0.87 & 0.56 & 0.85 & 0.60 \\ \hline   n_g=20  & …… & 0.62 & …… & 0.67 \\ \hline  n_g=30 & …… & 0.67 & …… & 0.71 \\ \hline  n_g=50 & …… & 0.73 & …… & 0.76 \\ \hline  \text{Cos1} & & & &  \\ \hline n_g=10  & 0.86 & 0.55 & 0.84 & 0.58 \\ \hline   n_g=20  & …… & 0.60 & …… & 0.64\\ \hline  n_g=30  & …… & 0.64 & …… & 0.68 \\ \hline  n_g=50  & …… & 0.70 & …… & 0.74\\ \hline  \text{Cos2} & & & &  \\ \hline n_g=10  & 0.88 & 0.58 & 0.86 & 0.63 \\ \hline   n_g=20  & …… & 0.65 & …… & 0.70 \\ \hline  n_g=30  & …… & 0.70 & …… & 0.75\\ \hline  n_g=50  & …… & 0.76 & …… & 0.79\\ \hline  \text{Cos3} & & & &  \\ \hline n_g=10  & 0.87 & 0.55 & 0.85 & 0.59 \\ \hline   n_g=20  & …… & 0.60 & …… & 0.66\\ \hline  n_g=30  & …… & 0.65 & …… & 0.70 \\ \hline  n_g=50  & …… & 0.71 & …… & 0.75\\ \hline  \text{Cos4} & & & &  \\ \hline n_g=10  & 0.87 & 0.57 & 0.85 & 0.61 \\ \hline   n_g=20  & …… & 0.64 & …… & 0.69 \\ \hline  n_g=30  & …… & 0.68 & …… & 0.73\\ \hline  n_g=50  & …… & 0.74 & …… & 0.77 \\ \hline 
\end{array}$}
\caption{PCC between peak height and steepness for different cosmological models, noise levels and the smoothing scales.}\label{table:PCC}
\end{table}

\subsection{Fisher analyses for the {\it High} samples}\label{sec:Fisher forecast for High sample}
Here we present the Fisher analyses for the {\it High} samples, and compare with the model predictions to understand further the differences of the two peak statistics.

\begin{figure*}
    \flushleft
    \includegraphics[ scale=0.285]{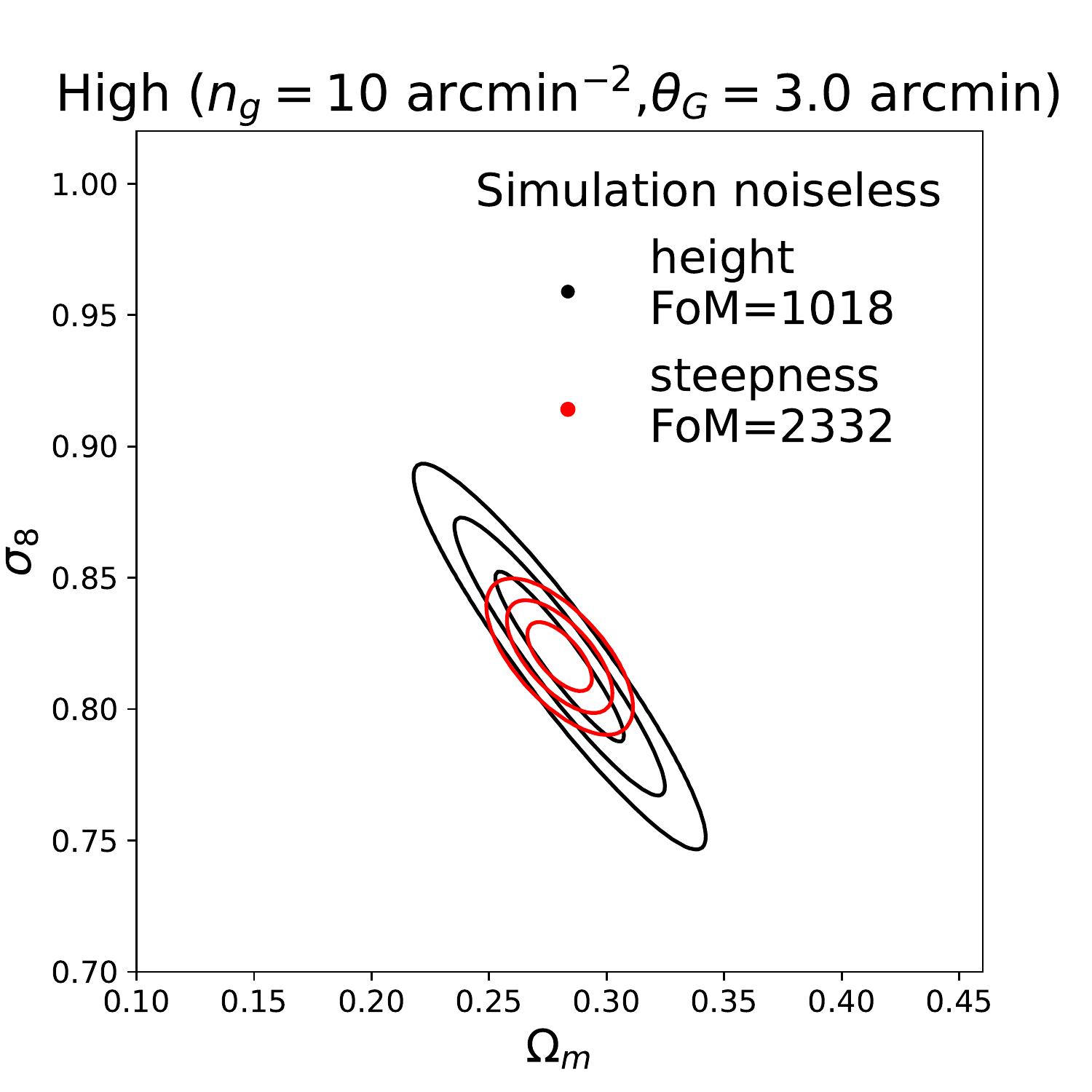}
    \includegraphics[ scale=0.285]{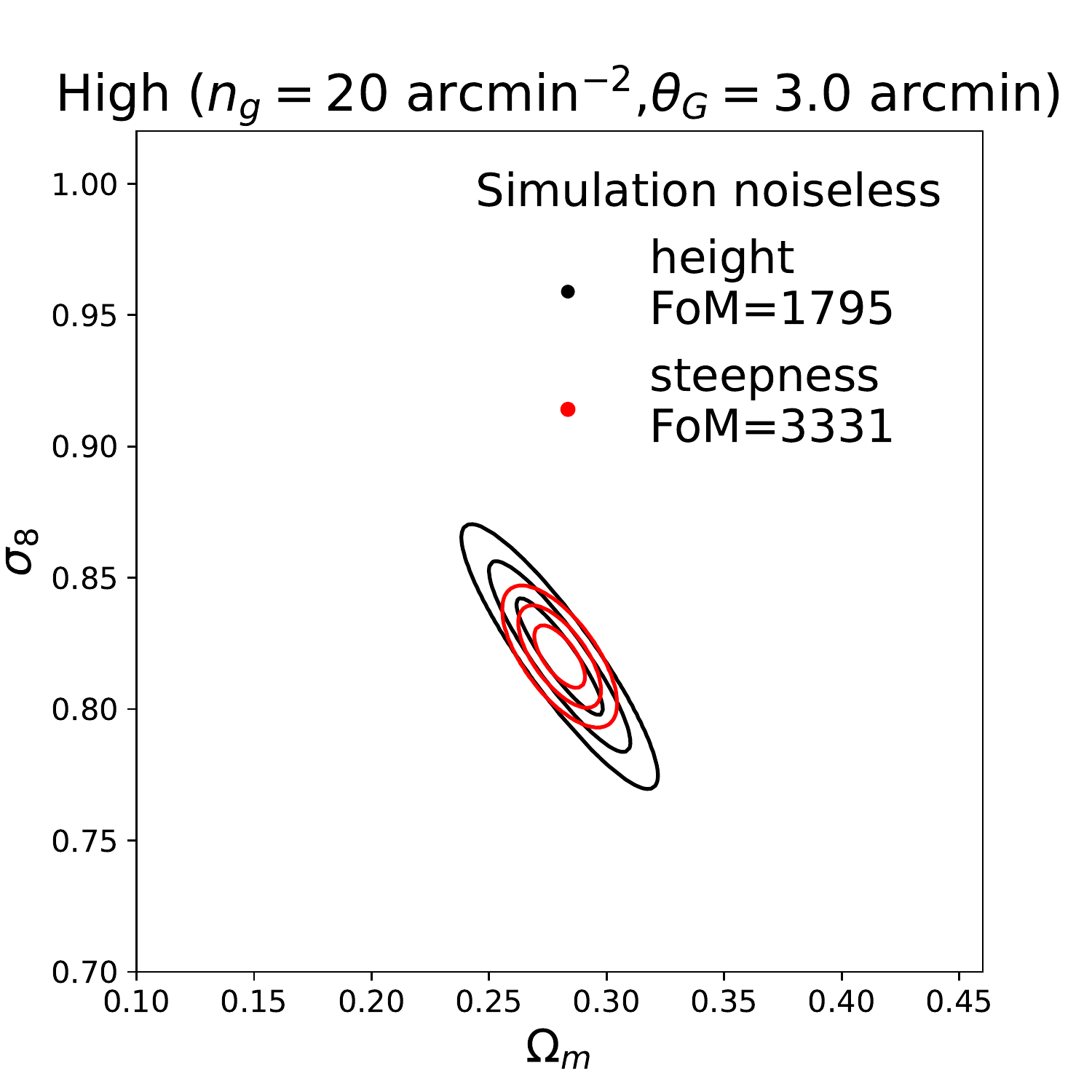}
    \includegraphics[ scale=0.285]{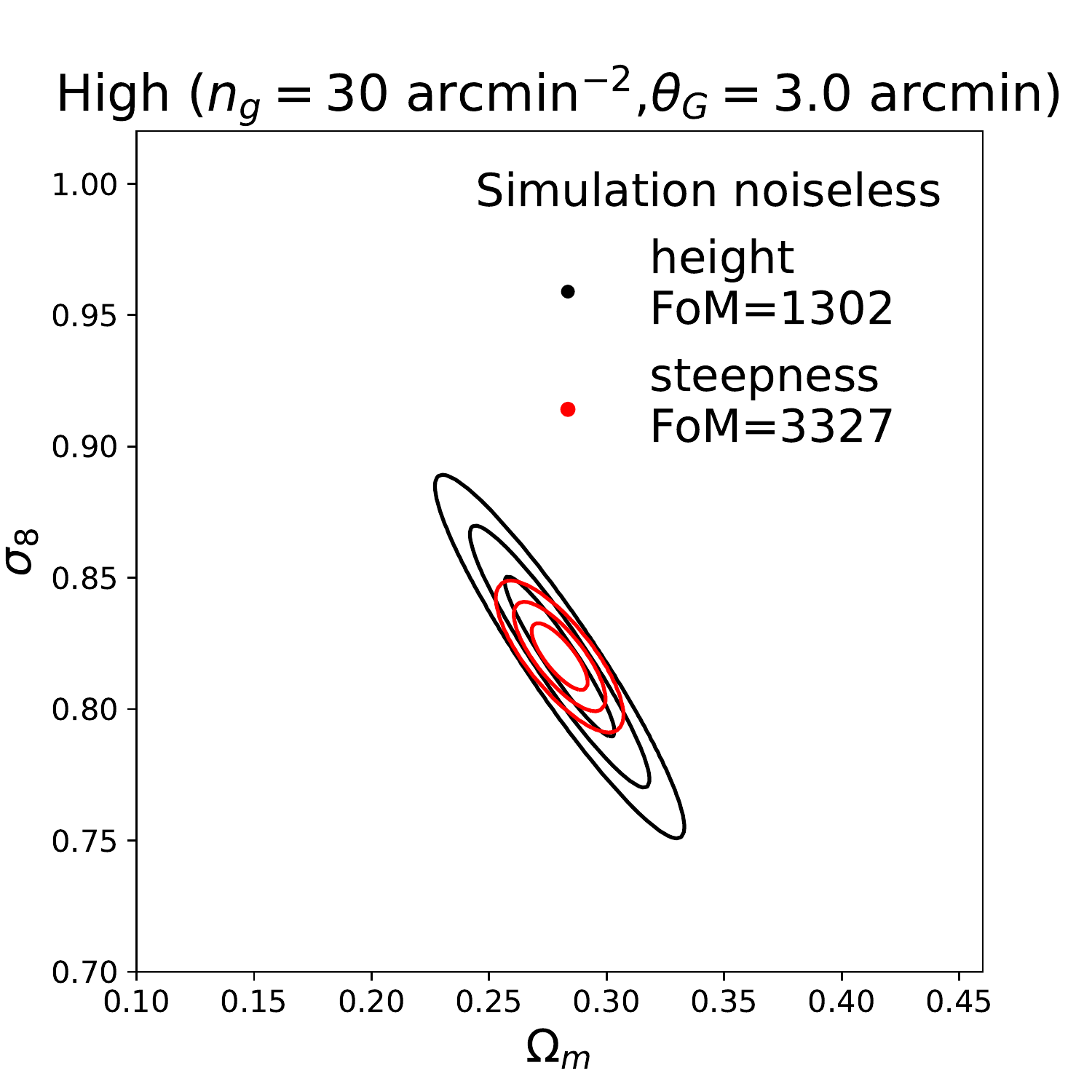}
    \includegraphics[ scale=0.285]{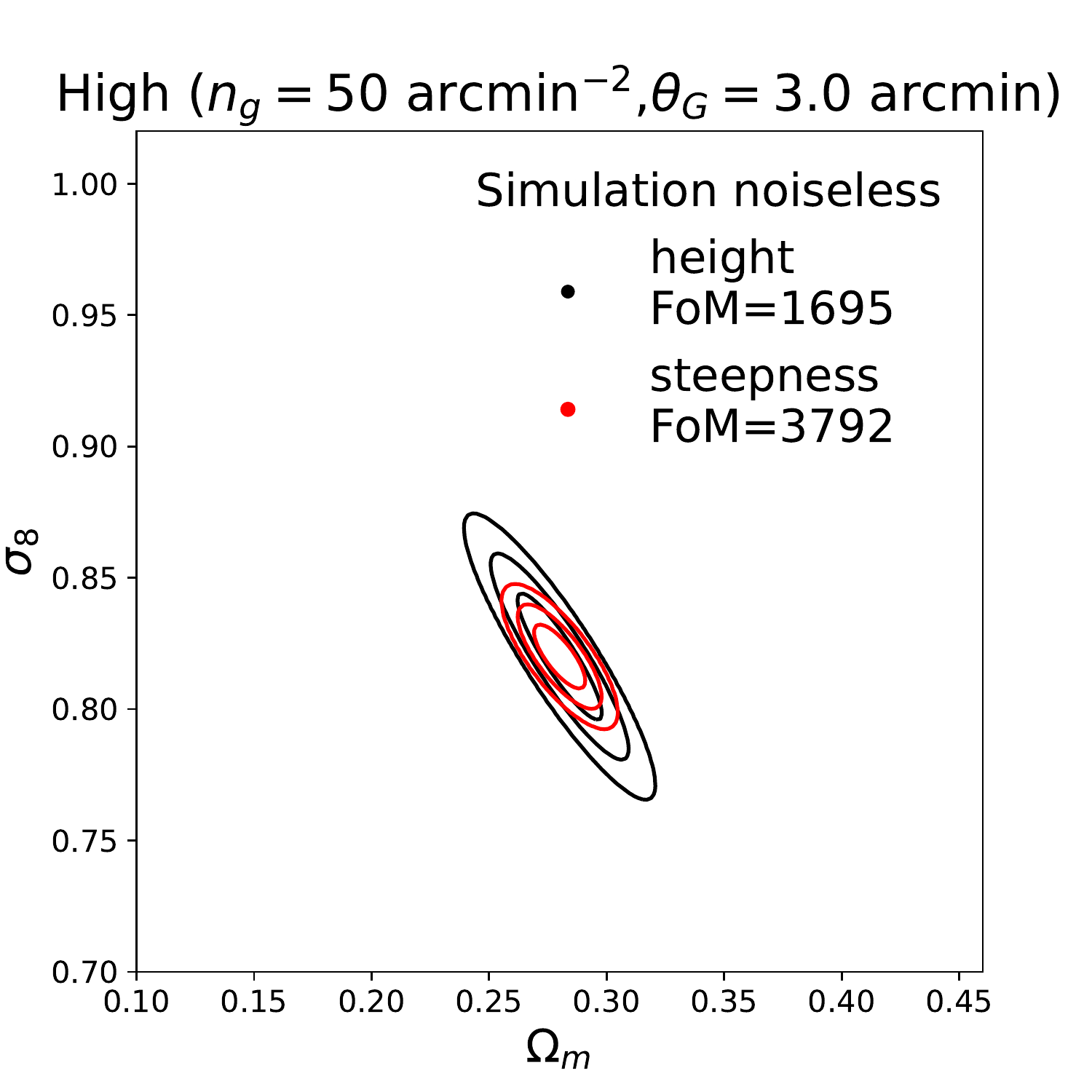}\\
    \flushleft
    \includegraphics[ scale=0.285]{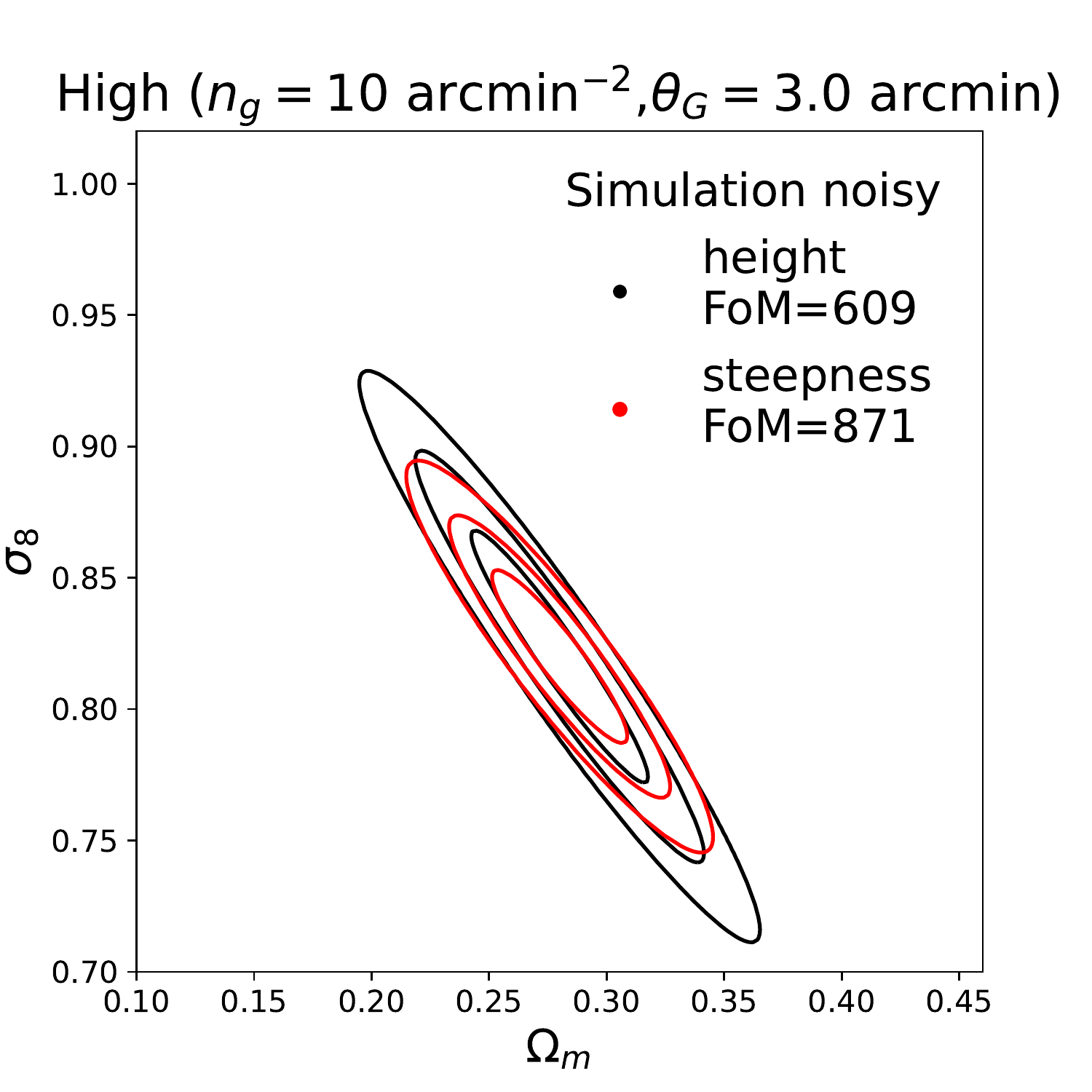}
    \includegraphics[ scale=0.285]{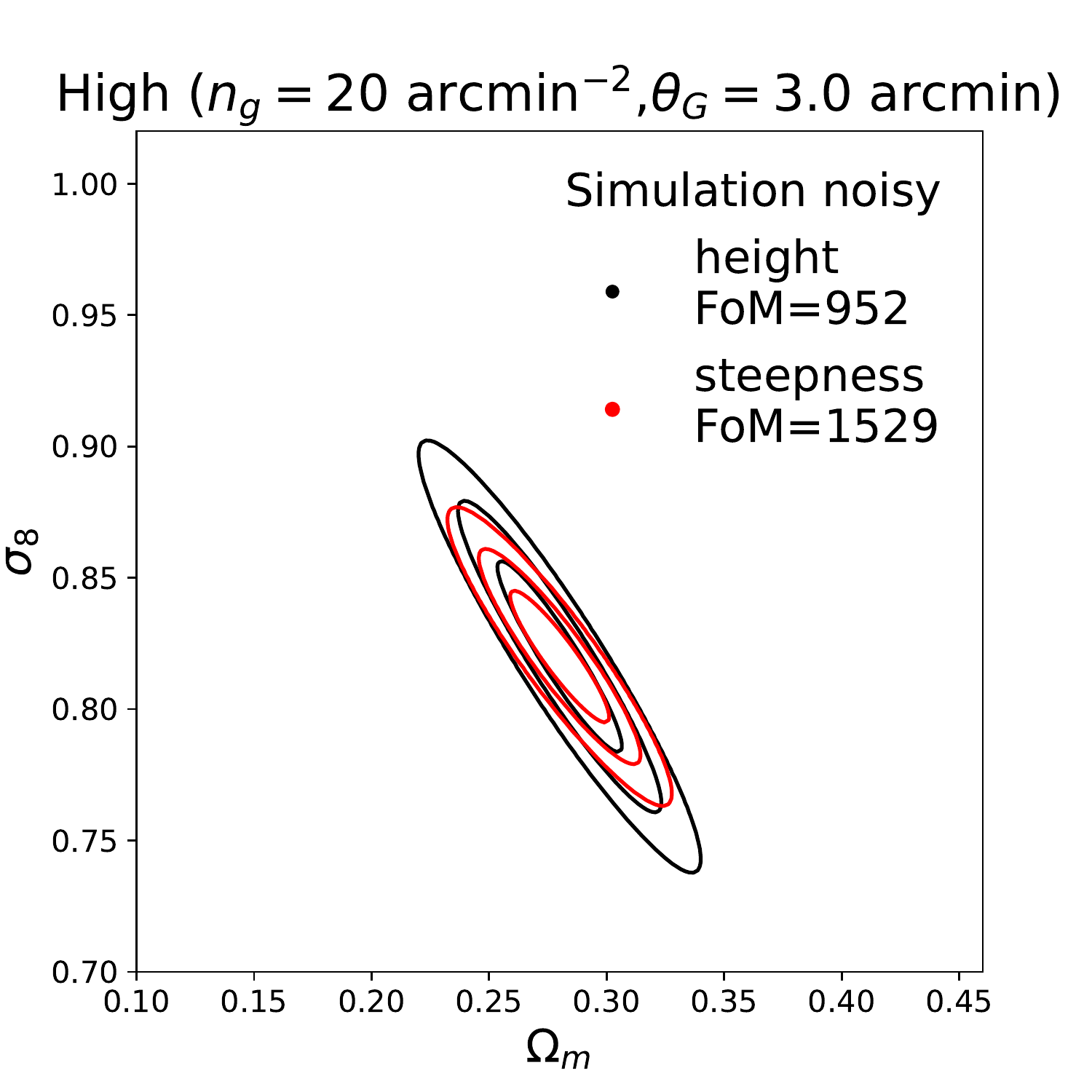}
    \includegraphics[ scale=0.285]{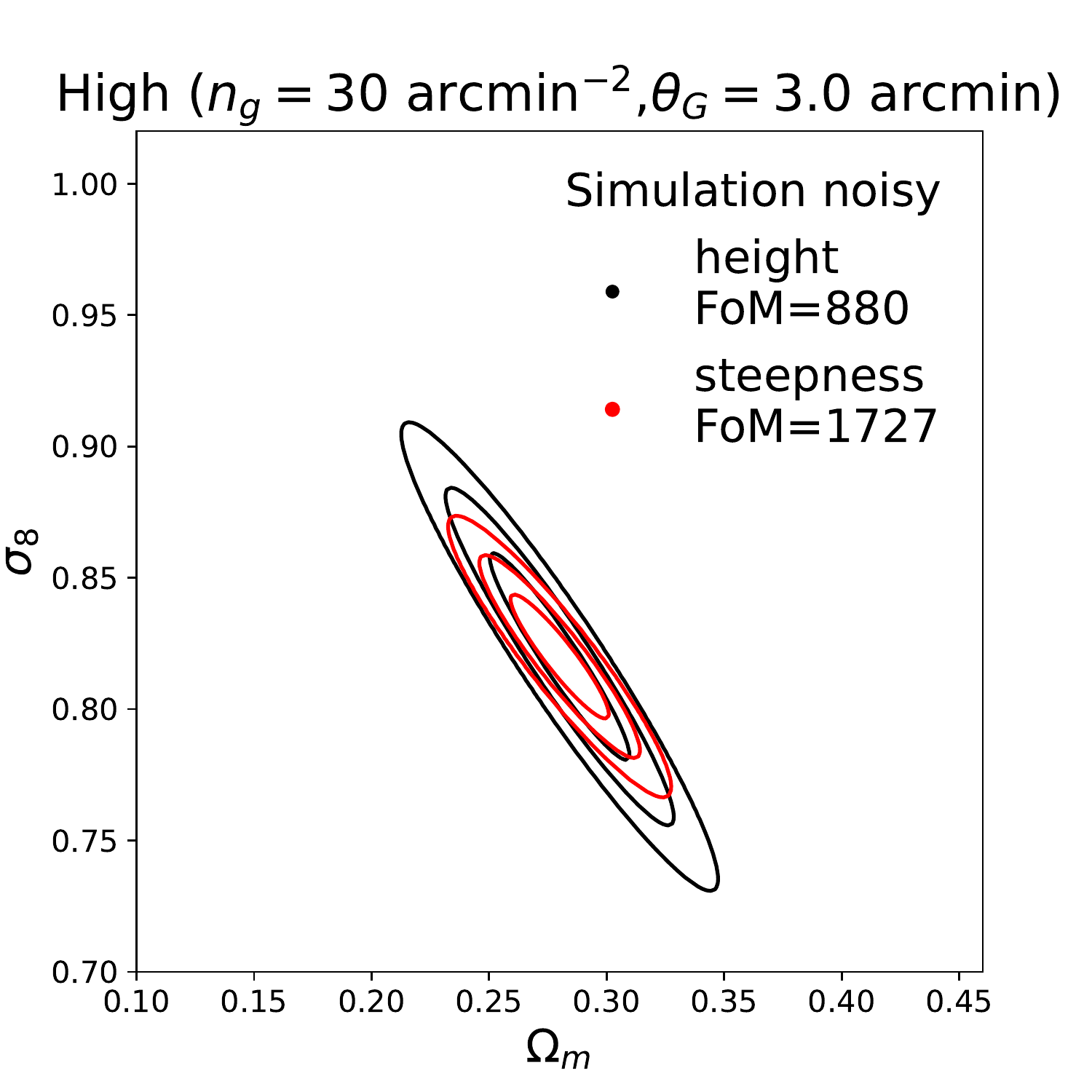}
    \includegraphics[ scale=0.285]{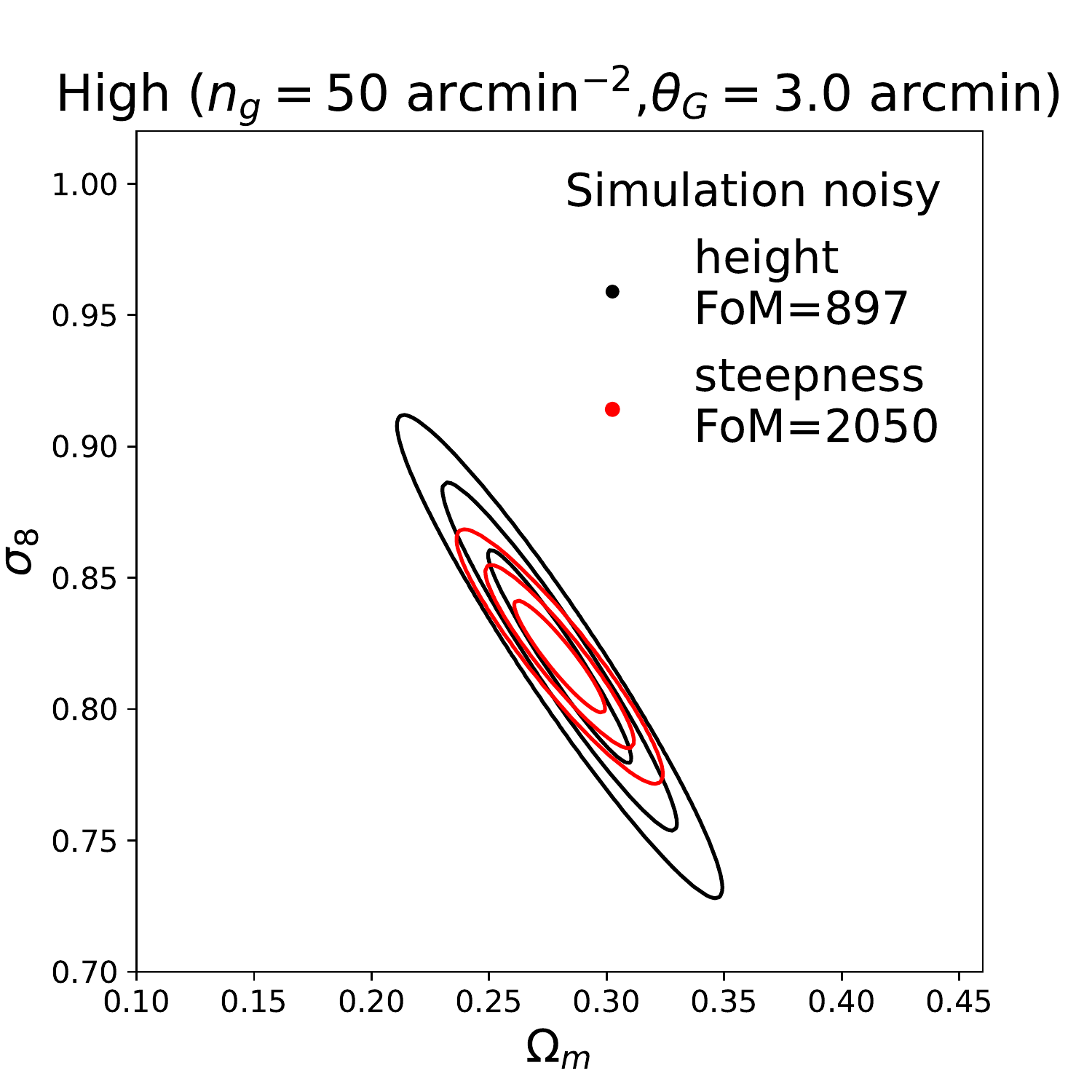}\\
    \flushleft
    \includegraphics[ scale=0.285]{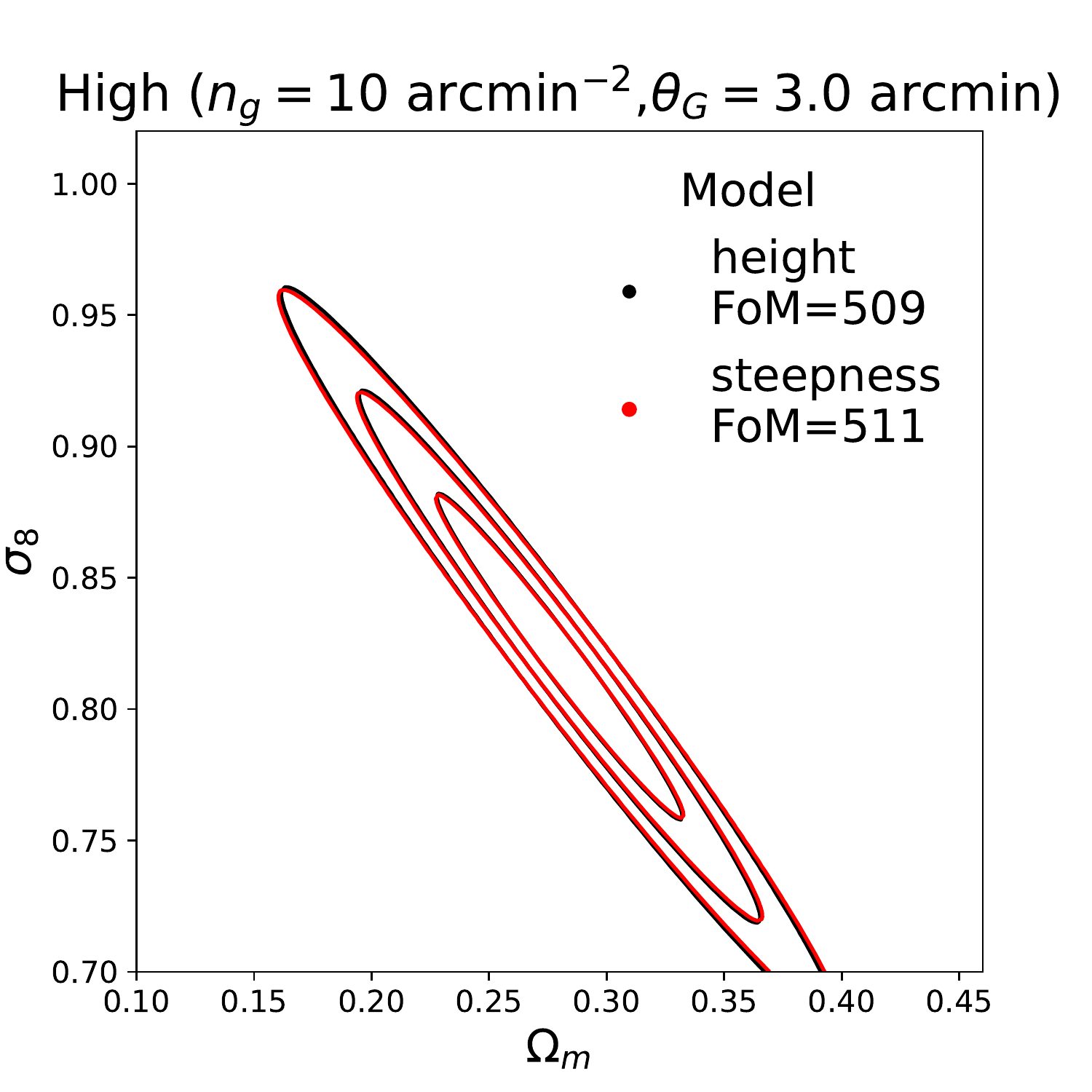}
    \includegraphics[ scale=0.285]{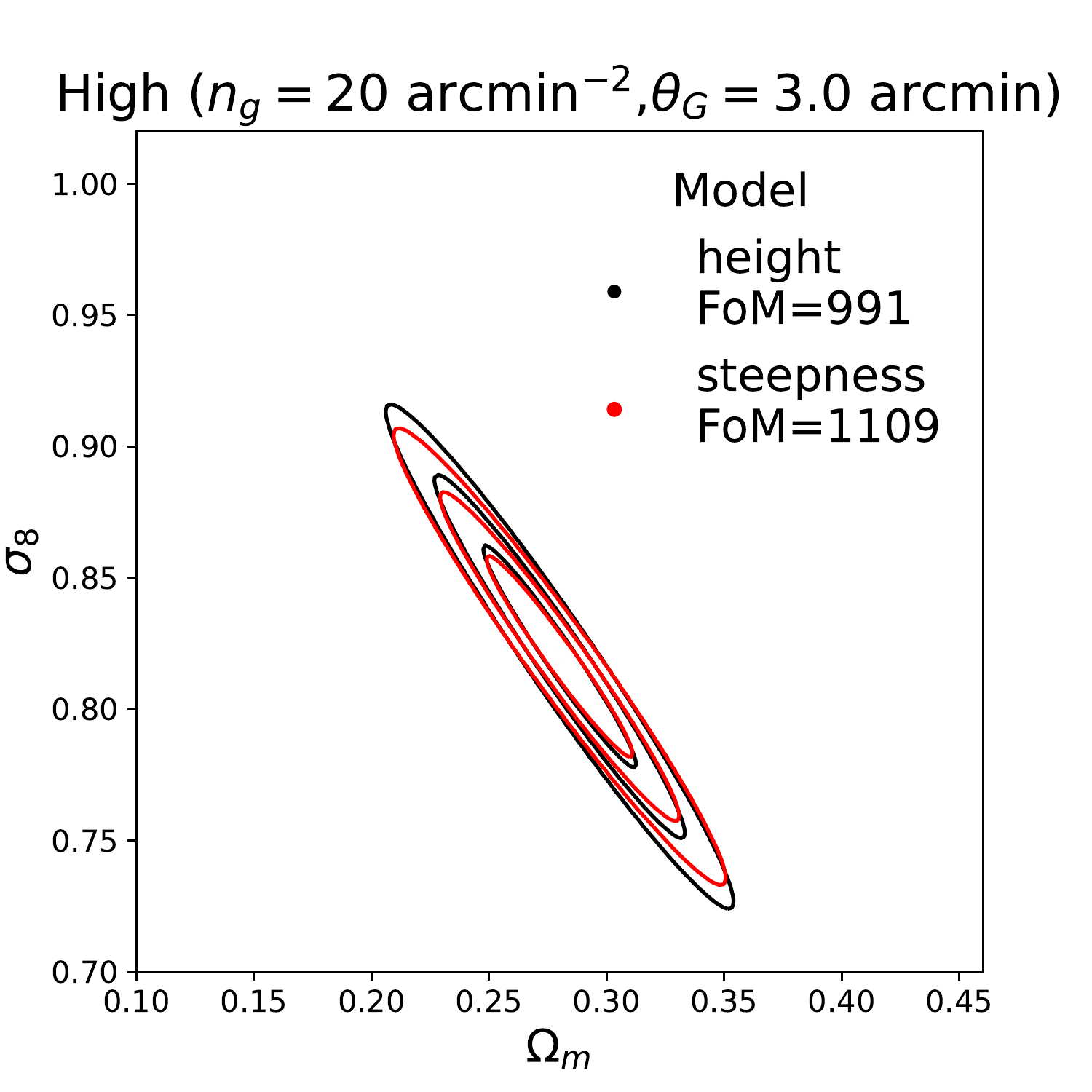}
    \includegraphics[ scale=0.285]{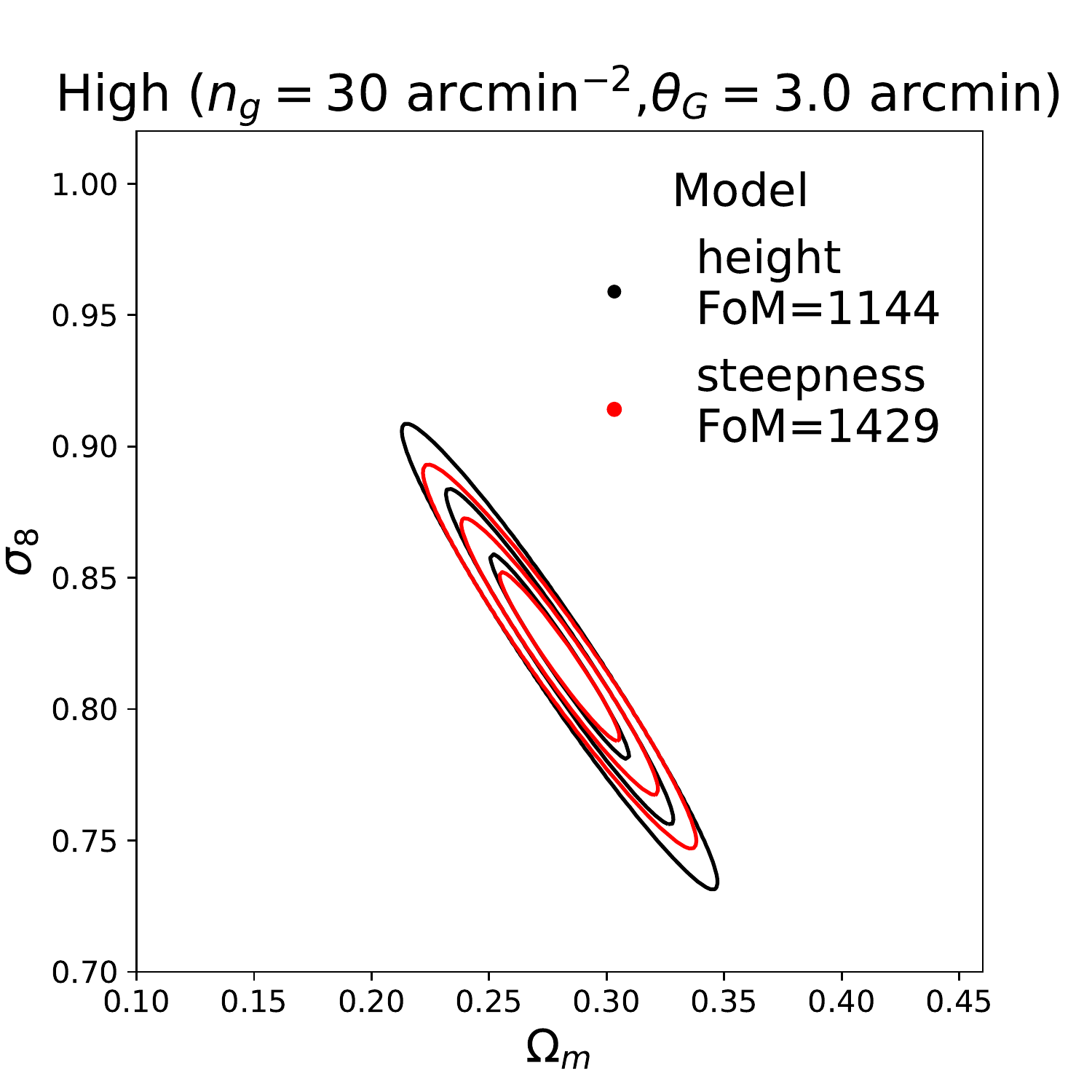}
    \includegraphics[ scale=0.285]{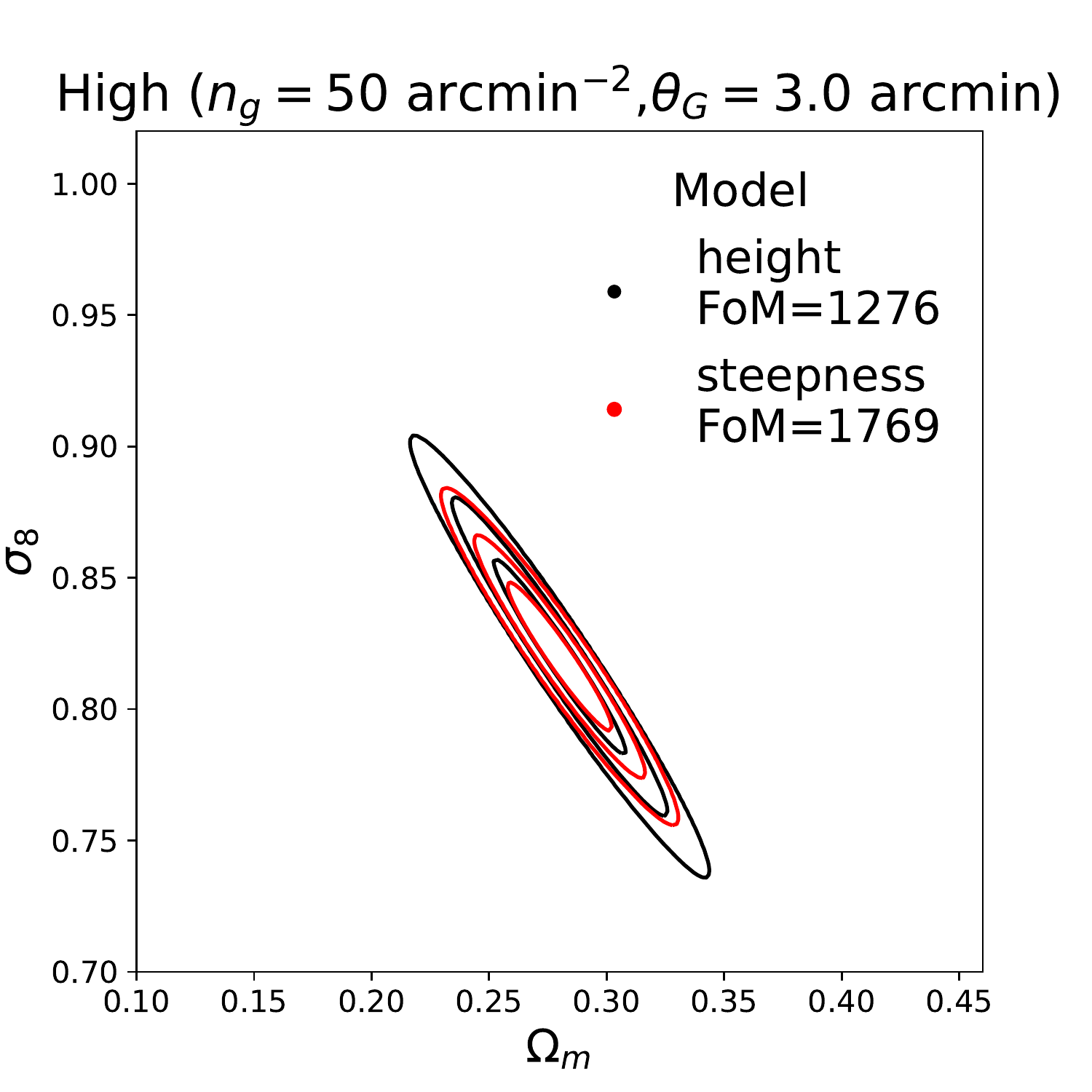}\\
    \flushleft
    \includegraphics[ scale=0.285]{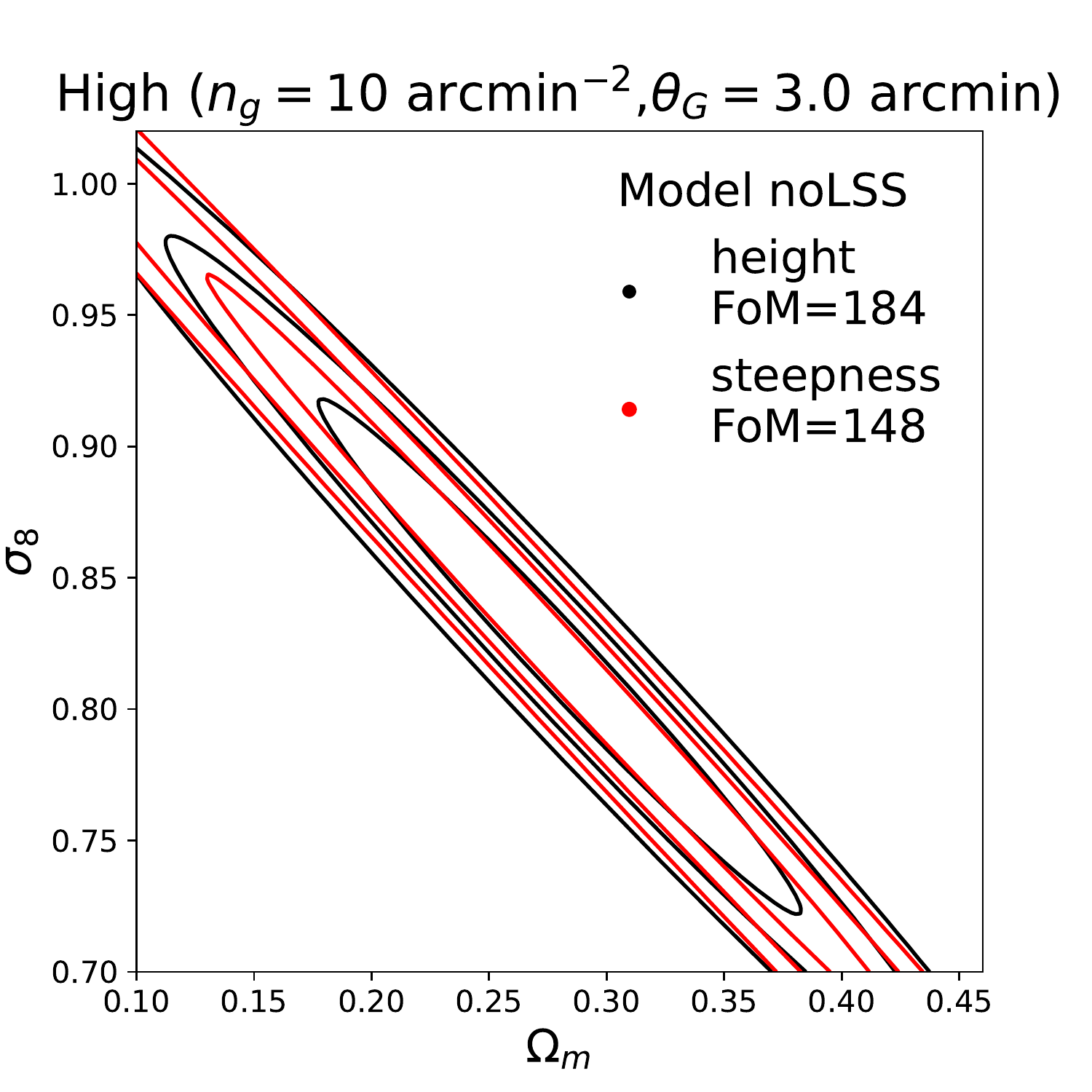}
    \includegraphics[ scale=0.285]{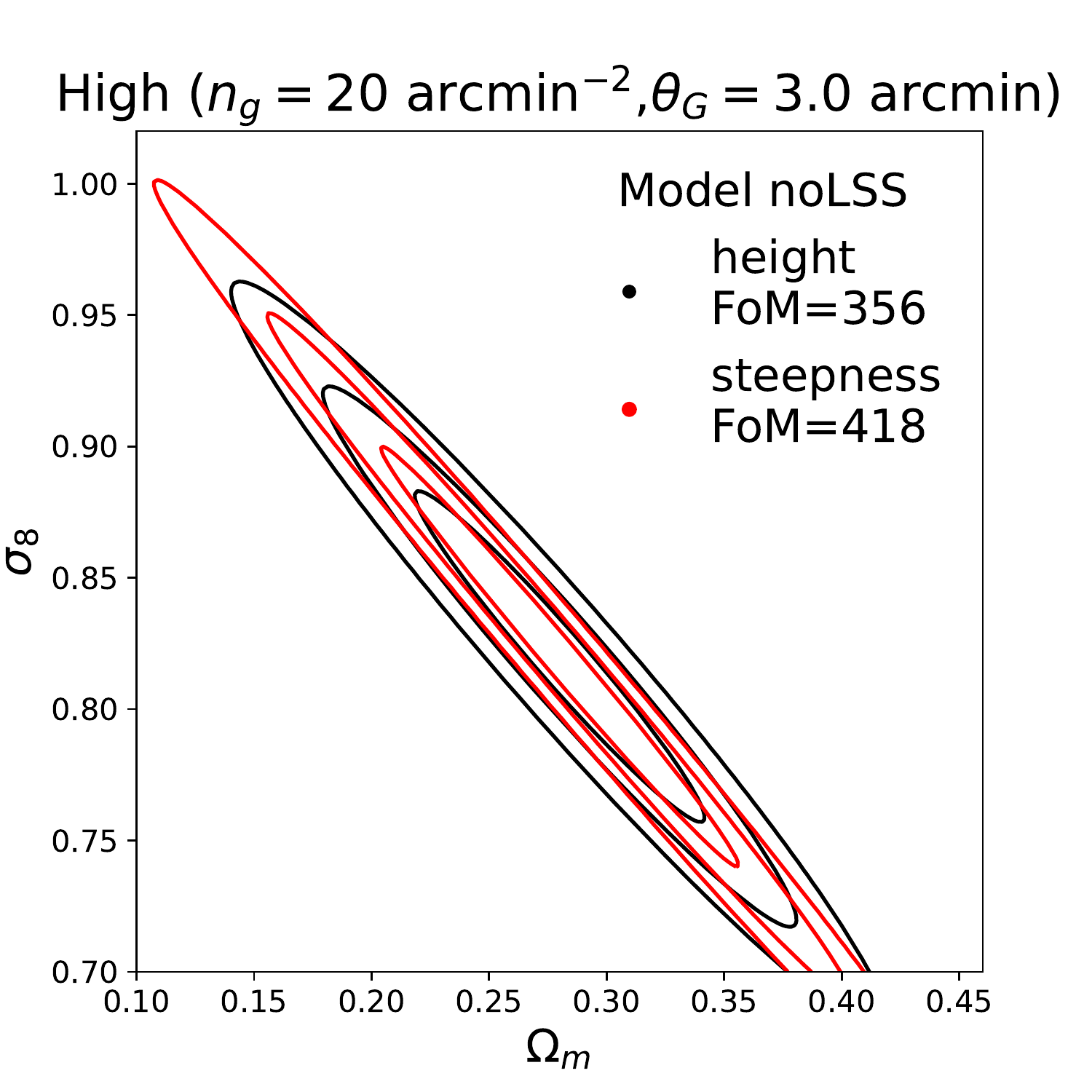}
    \includegraphics[ scale=0.285]{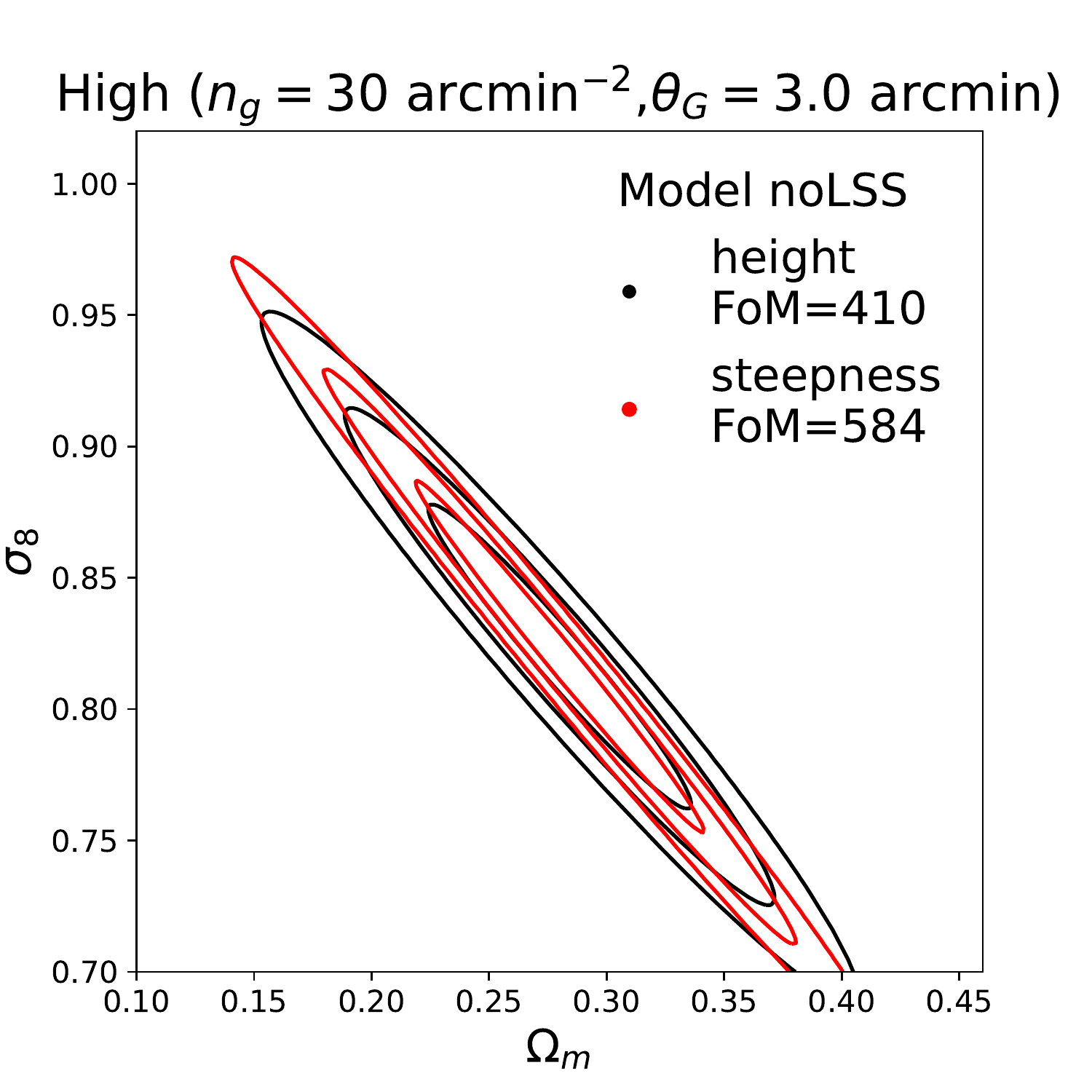}
    \includegraphics[ scale=0.285]{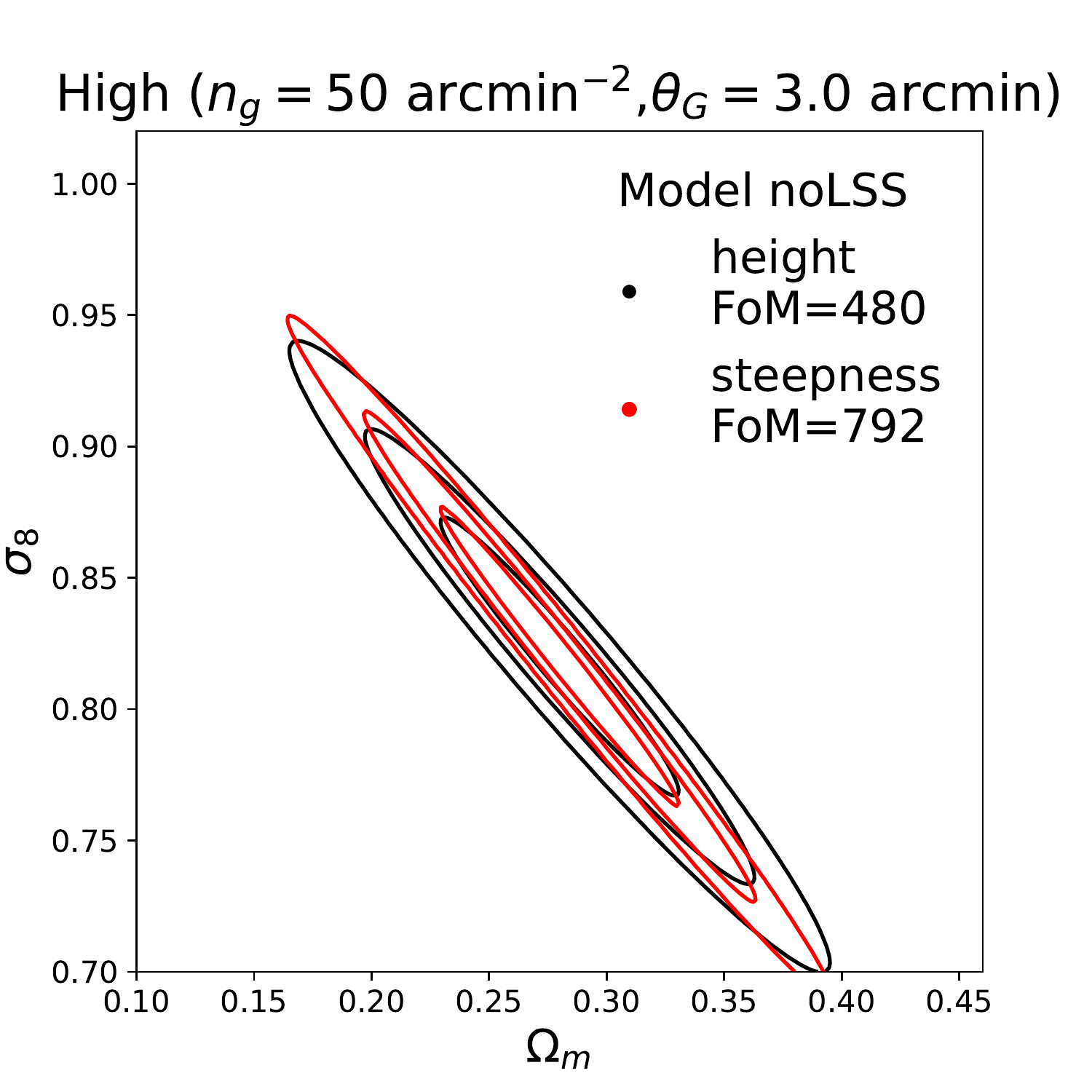}\\
    \caption{The Fisher posterior probability distributions in the $\Omega_{\rm m}-\sigma_8$ plane from the {\it High} samples with $\theta_G=3\hbox{ arcmin}$. The upper two rows are from simulations with the first and the second rows being the results of noiseless and noisy cases, respectively. Because of the different cuts for {\it High} samples with different $n_g$,
the noiseless results shown in the first row also contain four panels. The third row is from our theoretical model predictions corresponding to the noisy cases of the second row. The last row is also from the model calculations but setting $\sigma_{\text{LSS},i}=0$ artificially. The meanings of the contours are the same as in Figure \ref{fig:Fisher_all}. Noticing that the plotting range is larger than Figure \ref{fig:Fisher_all}, and the contours here are actually much larger than those in Figure \ref{fig:Fisher_all}.} 
    \label{fig:Fisher_high_3sm}
\end{figure*}

Figure \ref{fig:Fisher_high_3sm} shows the results with $\theta_G=3\hbox{ arcmin}$. The first and second rows are the noiseless and noisy results where the derivatives involved in the Fisher calculations are computed from the simulated peak counts of different cosmological models. 
The third row is the theoretical predictions of the noisy cases with the derivatives calculated from our high peak model. The last row is also from the model but with the large-scale structure effects artificially turned off.

In comparison with the results presented in Figure \ref{fig:Fisher_all} for the {\it All} samples, the contours here are larger showing that low peaks also contain cosmological information. For the {\it High} samples here, simulation results show that the steepness statistics (red) also tend to lead to tighter constraints than that of the peak height (black). The differences of the two decrease with the increase of the shape noise. 

The theoretical results in the third row are directly comparable with the simulation results shown in the second row. Qualitatively, we see good agreements between the trends showing in the two rows. With the increase of $n_g$ and thus the decrease of the shape noise, the steepness statistics deliver tighter cosmological constraints. The overall size and direction of the contours are also more or less consistent. Some quantitative differences between the simulation and theoretical results are seen. Because the constraints on ($\Omega_{\rm m}, \sigma_8$) are highly degenerate, the Fisher results are very sensitive to the values of the derivatives with respect to the cosmological parameters as seen from Eq.(\ref{eq:Fisher}). Even small differences between the values calculated from simulations and from the model can lead to notable differences of the contours. Nonetheless, the agreements seen here can help us to understand the physical reasons giving rise to the differences of the two peak statistics. 

By turning off the projection effects of large-scale structures setting $\sigma_{\rm{LSS},i}=0$ in our model calculations, we obtain the results shown in the last row. Here we see larger contours than the ones shown in the third row reflecting that the large-scale structure effects contain additional cosmological information. Without the contributions from $\sigma_{\rm{LSS},i}$, there is also a trend that with the decrease of the shape noise, the steepness statistics leads to better constraints than that of the height. As explained previously, this is due to the different dependences on the halo profile between the two statistics, resulting in different sensitivities on the halo mass function at different redshifts.

\begin{figure}
    \includegraphics[scale=0.5]{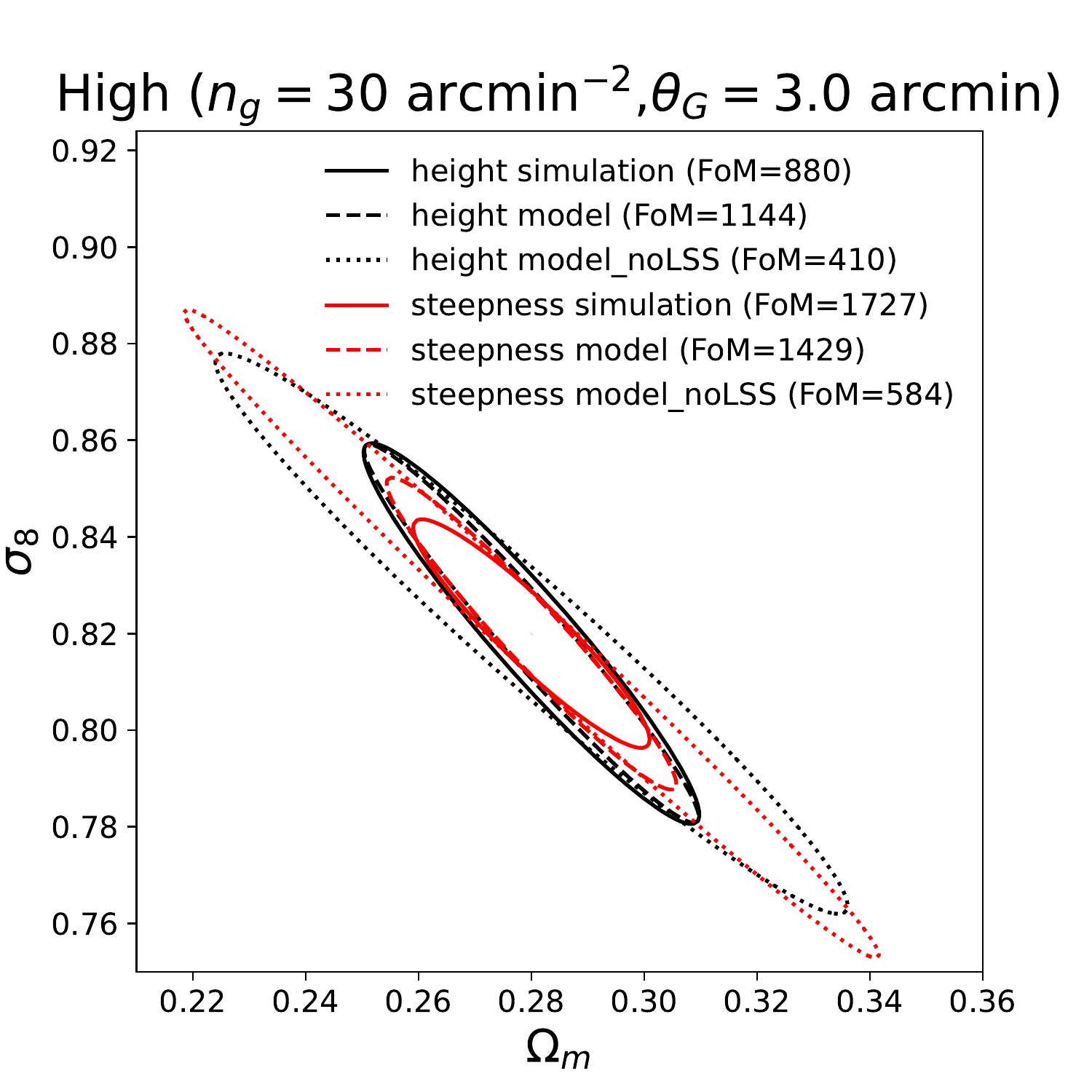}
    \caption{Comparison of the Fisher results from simulation and from the model predictions from high peaks. The solid, dashed and dotted contours correspond to the 68\% confidence level of the simulation, the model calculations with and without LSS contribution, respectively. The black and red contours are from the peak height and steepness statistics, respectively. The case is for the {\it High} sample with $n_g=30\hbox{ arcmin}^{-2}$ and $\theta_G=3\hbox{ arcmin}$.}
    \label{fig:Fisher_mod_vs_sim}
\end{figure}

To see more clearly the comparison of the model predictions and the simulation results, in Figure \ref{fig:Fisher_mod_vs_sim}, we show the Fisher results from high peaks for the case of $n_g=30\hbox{ arcmin}^{-2}$ and
$\theta_G=3\hbox{ arcmin}$. There the simulation results (solid) and the model predictions with (dashed) and without (dotted) LSS are presented. We can see the consistent trends between simulation results and the model predictions including LSS for both height and steepness statistics. Without LSS in the model, the contours are much larger. 

The results for the {\it High} samples with $\theta_G=2\hbox{ arcmin}$ are shown in Figure \ref{fig:Fisher_high_2sm}. With the similar trends as that of Figure \ref{fig:Fisher_high_3sm}, it is seen that the differences between the two peak statistics are less than their counterparts with $\theta_G=3\hbox{ arcmin}$ because
of the higher shape noise with the smaller smoothing scale given a same $n_g$.  

All together, our theoretical model analyses show that the different dependences on the halo profile and on the large-scale structure effects lead to the differences of the cosmological inferences between the two statistics for high peaks. For lower peaks, they mostly arise from the projection effects of large-scale structures and the shape noise. For the former, it contains cosmological information, and should be described as a random field although the Gaussian approximation adopted in our model for high peaks is not valid to predict low peaks. Thus for the {\it All} samples including low peaks, expectedly, it is the different dependences on the large-scale projection effects that play a dominant role in explaining the differences of the two peak statistics.

\section{Summary and Discussions}\label{chapter6}
In this study, we perform detailed comparisons of the peak statistics based on their height and steepness both from simulations and from our theoretical model. The main results are summarised as follows. 

\begin{itemize}
  \item [(1)]
   For peaks, their height and steepness show a significant positive correlation, which is stronger with the decrease of the shape noise. However, the correlation is not perfect even in the noiseless case, showing that the cosmological information embedded in the two peak statistics is not identical. With the help of our halo model for high peaks, we show that the different dependences on the halo density profile adding the random nature of the large-scale projection effects result in the not-perfect correlation between the peak height and steepness in the noiseless case. Including the shape noise further reduces the correlation.

  \item [(2)]
   From $\Delta\chi^2$ and Fisher analyses with simulated data, we systematically investigate the cosmological inferences from the two peak statistics by considering different number densities of galaxies $n_g$ and different smoothing scales $\theta_G$. The results show that for both the {\it All} samples and {\it High} samples, the steepness statistics tend to give better cosmological constraints than that of the peak height statistics. This advantage increases with the increase of $n_g$ and $\theta_G$ and thus the decrease of the shape noise, consistent qualitatively with the conclusion from \cite{2019NatAs...3...93R}.

  \item [(3)]
  To investigate the physical origin of the differences of the two peak statistics, we perform Fisher studies using our model for high peaks and compare with that from simulated high peaks. Again, the different dependence on the halo density profile matters. Particularly, for the steepness, it is associated with the second derivatives $K_{H}^{ij}$ that depends sensitively on the M-c relation of haloes. Even without considering the cosmological dependence of the M-c relation, its redshift dependence makes the two peak statistics sensitive to the halo mass function at different redshifts, which in turn affect their cosmological dependences. Furthermore, the projection effects from large-scale structures also contribute to the differences because of the different dependences on $\sigma_{\rm{LSS},i}$ of the two peak statistics. For the {\it All} samples where the low peaks are dominantly from the random fields of large-scale structures and the shape noise rather than from massive haloes, the different dependences on the large-scale projection effects play important roles in explaining the differences of the two peak statistics.  

\end{itemize}

We note again that comparing with the simulation results, our model for high peak steepness statistics shows a less accuracy than the height statistics. This reflects their different sensitivities to the physical inputs in the model, including the halo profile, the project effects of large-scale structures, etc., as discussed above. To derive high precision cosmological constraints from observed high peak steepness statistics, we therefore need to further improve our theoretical model. On the other hand, the sensitivities of the steepness statistics to the physical quantities also provide a possibility to constrain them together with the cosmological parameters from observed peak counts. We will pursue along this line in our future studies.

It is also pointed out that our studies here do not take into account possible effects from different systematics. They may affect the two statistics differently. For example, the mask effects from bad data have been analysed for the peak height statistics in \cite{2014ApJ...784...31L}. There we find that the very low number density of galaxies near masks affects the peak height statistics significantly. By excluding certain regions around them can effectively mitigate the mask effects at the expense of a fraction of the survey area \citep{2015MNRAS.450.2888L, 2018MNRAS.474.1116S}. However, it is not clear whether the same exclusions are sufficient for the peak steepness statistics, which deserves further careful studies. 

Furthermore, the boost effect from cluster member galaxies in a shear catalogue can lead to a more sophisticated effect on the peak steepness than on the height. This is because these member galaxies are not uniformly distributed but typically follow a profile that increases towards the central part of their host clusters. Thus the inclusion of them in the analyses can change the peak profiles originated from the clusters, which should depend sensitively on the spatial distribution of the member galaxies. For the peak height statistics, by estimating the boost factor using known clusters in the survey area, we can include the dilution effect in our model to control the bias induced from it \citep[e.g.,][]{2018MNRAS.474.1116S}. We do not expect that such a correction is directly applicable for the steepness statistics because of the profile dependence discussed above. Similar complications exist for the effects from galaxy intrinsic alignments. Studies have shown that for peaks, especially high peaks, the intrinsic alignments of cluster member galaxies play an important role to affect their statistics \citep[]{2016MNRAS.463.3653K,2022ApJ...940...96Z}. Thus the IA effect couples with the dilution effect, and should also depend on the member galaxy distribution within their host clusters. 

To understand the systematic effects on the peak steepness statistics is crucially important for its applicability to real observational data, 
and that will be our major efforts in the forthcoming studies.

\section*{Acknowledgements}


We sincerely thank the referee for the encouraging and detailed comments, which help us to improve the paper considerably. This study is supported by the NSFC grant No. 11933002. XKL and ZHF acknowledge the supports from the NSFC grant No. U1931210 and a grant from CAS Interdisciplinary Innovation Team. ZHF is also supported by the grant from the China Manned Space Projects with No. CMS-CSST-2021-A01. XKL also acknowledges the supports from NSFC of China under Grant No. 12173033, YNU Grant No. C176220100008, and the research grants from the China Manned Space Project with No. CMS-CSST-2021-B01. The cosmological simulations were mainly conducted on the Yunnan University Astronomy Supercomputer.

\section*{Data Availability}
The data underlying this article will be shared on reasonable request to the corresponding author.




\bibliographystyle{mnras}
\bibliography{peak_steepness} 




\appendix

\section{An approximation of $F(x_N)$}\label{sec:fx series}
In our model predictions with Eq.(\ref{eq:height number density}) and Eq.(\ref{eq:steepness number density}), the function $F(x_N)$ is involved, which is given by Eq.(\ref{eq:fx}). Here we provide a numerical approximation for its calculation.

We rewrite the definition of $F(x_N)$
\begin{equation}
\begin{aligned}
F\left(x_N\right)=& \exp(-C^2) \\
& \times \int_{0}^{1 / 2} d e_{N} \  8\left(x_N^{2} e_{N}\right) x_N^{2}\left(1-4 e_{N}^{2}\right) \exp \left(-4 x_N^{2} e_{N}^{2}\right) \\
& \times \int_{0}^{\pi} \frac{d \theta_{N}}{\pi} \exp \left[-4 x_N e_{N} C \cos 2 \theta_{N}\right],
\end{aligned}
\end{equation}
where $C=(K_H^{11}-K_H^{22})/\sigma_2$. Assuming spherical haloes, we have $C=(K_H^{rr}-K_H^r/r)/\sigma_2$. It is noted that the integral for $e_N$ has an upper limit of $1/2$, which is smaller than unity. Thus an approximation using the Taylor expansion is feasible. 
For that, we define a function
\begin{equation}
\begin{aligned}
f\left(y\right)=& \exp(-C^2) \\
& \times \int_{0}^{y} d e_{N} \  8\left(x_N^{2} e_{N}\right) x_N^{2}\left(1-4 e_{N}^{2}\right) \exp \left(-4 x_N^{2} e_{N}^{2}\right) \\
& \times \int_{0}^{\pi} \frac{d \theta_{N}}{\pi} \exp \left[-4 x_N e_{N} C \cos 2 \theta_{N}\right]
\end{aligned}
\end{equation}
and perform the Taylor expansion around $y=0$. After the expansion, we set $y=1/2$ and obtain $F(x_N)$ expressed by a summation of a series of terms. Specifically, it is given by 
\begin{equation}
\begin{aligned}
F(x_N)=&e^{-C^2}\bigg[(x_N^2-1+e^{-x_N^2})\\
              & +\sum_{m=1}^{\infty}C^{2m}\sum_{n=m+2}^{\infty} (-1)^m\frac{(-x_N^2)^n}{n!} \binom{n}{m} \frac{(n-m)(n-m-1)}{n(n-1)m!}\bigg].
\end{aligned}
\end{equation}
After a mathematical manipulation, we finally obtain the following expression for $F(x_N)$
\begin{equation}\label{eq:fx series}
F(x_N)=x_N^2-C^2-1+(1+C^2)e^{-x_N^2}+C^2x_N^2e^{-x_N^2}+\sum_{k=0}^{\infty}G_k\frac{x_N^{2(k+2)}}{(k+2)!}e^{-x_N^2},
\end{equation}
where 
\begin{sequation}
G_k=C^2-(k+1)-e^{-C^2}\left[\sum_{m=0}^{k}(C^2-k-1)\frac{C^{2m}}{m!}-(k+1)\frac{C^{2(k+1)}}{(k+1)!}\right],
\end{sequation}
which satisfies the following recurrence relation
\begin{sequation}
G_{k-1}-G_k=1-e^{-C^2}\sum_{m=0}^{k}\frac{C^{2m}}{m!}.
\end{sequation}

For the numerical calculations, we find that truncating the summation Eq.(\ref{eq:fx series}) at a value of $k>C^2$ gives rise to a good enough accuracy for $F(x_N)$. We apply this approximation in our model calculations.  

\section{Steepness operator}\label{sec:Fisher L vs L2}

In our analyses, we apply the discrete operator Eq.(\ref{steepness}) to calculate the steepness of a peak in 
a convergence map. Here we show mathematically that it indeed gives rise to an estimate of $-(K_N^{11}+K_N^{22})$ in 
unit of pixel$^{-2}$. Specifically, we have
\begin{equation}
\begin{aligned}
\left[S(x,y)\right]=&4 K_N(x,y)\\
&-\left[K_N(x+h,y)+K_N(x-h,y)\right]\\
&-\left[K_N(x,y+h)+K_N(x,y-h)\right],
\end{aligned}
\end{equation}
where $h=1$ pixel. Under the second-order Taylor expansion, we have
\begin{equation}
\begin{aligned}
K_N(x+h,y)&+K_N(x-h,y)\\
&\approx K_N(x,y)+K_N^{1}(x,y)h+\frac{1}{2}K_N^{11}(x,y)h^2\\
             &+K_N(x,y)-K_N^{1}(x,y)h+\frac{1}{2}K_N^{11}(x,y)h^2\\
             &=2K_N(x,y)+K_N^{11}(x,y)h^2,
\end{aligned}
\end{equation}
where $K_N^{1}$ and $K_N^{11}$ are the first- and second-order derivatives of $K_N(x,y)$ with respect to $x$. Similarly,
\begin{equation}
K_N(x,y+h)+K_N(x,y-h)
             =2K_N(x,y)+K_N^{22}(x,y)h^2.
\end{equation}
We therefore obtain
\begin{equation}
\left[S(x,y)\right]\approx-[K_N^{11}+K_N^{22}]h^2.
\end{equation}
With $h=1$ pixel, $[S]$ is an estimate of $-[K_N^{11}+K_N^{22}]$ in unit of pixel$^{-2}$.

\begin{figure}
    \includegraphics[scale=0.5]{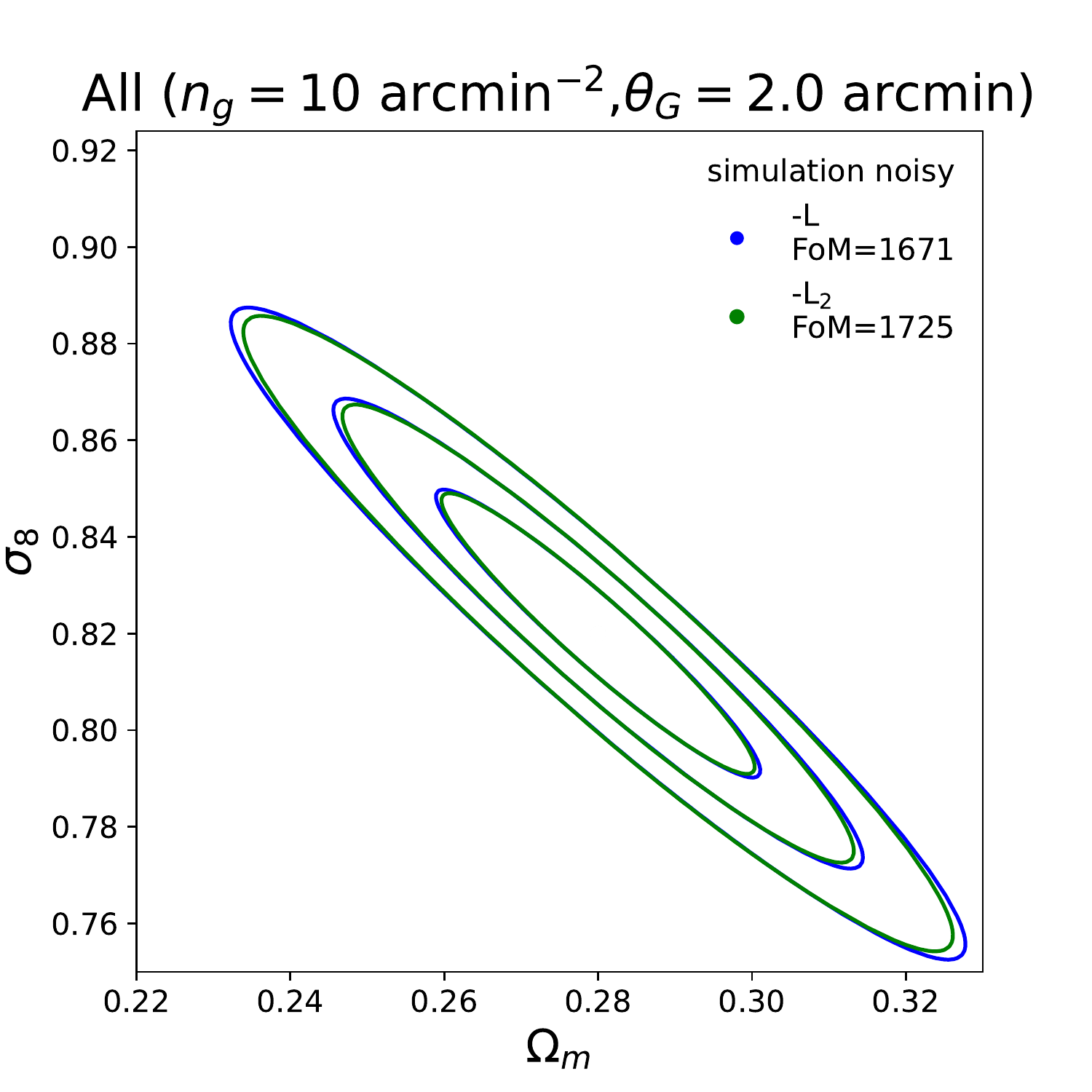}
    \caption{Comparisons between the Fisher results from the peak steepness statistics with $-L$ (blue) and $-L_2$ (green) calculations. They are noisy case from the {\it All} sample with $n_g=10\hbox{ arcmin}^{-2}$ and $\theta_G=2\hbox{ arcmin}$. The meanings of the contours are the same as in Figure \ref{fig:Fisher_all}.}
    \label{fig:Fisher_L1vsL2}
\end{figure}

To further show the impact of different steepness calculations, $-L$ and $-L_2$, we present in Figure \ref{fig:Fisher_L1vsL2} the corresponding Fisher results for the noisy case of $n_g=10\hbox{ arcmin}^{-2}$ and $\theta_G=2\hbox{ arcmin}$ from {\it All} sample. It is seen that they are nearly identical, showing a negligible effect from the different steepness operators when the smoothing scale is much larger than the pixel scale.

\section{The influence of the discrepancy between the results of simulations and the model in small x peaks}\label{modeltest}
\begin{figure}
    \includegraphics[scale=0.5]{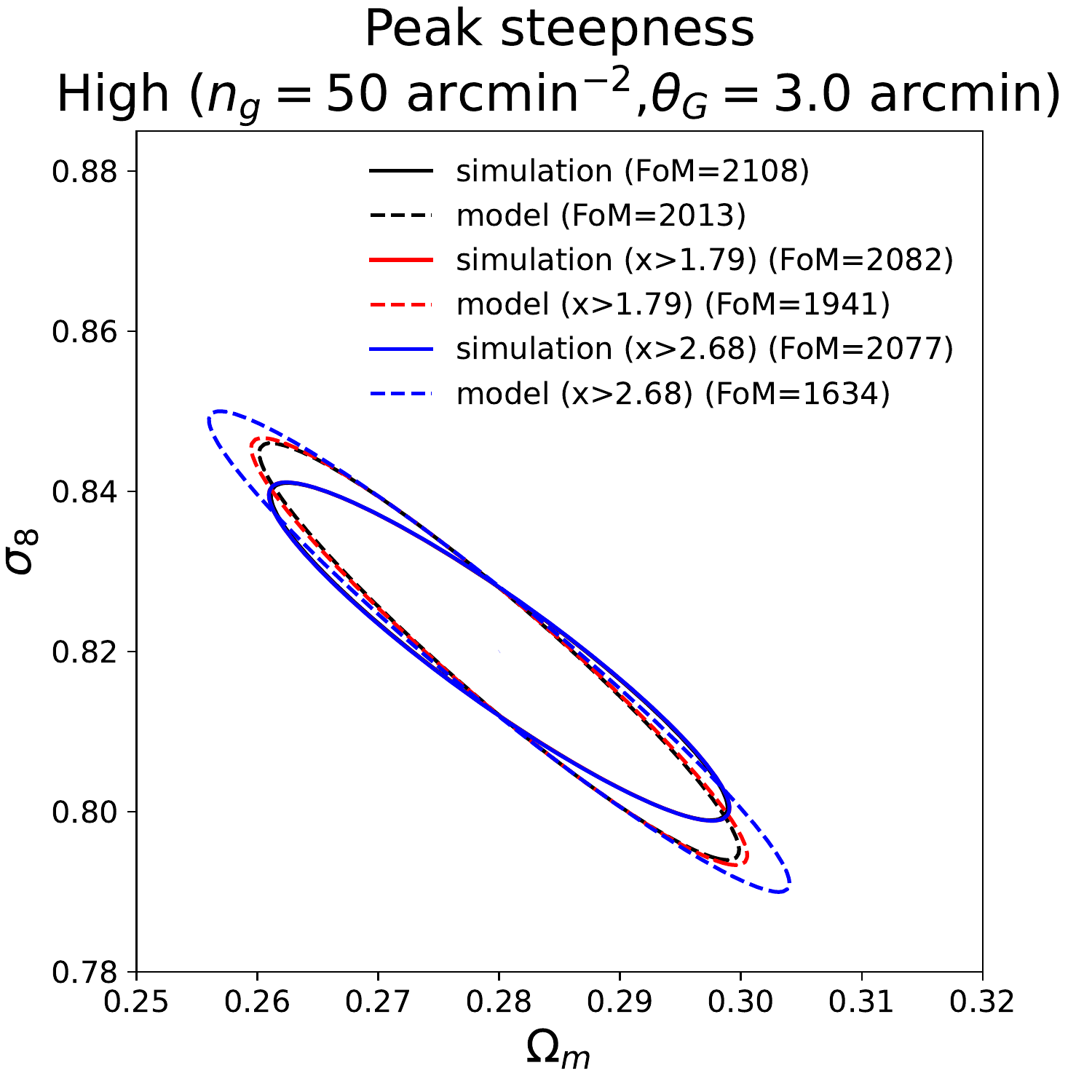}
    \caption{Comparisons between the Fisher results (68\% confidence level) from the peak steepness statistics from the original $\it{High}$ sample (black) with $n_g=50$ arcmin$^{-2}$ and $\theta_G=3$ arcmin and the results excluding the counts of the first one (red) or two bins (blue) at $x<5$ (see the lower right panel of Figure \ref{fig:peak model distribution}). The solid and dashed lines are for the simulation and the model results, respectively.}
    \label{fig:Fisher_ng50_3sm_test}
\end{figure}

In the lower right panel of Figure \ref{fig:peak model distribution} with $n_g=50\hbox{ arcmin}^{-2}$ and $\theta_G=3\hbox{ arcmin}$, relatively large discrepancies between simulation results and the model calculations at $x<5$ are seen. We analyse if the disagreements can affect our conclusions about the two peak statistics significantly. Here in accord with the lower right panel of Figure 8, we divide the first $x$ bin of $[0.0,2.0]\times \sqrt{5}$ shown in Table \ref{table:binning for High sample} into four narrow bins of $[0.0,0.8]\times \sqrt{5}, [0.8,1.2]\times \sqrt{5}, [1.2,1.6]\times \sqrt{5}$ and $[1.6,2.0]\times \sqrt{5}$, and perform Fisher analyses accordingly. In Figure \ref{fig:Fisher_ng50_3sm_test}, we show the test results to compare the peak steepness Fisher contours ($68\%$ levels) from the full {\it High} sample and those without the contributions from the first one or two narrow bins. The corresponding samples include peaks with steepness $x>0$, $x>1.79$ and $x>2.68$ respectively. It is seen that for both the results of simulations and the model, the Fisher contours do not change considerably. It indicates that the low-x peaks contain relatively minor cosmological information, and the discrepancies there between simulation results and the model predictions should not affect our conclusions qualitatively. As one of our major efforts in future, we will further improve the model for the steepness statistics.

\section{The Fisher results from the {\it High} samples with $\theta_G=2$ arcmin}
\begin{figure*}
    \flushleft
    \includegraphics[ scale=0.285]{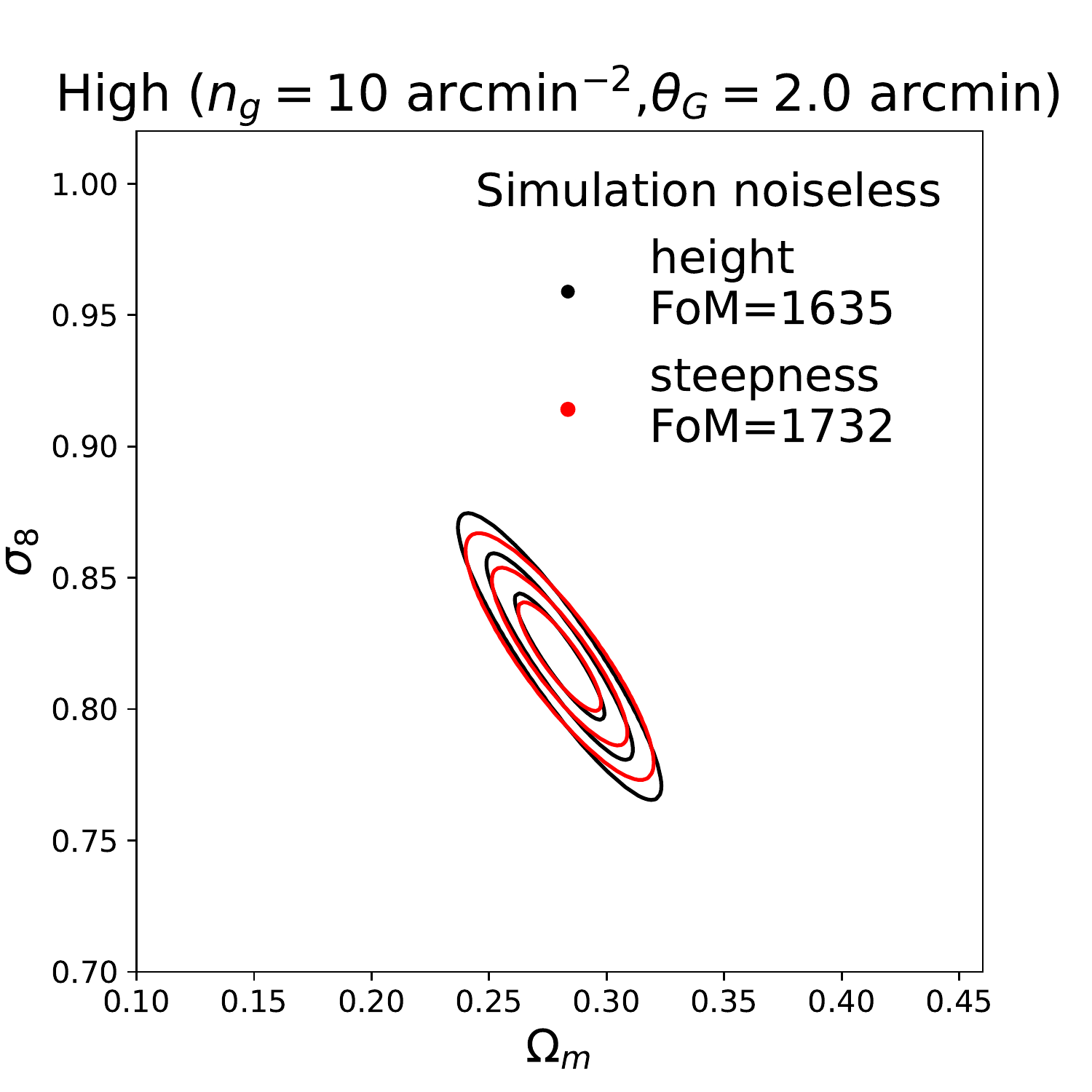}
    \includegraphics[ scale=0.285]{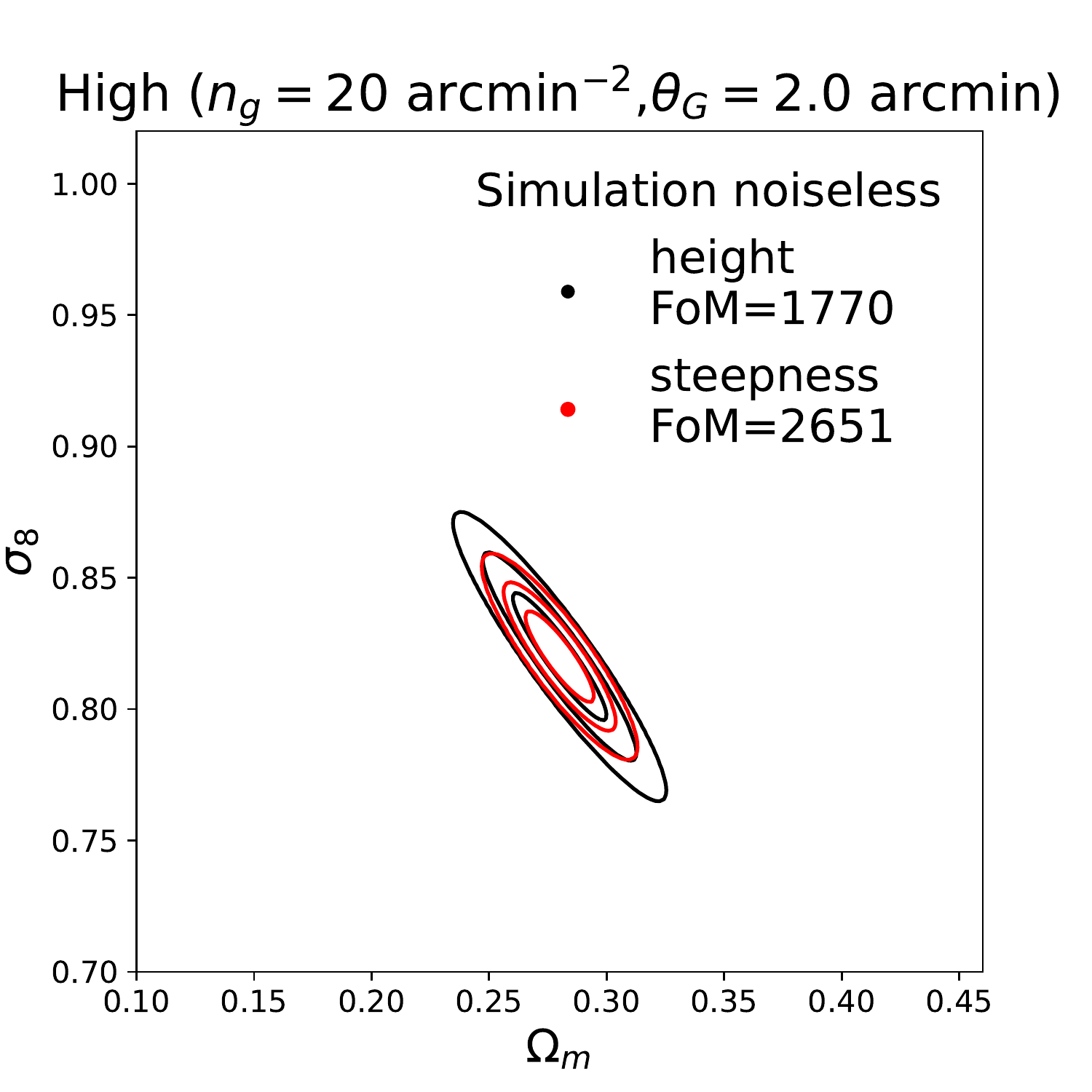}
    \includegraphics[ scale=0.285]{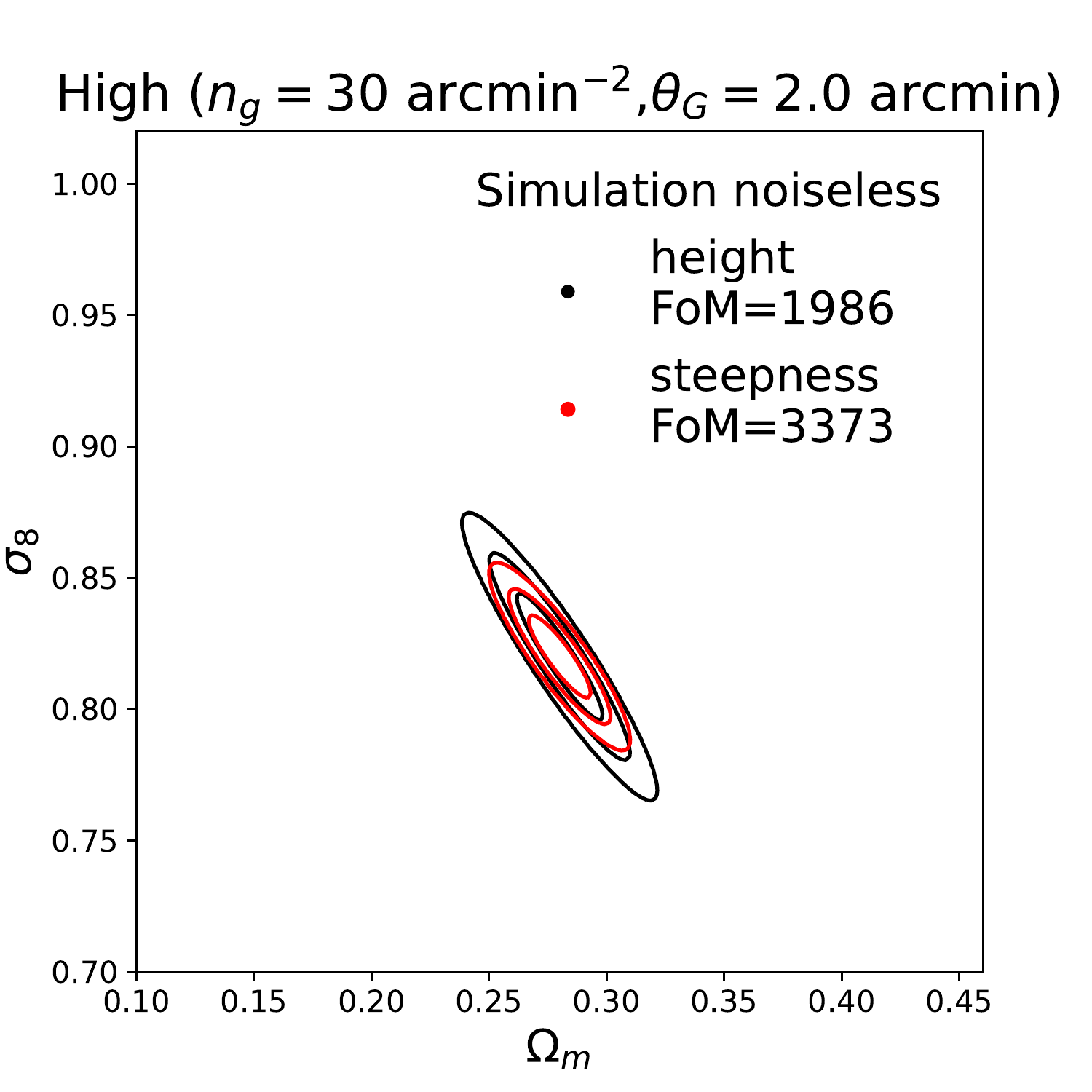}
    \includegraphics[ scale=0.285]{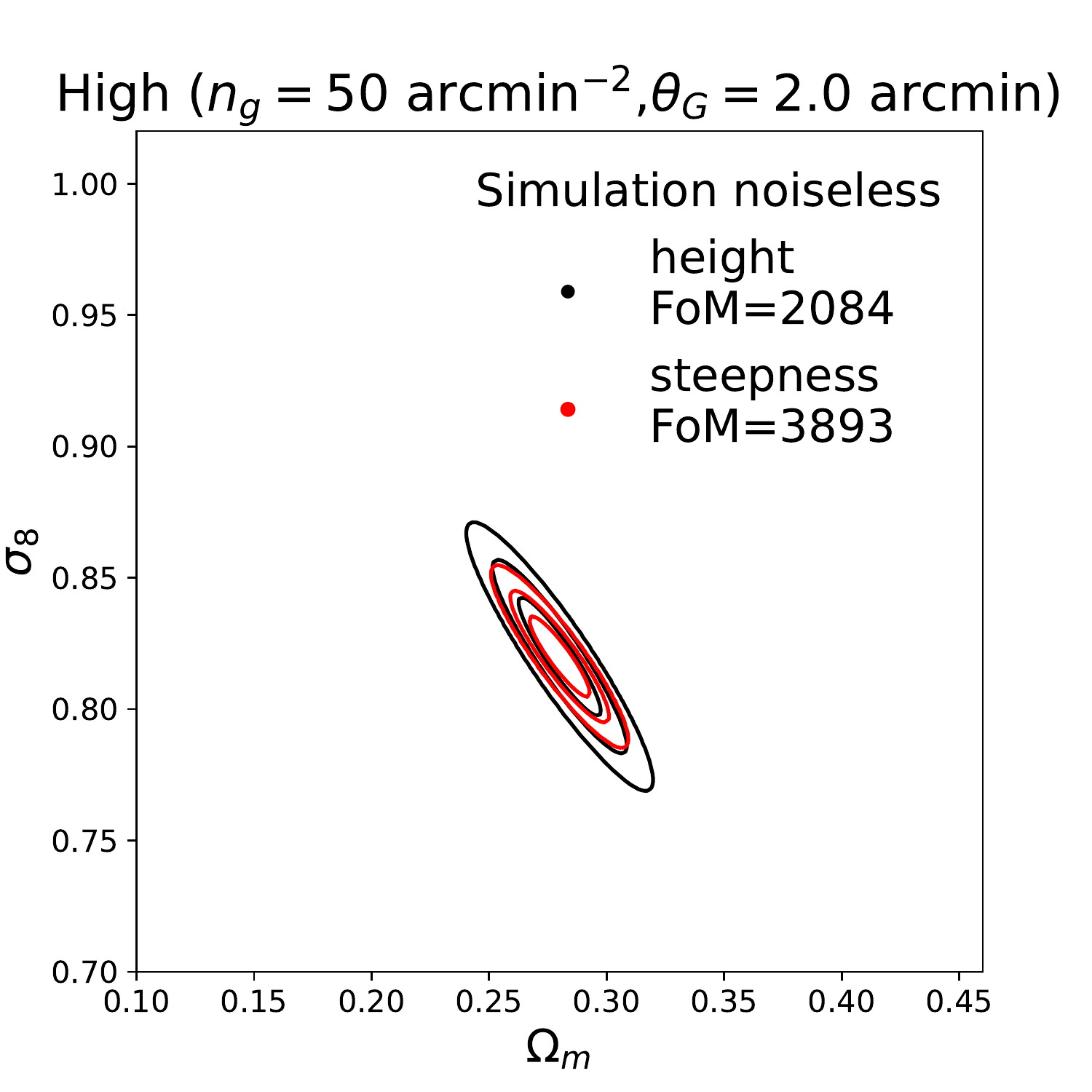}\\
    \flushleft
    \includegraphics[ scale=0.285]{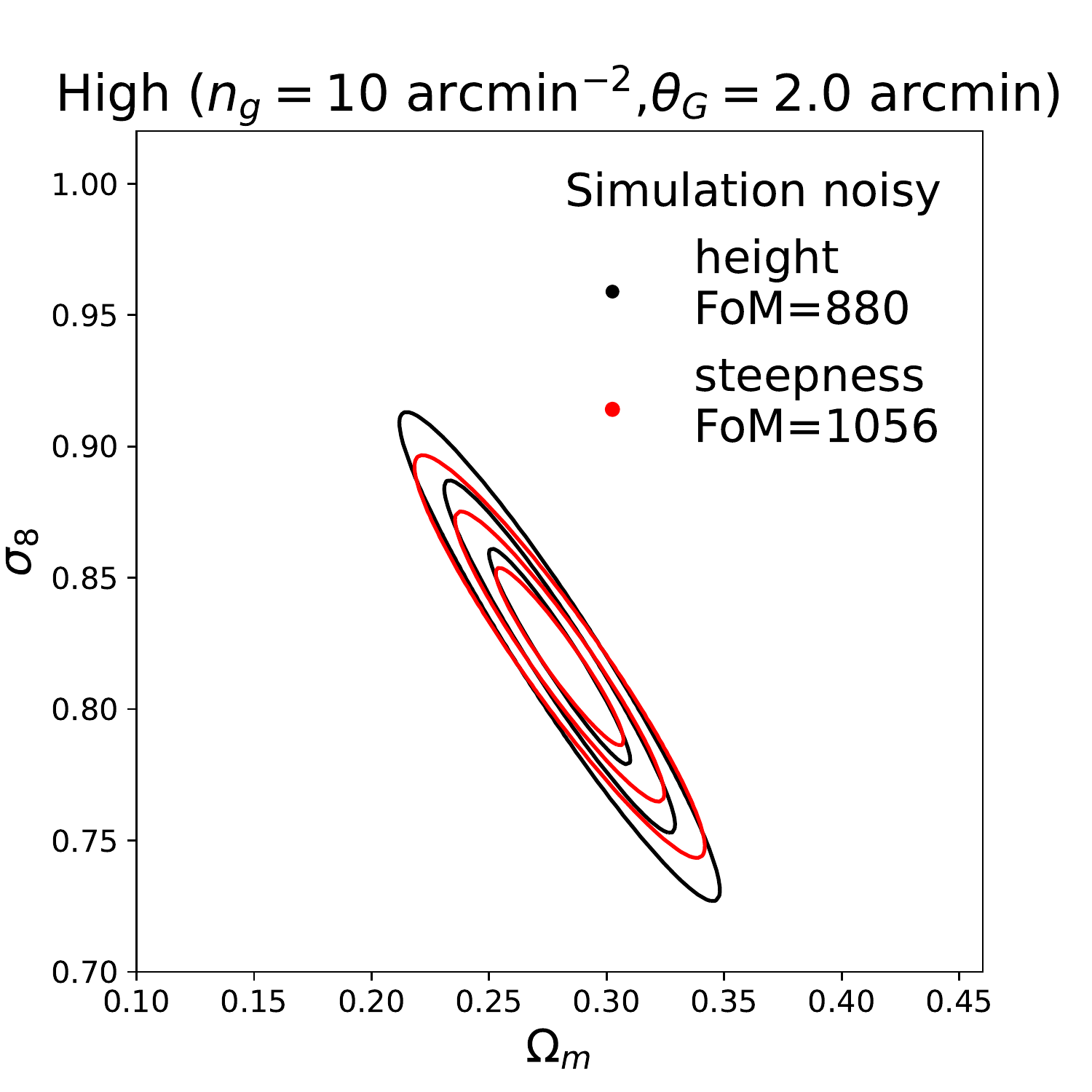}
    \includegraphics[ scale=0.285]{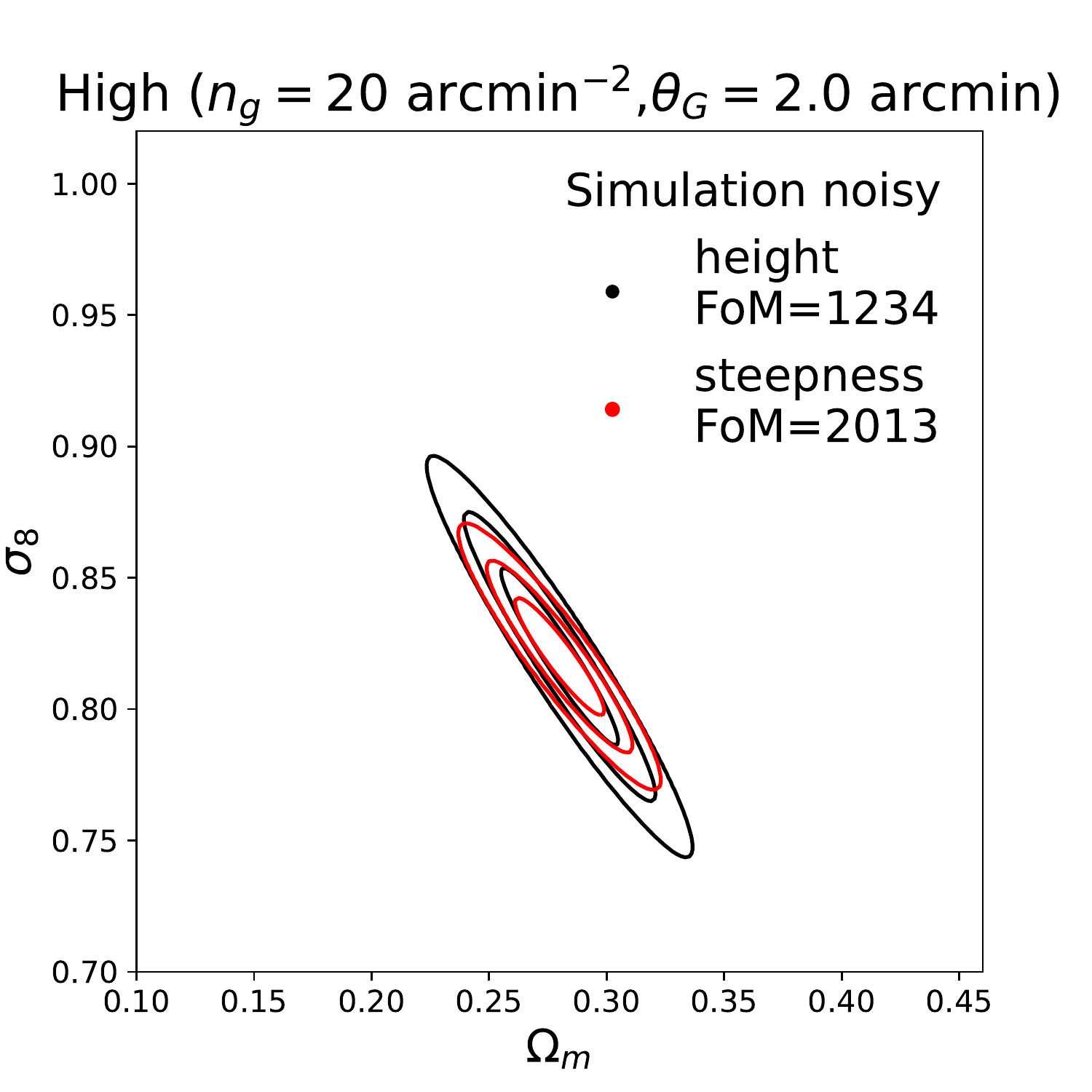}
    \includegraphics[ scale=0.285]{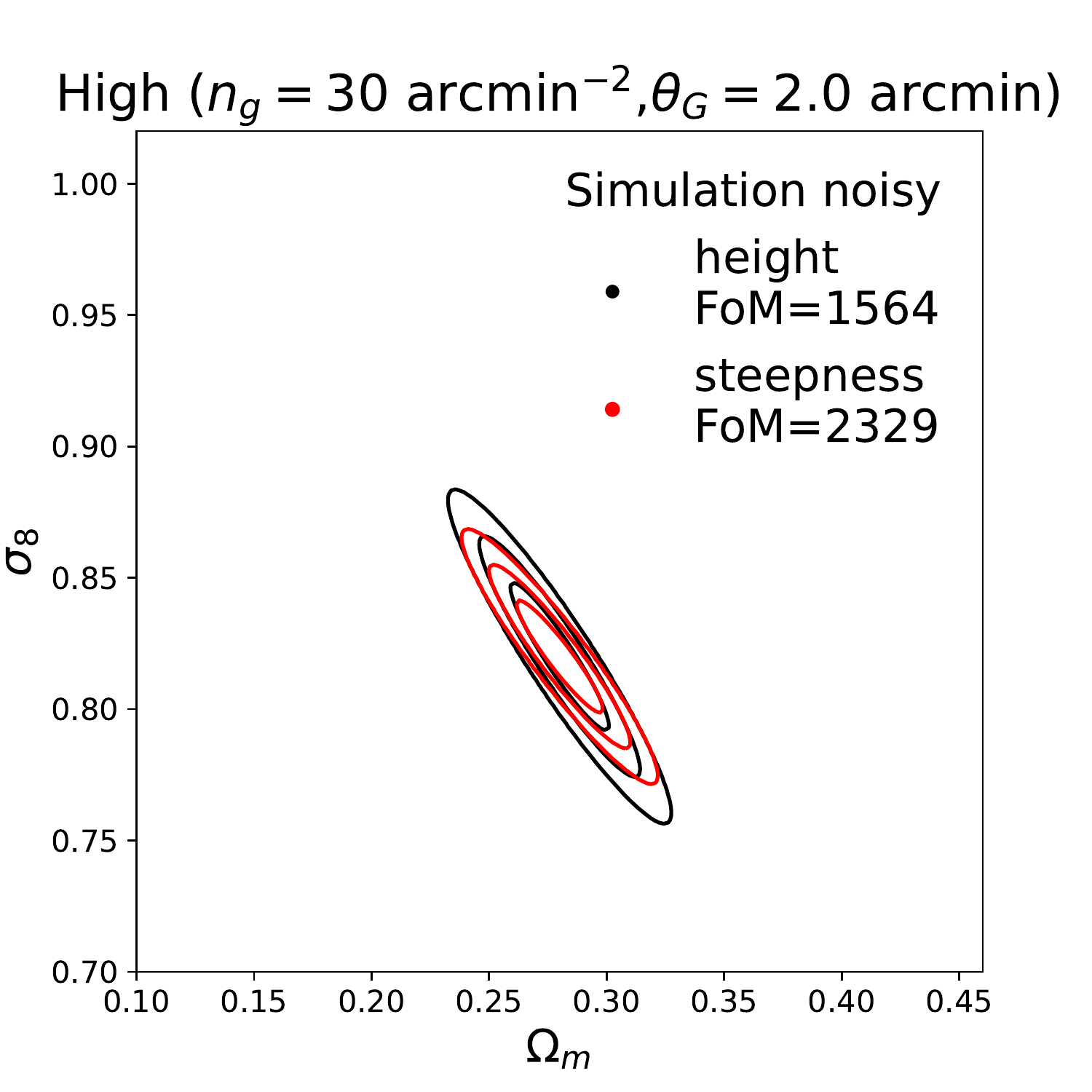}
    \includegraphics[ scale=0.285]{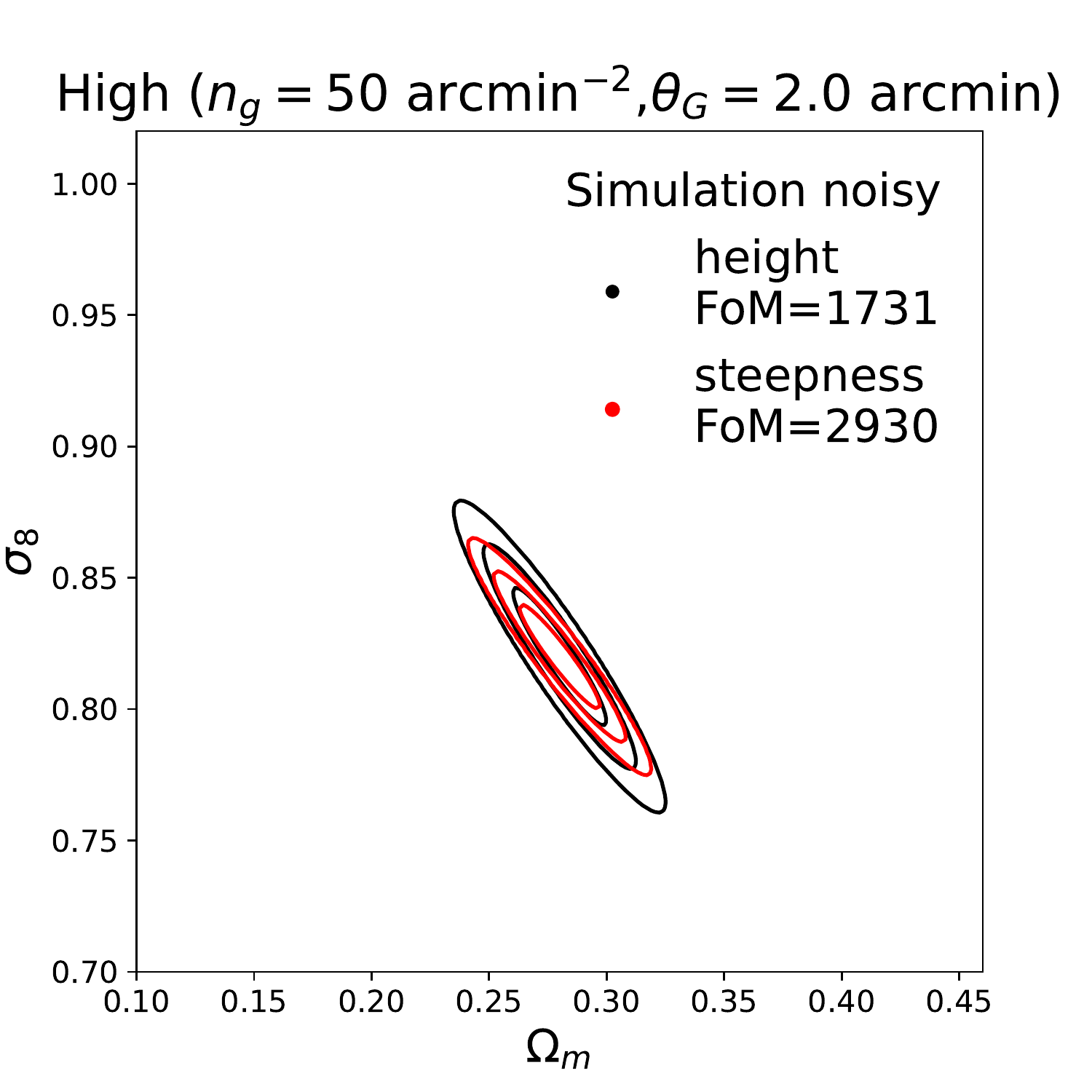}\\
    \flushleft
    \includegraphics[ scale=0.285]{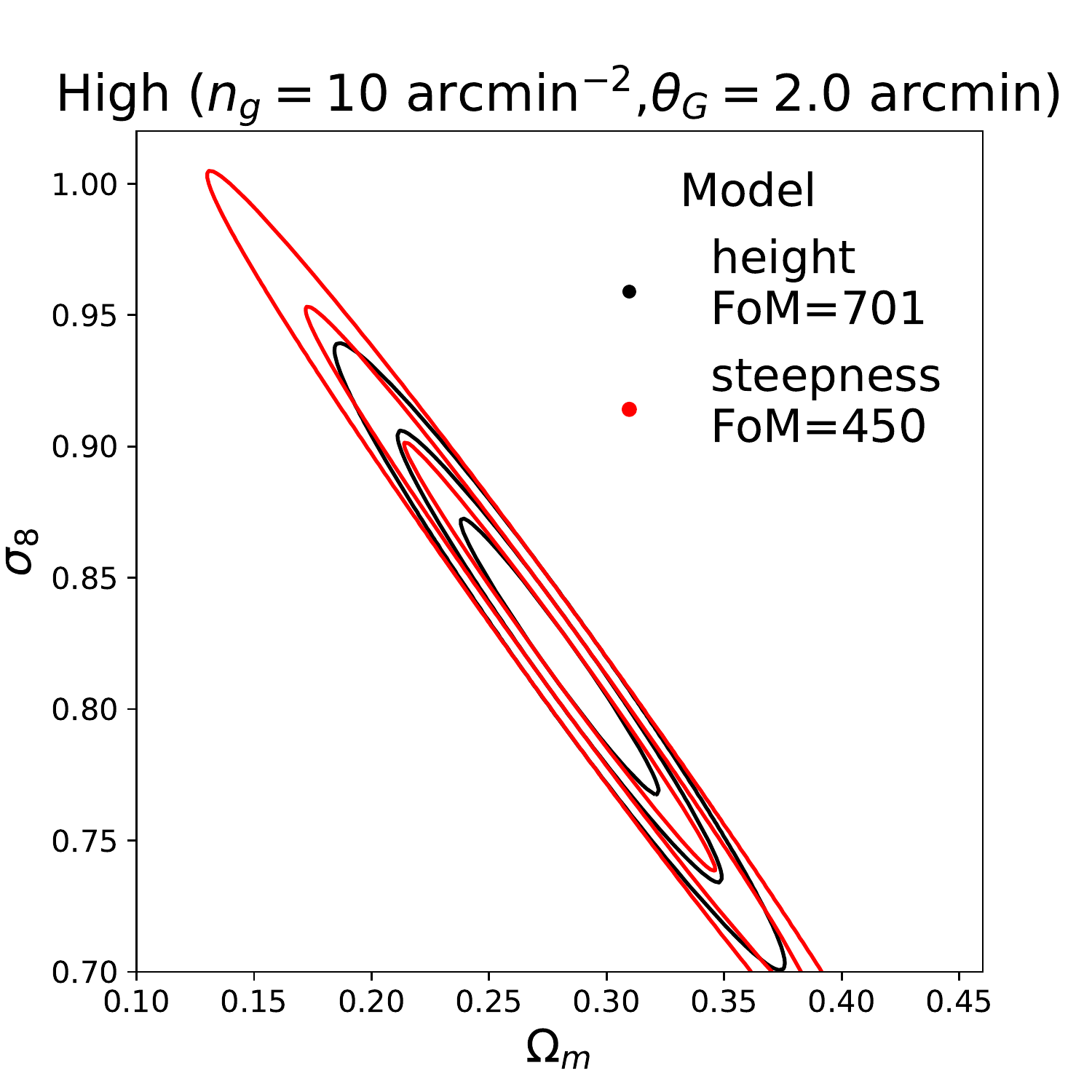}
    \includegraphics[ scale=0.285]{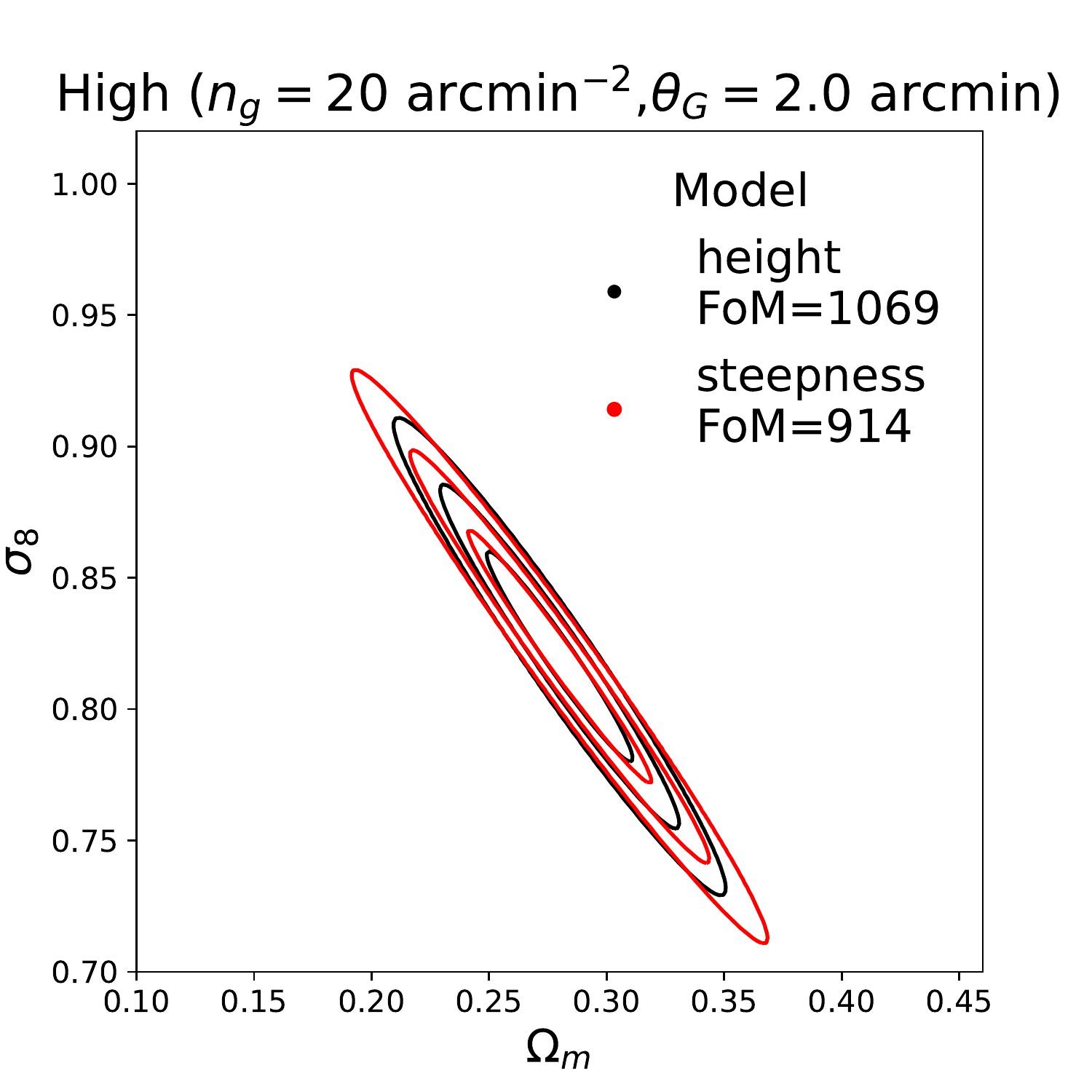}
    \includegraphics[ scale=0.285]{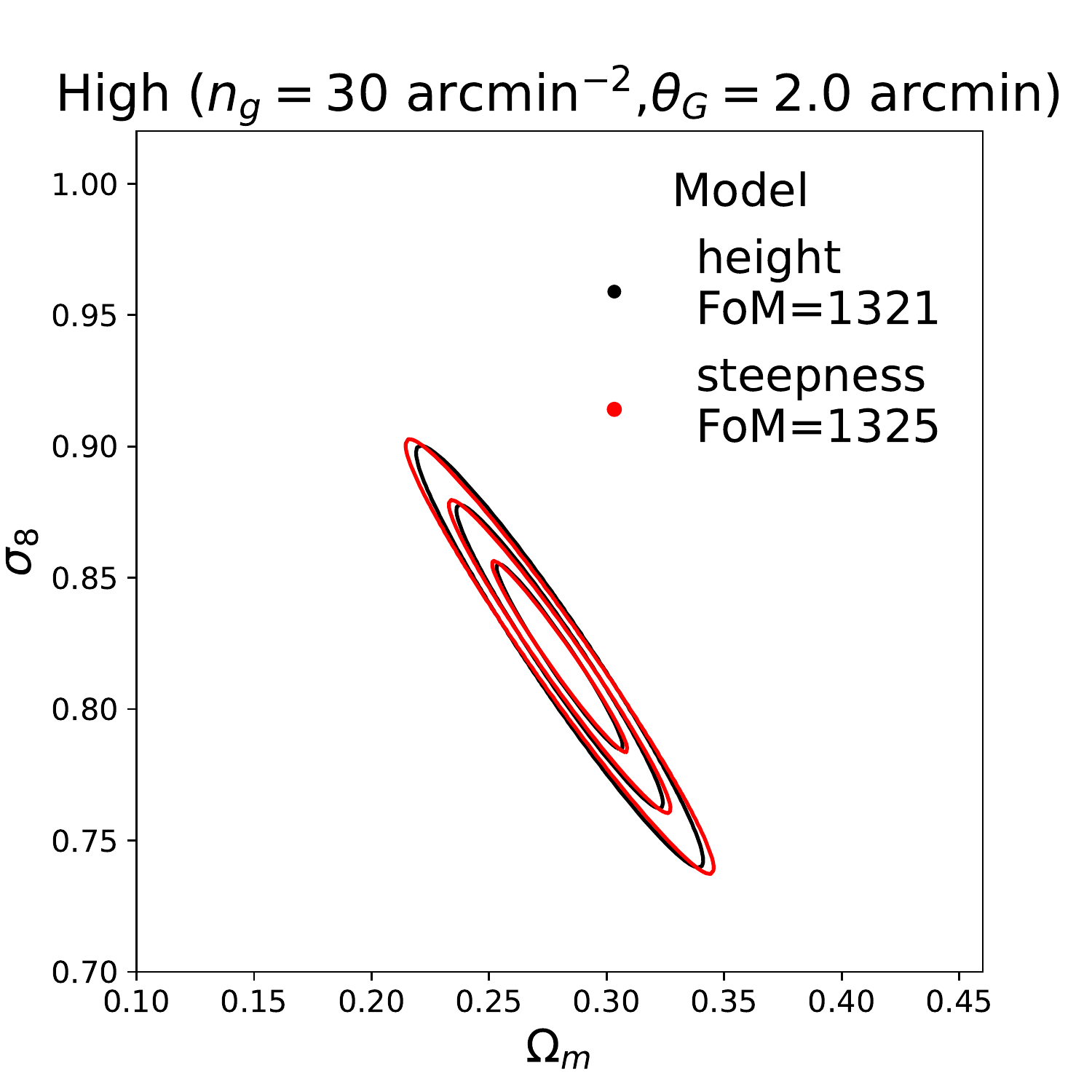}
    \includegraphics[ scale=0.285]{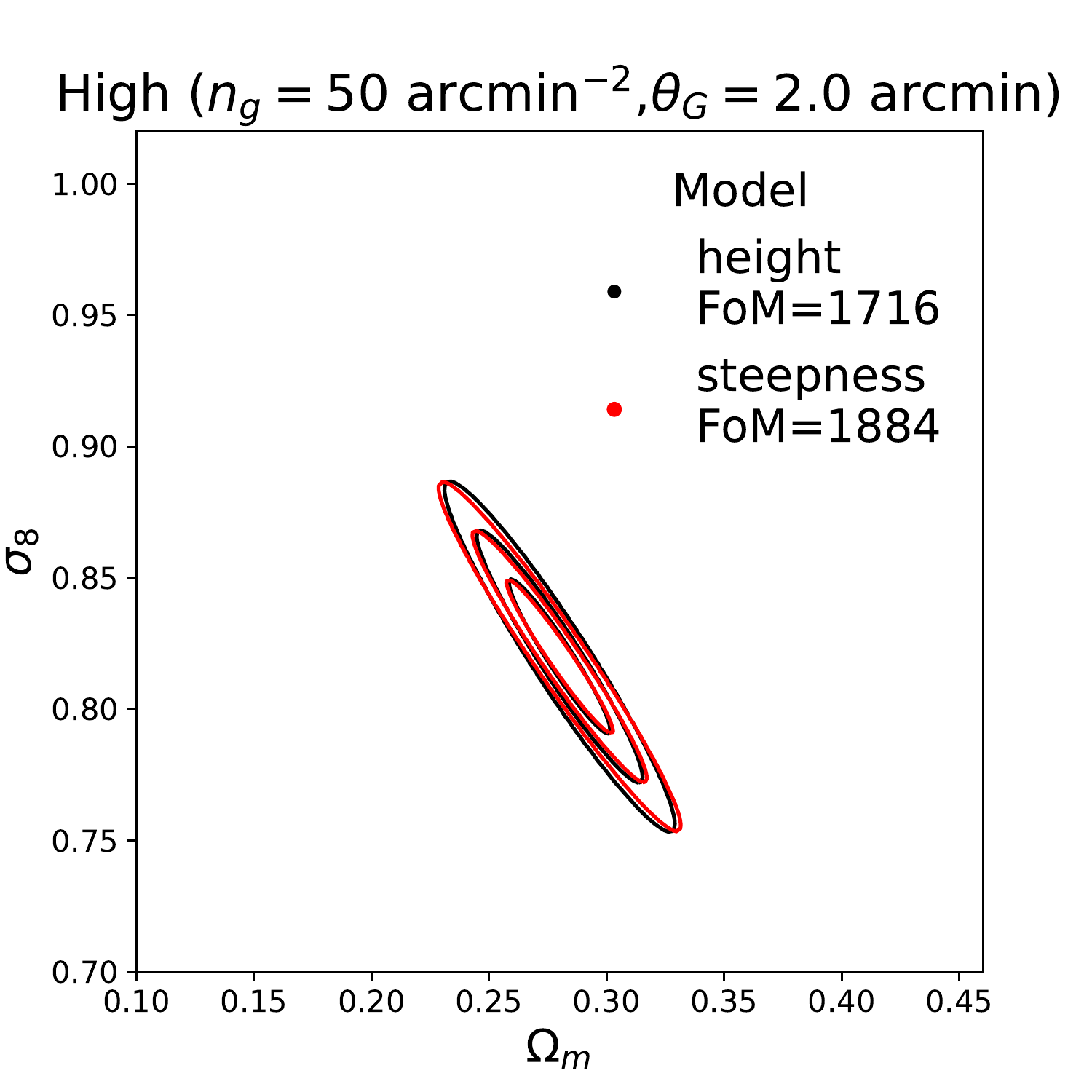}\\
    \flushleft
    \includegraphics[ scale=0.285]{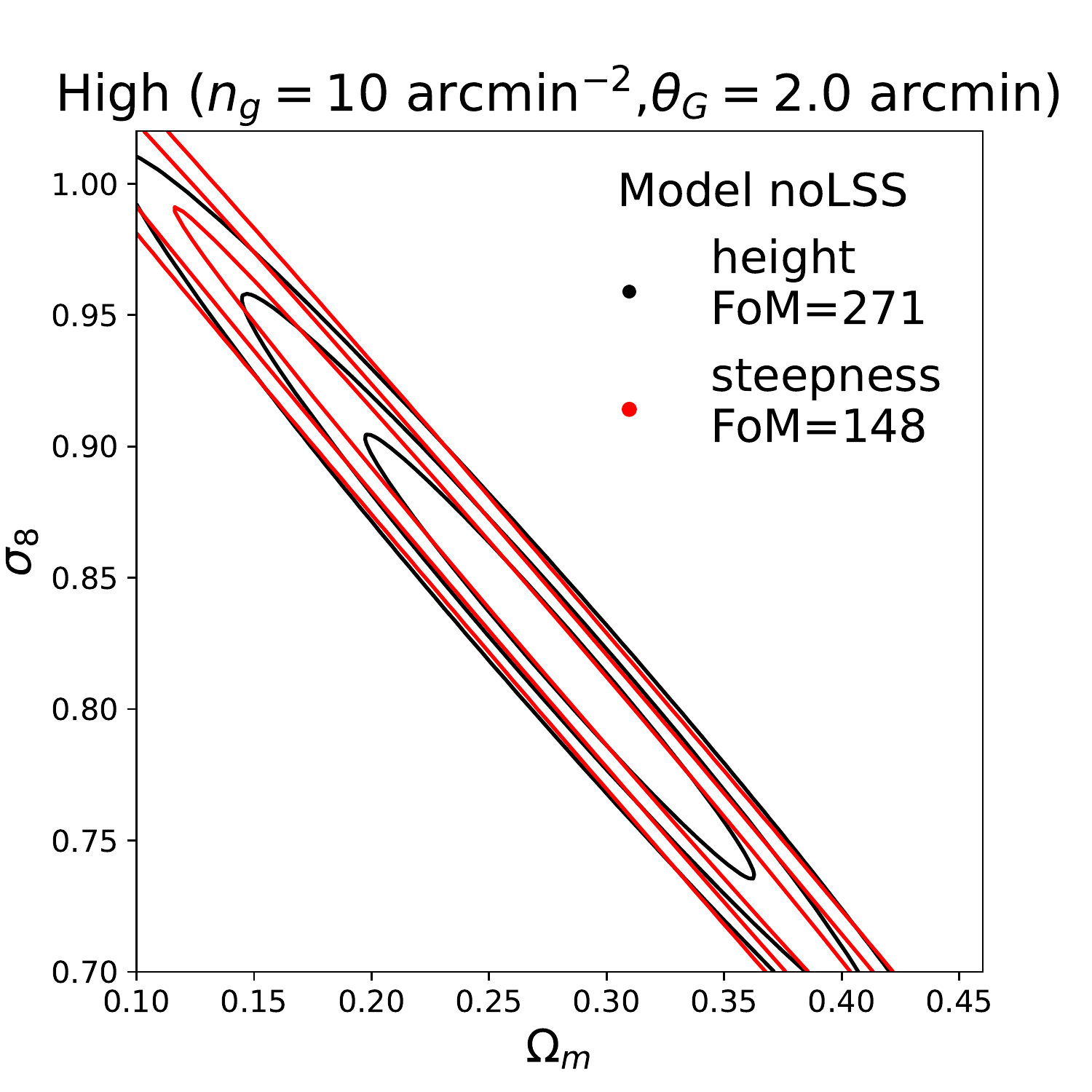}
    \includegraphics[ scale=0.285]{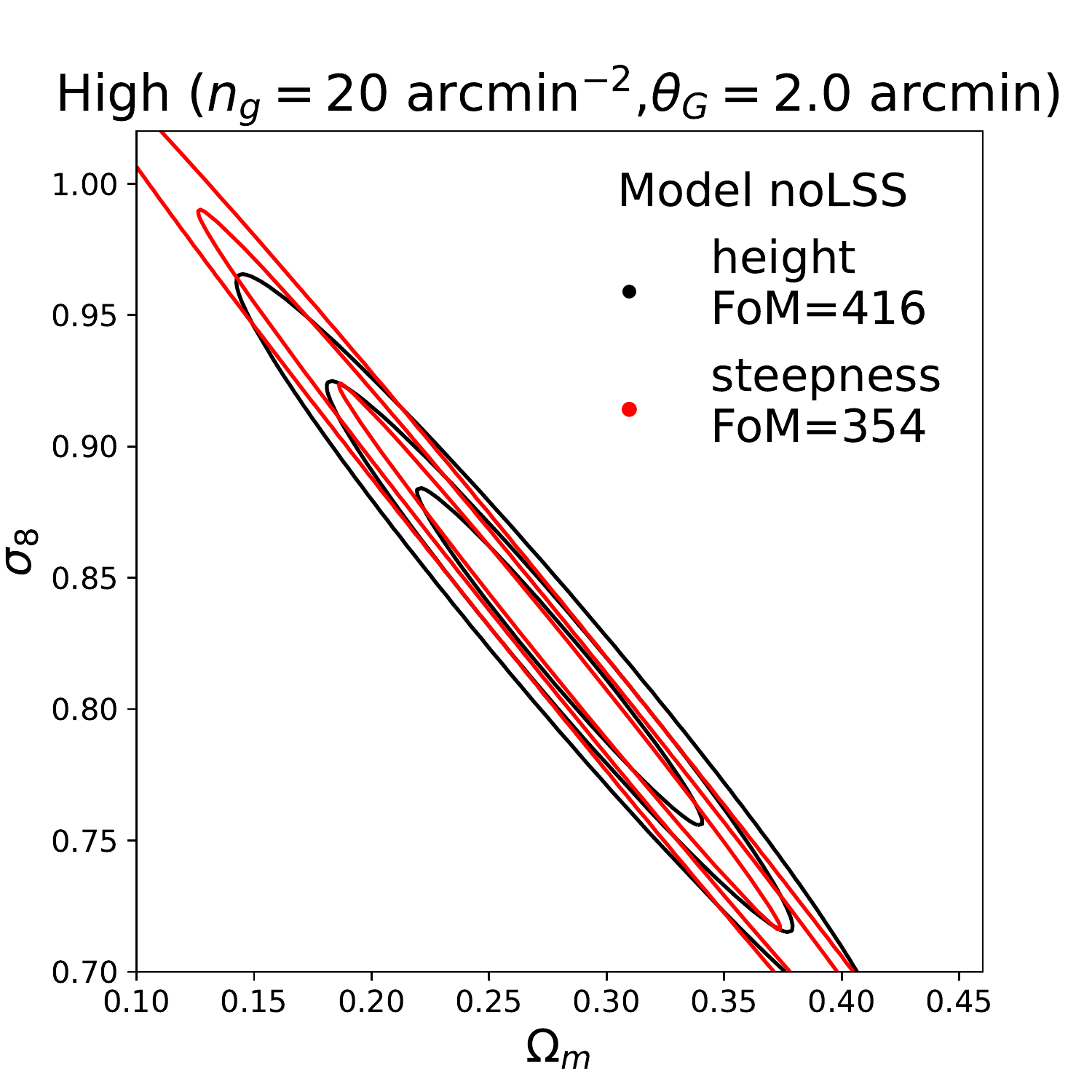}
    \includegraphics[ scale=0.285]{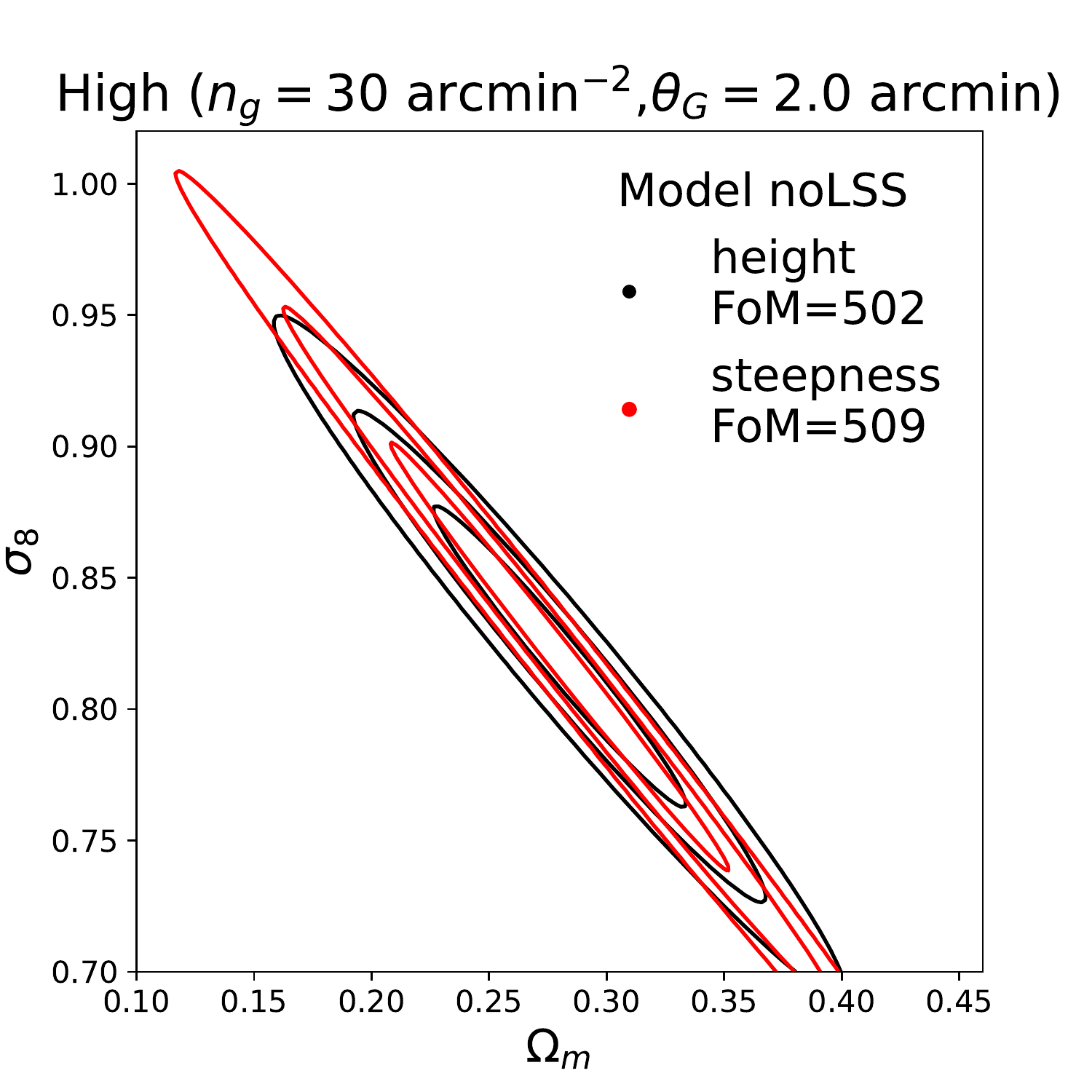}
    \includegraphics[ scale=0.285]{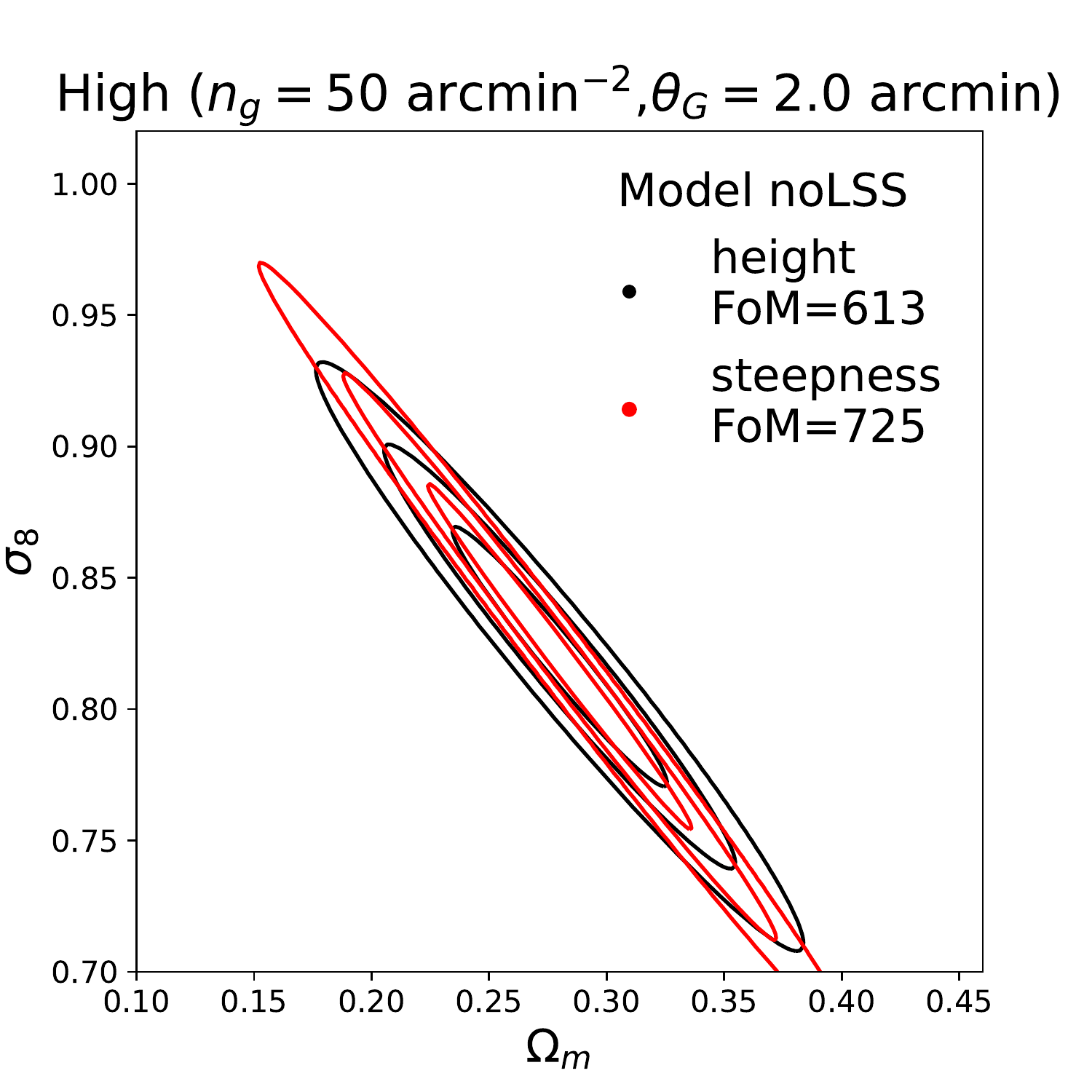}\\
    \caption{Similar to Figure \ref{fig:Fisher_high_3sm}, but with $\theta_G=2\hbox{ arcmin}$.} 
    \label{fig:Fisher_high_2sm}
\end{figure*}

For a comparison and the completeness, we show the Fisher analyses results for the {\it High} samples with $\theta_G=2\hbox{ arcmin}$ in Figure \ref{fig:Fisher_high_2sm}.
We see a similar trend as that in Figure \ref{fig:Fisher_high_3sm} with $\theta_G=3\hbox{ arcmin}$. On the other hand, the differences between the two peak statistics are systematically less here. This is because given a $n_g$, the shape noise is larger
for the smaller smoothing scale, which suppresses the advantage of the peak steepness statistics over its height counterpart as we discussed in the main text. 

\bsp	
\label{lastpage}
\end{document}